\documentclass[twocolumn,aps,prc,superscriptaddress,showpacs,floatfix,longbibliography,nofootinbib]{revtex4-1}
\usepackage{url}
\usepackage{cancel}
\usepackage[colorlinks,linkcolor=blue,citecolor=blue,filecolor=black,urlcolor=blue]{hyperref}
\usepackage{epsfig,graphics}
\usepackage{graphicx}
\usepackage{dcolumn}
\usepackage{bm}
\usepackage[usenames]{color}
\usepackage{amssymb}
\usepackage{amsmath}
\usepackage{multirow}
\usepackage{float}
\usepackage{harpoon}
\usepackage{MnSymbol}
\usepackage{appendix}
\usepackage{color}
\usepackage{hyperref}
\usepackage{cleveref}

\begin{document}

\title{Bayesian inference of neutron-star observables based on effective nuclear interactions}
\author{Jia Zhou}
\affiliation{Shanghai Institute of Applied Physics, Chinese Academy of Sciences, Shanghai 201800, China}
\affiliation{University of Chinese Academy of Sciences, Beijing 100049, China}
\author{Jun Xu}\email[Correspond to\ ]{junxu@tongji.edu.cn}
\affiliation{School of Physics Science and Engineering, Tongji University, Shanghai 200092, China}
\affiliation{Shanghai Advanced Research Institute, Chinese Academy of Sciences, Shanghai 201210, China}
\affiliation{Shanghai Institute of Applied Physics, Chinese Academy of Sciences, Shanghai 201800, China}
\author{Panagiota Papakonstantinou
}
\affiliation{Rare Isotope Science Project, Institute for Basic Science, Daejeon 34000, Korea}
\begin{abstract}
Based on the Skyrme-Hartree-Fock model (SHF) as well as its extension (the Korea-IBS-Daegu-SKKU (KIDS) model) and the relativistic mean-field (RMF) model, we have studied the constraints on the parameters of the nuclear matter equation of state (EOS) from adopted astrophysical observables using a Bayesian approach. While the masses and radii of neutron stars generally favors a stiff isoscalar EOS and a moderately soft nuclear symmetry energy, model dependence on the constraints is observed and mostly originates from the incorporation of higher-order EOS parameters and difference between relativistic and non-relativistic models. At twice saturation density, the value of the symmetry energy is constrained to be $48^{+15}_{-11}$ MeV in the standard SHF model, $48^{+8}_{-15}$ MeV in the KIDS model, and $48^{+5}_{-6}$ MeV in the RMF model, around their maximum {\it a posteriori} values within $68\%$ confidence intervals. Our study helps to obtain a robust constraint on nuclear matter EOS, and meanwhile, to understand the model dependence of the results.
\end{abstract}
\maketitle

\section{Introduction}
\label{sec:intro}

Compact stars are natural laboratories for investigating properties of dense nuclear matter. Observables of neutron stars, such as their masses, radii, as well as the gravitational waves emitted from the mergers of binary stars, are helpful for understanding the equation of state (EOS) of nuclear matter in both the isoscalar and isovector channels~\cite{Li:2021thg,Lattimer:2021emm}, or in other words, the binding energy per nucleon $E_{SNM}$ in isospin symmetric nuclear matter and the energy excess due to the finite isospin asymmetry characterized by the nuclear symmetry energy $E_{sym}$. For example, the mass of the neutron star is determined by the stiffness of the nuclear matter EOS, and the radius of the neutron star is closely related to the nuclear symmetry energy~\cite{Lattimer:2006xb}. Thanks to the pioneer studies by nuclear physicists, $E_{SNM}(\rho)$ and $E_{sym}(\rho)$ around the saturation density $\rho_0$ are better constrained, compared to those at suprasaturation densities. For instance, the incompressibility $K_0$ characterizing the stiffness of $E_{SNM}(\rho)$ is constrained within $220-260$ MeV from studies on the isoscalar giant monopole resonance (ISGMR)~\cite{PhysRevLett.109.092501,PhysRevC.97.025805,Shlomo2006,Colo:2013yta,Garg:2018uam}, and the value $E_{sym}^0$ and the slope parameter $L$ of the nuclear symmetry energy at the saturation density are constrained respectively within $E_{sym}^0=31.7 \pm 3.2$ MeV and $L=58.7 \pm 28.1$ MeV from surveying dozens of analyses~\cite{LI2013276,RevModPhys.89.015007}. Neutron star observables may help to constrain better higher-order EOS parameters characterizing $E_{SNM}$ and $E_{sym}$ at suprasaturation densities.

It is encouraging to see that data on neutron star properties have been emerging in recent years, providing opportunities to constrain the nuclear matter EOS in the multimessage era of nuclear physics. From relativistic Shapiro time delay, the mass of PSR J0740+6620 was measured to be $2.14^{+0.10}_{-0.09}M_{\odot}$~\cite{Cromartie_2019} and later refined to be $2.08^{+0.07}_{-0.07}M_{\odot}$~\cite{Fonseca_2021}, providing a large maximum mass to rule out soft EOS of neutron star matter. Based on data collected by Neutron Star Interior Composition Explorer (NICER), the radius of PSR J0740+6620 was further measured to be $R=13.7^{+2.6}_{-1.5}$ km in Ref.~\cite{Miller_2021} and $R=12.39^{+1.30}_{-0.98}$ km in Ref.~\cite{Riley_2021}. For canonical neutron stars, their radii are estimated to be within $R_{1.4}=10.62-12.83$ km inferred from photospheric radius expansion bursts and thermal emissions~\cite{Lattimer:2014sga}. The more recent measurements of PSR J0035+451 by NICER gave a mass of $1.44^{+0.15}_{-0.14}M_{\odot}$ and a radius of $R=13.02^{+1.24}_{-1.06}$ km in Ref.~\cite{Miller_2019}, and a mass of $1.34^{+0.15}_{-0.16}M_{\odot}$ and a radius of $R=12.71^{+1.14}_{-1.19}$ km in Ref.~\cite{Riley_2019}, with the deduced radius slightly larger than that from Ref.~\cite{Lattimer:2014sga} while there are significant overlaps. Besides, the analysis of GW170817 by the LIGO/Virgo Collaboration has found that the tidal deformability from the neutron star merger is constrained within $\Lambda_{1.4} = 190^{+390}_{-120}$~\cite{PhysRevLett.121.161101} for canonical neutron stars.

Observations of neutron stars have been used to constrain the nuclear matter EOS based on various models, e.g., the parameterized EOSs such as those directly using EOS parameters~\cite{Margueron:2017eqc,Xie:2019sqb,Xie:2020tdo}, using polytropic EOSs~\cite{Read:2009yp,Ozel:2009da,Steiner:2010fz,Greif:2020pju,Al-Mamun:2020vzu}, and speed-of-sound models~\cite{Tews:2018kmu,Annala:2019puf}, and non-parametric models such as those using spectral methods~\cite{Lindblom:2010bb} or Gaussian processes~\cite{Essick:2021kjb}, as well as the chiral effective field theory~\cite{Lim:2018bkq,Raaijmakers:2021uju,Malik:2022zol,Patra:2022yqc}. In the present study, we investigate the constraints on the EOS from neutron star observables based on widely used effective nuclear models, i.e., the Skyrme-Hartree-Fock (SHF) model as well as its extension (the Korea-IBS-Daegu-SKKU (KIDS) model), and the relativistic mean-field (RMF) model. Compared with the parameterized EOS, these phenological models, which start from an effective nuclear interaction or a Lagrangian and give the nuclear matter EOS based on the mean-field approximation, have a more clear and better-defined theoretical basis. Another advantage of employing effective nuclear interaction models is that one can study neutron stars, heavy-ion reactions, and nuclear structures based on the same model with well-developed approaches, thus helpful for constraining the nuclear force and the nuclear matter EOS from high to low densities. In order to constrain quantitatively different EOS parameters and investigate their correlations under the constraints of multiple neutron star observables, we employ the Bayesian analysis in the present study. In previous studies~\cite{Chen:2010qx,PhysRevC.105.044305,PhysRevC.90.044305}, coefficients in these effective models have been successfully expressed inversely in terms of EOS parameters, so we are able to do Bayesian sampling in the space of EOS parameters rather than in that of model coefficients, making the Bayesian analysis more effective. We have also investigated the effect of the neutron star crust on the observables as well as its impact on the constraint of EOS parameters. While considerable model dependence on the final constraints is observed, a stiff $E_{SNM}$ and a moderately soft $E_{sym}$ at suprasaturation densities are favored by the adopted astrophysical observables based on the SHF model as well as its extension and the RMF model.

The rest part of this manuscript is organized as follows. Section~\ref{sec:theory} reviews briefly the theoretical framework, including the formulism of the SHF model as well as its extension and the RMF model, the way to calculate neutron star observables based on effective interactions, and the Bayesian analysis method. Section~\ref{sec:results} discusses the detailed constraints and correlations on EOS
parameters from neutron star observables after a sensitivity analysis. Finally, we conclude and outlook in Sec.~\ref{sec:summary}.

\section{Theoretical framework}
\label{sec:theory}

In the present study, compact stars are assumed to consist of only nucleons and leptons, and their properties are obtained from the EOS of neutron star matter based on the non-relativistic SHF model and the RMF model. We choose a standard energy-density functional (EDF) of the SHF model as in Ref.~\cite{Chen:2010qx} as well as an extension for the density-dependent term, which was named as the KIDS model~\cite{PhysRevC.97.014312}. For the RMF model, we choose the Lagrangian form as in Ref.~\cite{PhysRevC.90.044305}. The chosen EDFs of the SHF, KIDS, and RMF models allow us to express model parameters inversely in terms of EOS parameters, so we are able to change a single physics quantity at one time while keeping the values of other quantities unchanged. The core-crust transition density is consistently calculated for a given set of EOS parameters based on the effective nuclear interaction, and different EOSs are used in the liquid core, inner crust, and outer crust of the neutron star, from which the neutron star properties can be obtained by solving the Tolman-Oppenheimer-Volkoff (TOV) equation as well as the coupled differential equation for the calculation of the tidal deformability. The EOS parameters are then constrained by comparing the resulting neutron star properties with the adopted astrophysical observables based on a Bayesian approach.

\subsection{Definition of EOS parameters}

In this subsection, we briefly remind the reader the definition of the EOS parameters that characterize the density dependence of $E_{SNM}(\rho)$ and $E_{sym}(\rho)$. The binding energy per nucleon in isospin asymmetric nuclear matter with nucleon density $\rho=\rho_n+\rho_p$ and isospin asymmetry $\delta = (\rho_n-\rho_p)/\rho$ can be expressed as
\begin{equation}
E(\rho,\delta) = E_{SNM}(\rho) + E_{sym}(\rho) \delta^2 + O(\delta^4),
\end{equation}
where the symmetry energy is defined as
\begin{equation}
E_{sym}(\rho) = \frac{1}{2} \left[\frac{\partial^2 E(\rho,\delta)}{\partial \delta^2}\right]_{\delta=0}.
\end{equation}
The higher-order $\delta$ terms are generally much smaller, so the EOS is mostly dominated by $E_{SNM}(\rho)$ and $E_{sym}(\rho)$. Both $E_{SNM}(\rho)$ and $E_{sym}(\rho)$ contain contributions from the kinetic part and the potential part. While the kinetic part is calculated from the quasi-particle assumption, the potential part depends on the EDFs.

Around the saturation density $\rho_0$, $E_{SNM}(\rho)$ and $E_{sym}(\rho)$ can be expanded in the power of $\chi = \frac{\rho-\rho_0}{3\rho_0}$ as
\begin{eqnarray}
E_{SNM}(\rho) &=& E_{SNM}(\rho_0) + \frac{K_0}{2!} \chi^2 + \frac{Q_0}{3!} \chi^3 + O(\chi^4), \nonumber \\
E_{sym}(\rho) &=& E_{sym}(\rho_0) + L \chi + \frac{K_{sym}}{2!} \chi^2 + \frac{Q_{sym}}{3!} \chi^3 + O(\chi^4). \nonumber
\end{eqnarray}
In the above, the linear term in the expansion of $E_{SNM}(\rho)$ vanishes due to zero pressure of SNM at $\rho_0$. The independent EOS parameters relevant in the present study are the saturation density $\rho_0$, the binding energy $E_0$, the incompressibility $K_0$, and the skewness parameter $Q_0$ of SNM at $\rho_0$, the symmetry energy $E_{sym}^0$ and its slope parameter $L$, curvature parameter $K_{sym}$, and skewness parameter $Q_{sym}$ at $\rho_0$, and they are defined respectively as
\begin{eqnarray}
&&\left[\frac{\partial E_{SNM}(\rho)}{\partial \rho}\right]_{\rho=\rho_0} = 0,\\
&&E_0 \equiv E_{SNM}(\rho_0),\\
&&K_0 = 9\rho_0^2 \left[\frac{\partial^2 E_{SNM}(\rho)}{\partial \rho^2}\right]_{\rho=\rho_0},\\
&&Q_0 = 27\rho_0^3 \left[\frac{\partial^3 E_{SNM}(\rho)}{\partial \rho^3}\right]_{\rho=\rho_0},\\
&&E_{sym}^0 \equiv E_{sym}(\rho_0),\\
&&L = 3\rho_0 \left[\frac{\partial E_{sym}(\rho)}{\partial \rho}\right]_{\rho=\rho_0},\\
&&K_{sym} = 9\rho_0^2 \left[\frac{\partial^2 E_{sym}(\rho)}{\partial \rho^2}\right]_{\rho=\rho_0},\\
&&Q_{sym} = 27\rho_0^3 \left[\frac{\partial^3 E_{sym}(\rho)}{\partial \rho^3}\right]_{\rho=\rho_0}.
\end{eqnarray}

\subsection{Skyrme-Hartree-Fock model}

Neglecting the spin-orbit interaction, the effective interaction between nucleons at $\vec{r}_1$ and $\vec{r}_2$ in the standard SHF model can be expressed as
\begin{eqnarray}\label{v12}
v^{SHF}(\vec{r}_1,\vec{r}_2) &=& t_0(1+x_0P_\sigma)\delta(\vec{r}) \notag \\
&+& \frac{1}{2} t_1(1+x_1P_\sigma)[{\vec{k}'^2}\delta(\vec{r})+\delta(\vec{r})\vec{k}^2] \notag\\
&+&t_2(1+x_2P_\sigma)\vec{k}' \cdot \delta(\vec{r})\vec{k} \notag\\
&+&\frac{1}{6}t_3(1+x_3P_\sigma)\rho^\alpha(\vec{R})\delta(\vec{r}).
\end{eqnarray}
In the above, $\vec{r}=\vec{r}_1-\vec{r}_2$ and $\vec{R}=(\vec{r}_1+\vec{r}_2)/2$ are respectively the relative and central coordinates for the two nucleons, $\vec{k}=(\nabla_1-\nabla_2)/2i$ is the relative momentum operator and $\vec{k}'$ is its complex conjugate acting on the left, and $P_\sigma=(1+\vec{\sigma}_1 \cdot \vec{\sigma}_2)/2$ is the spin exchange operator.

Based on the Hartree-Fock approach, the above effective interaction leads to the following energy density for uniform nuclear matter
\begin{equation}\label{eshf}
\epsilon = \epsilon_k + \epsilon_0 + \epsilon^{SHF}_\rho + \epsilon_{eff},
\end{equation}
where the kinetic energy density $\epsilon_k$ as well as the potential energy density $\epsilon_0$ from the zero-range interaction, $\epsilon^{SHF}_\rho$ from the density-dependent interaction, and $\epsilon_{eff}$ from the momentum-dependent interaction can be expressed respectively as
\begin{eqnarray}
\epsilon_k &=& \frac{\tau}{2m}, \notag\\
\epsilon_0 &=& \frac{t_0}{4}[(2+x_0)\rho^2-(2x_0+1)(\rho_n^2+\rho_p^2)], \notag\\
\epsilon^{SHF}_\rho &=& \frac{t_3\rho^\alpha}{24} [(2+x_3)\rho^2-(2x_3+1)(\rho_n^2+\rho_p^2)], \notag\\
\epsilon_{eff} &=& \frac{1}{8}[t_2(2x_2+1)-t_1(2x_1+1)](\tau_n\rho_n+\tau_p\rho_p) \notag\\
&+&\frac{1}{8}[t_1(2+x_1)+t_2(2+x_2)]\tau\rho,\notag
\end{eqnarray}
with $m$ being the bare nucleon mass and $\tau=\tau_n+\tau_p$ being the total kinetic density. For nucleons with isospin index $q=n,p$ in a cold static nuclear matter, the kinetic density is $\tau_q=p_{Fq}^5/10\pi^2$, with $p_{Fq}=(3\pi^2\rho_q)^{1/3}$ being the Fermi momentum. The parameters $t_0$, $t_1$, $t_2$, $t_3$, $x_0$, $x_1$, $x_2$, $x_3$, and $\alpha$ can be solved inversely from the macroscopic quantities~\cite{Chen:2010qx}, i.e., the saturation density $\rho_0$, the binding energy $E_0$ at $\rho_0$, the incompressibility $K_0$, the isoscalar and isovector nucleon effective mass $m_s^\star$ and $m_v^\star$ at the Fermi momentum in normal nuclear matter, the value $E_{sym}^0$ and the slope parameter $L$ of the symmetry energy at $\rho_0$, and the isoscalar and isovector density gradient coefficient $G_S$ and $G_V$.

As an extension of the above standard SHF EDF, the density-dependent term in the effective interaction [Eq.~(\ref{v12})] is replaced by the following form in the KIDS model
\begin{equation}
v^{KIDS}_\rho (\vec{r}_1,\vec{r}_2) = \frac{1}{6}\sum_{i=1}^{3} (t_{3i}+y_{3i}P_\sigma)\rho^{i/3}(\vec{R})\delta(\vec{r}),
\end{equation}
and the energy density is modified accordingly to
\begin{equation} \label{ekids}
\epsilon = \epsilon_k + \epsilon_0 + \epsilon^{KIDS}_\rho + \epsilon_{eff},
\end{equation}
where
\begin{equation}
\epsilon^{KIDS}_\rho = \sum_{i=1}^{3} \left[\frac{1}{16} t_{3i} \rho^{2+i/3} - \frac{1}{48} (t_{3i}+2y_{3i})\rho^{i/3}\rho_3^2 \right]
\end{equation}
is the contribution from the density-dependent interaction, with $\rho_3=\rho_n-\rho_p$ being the isovector density. Compared to the standard SHF model, the additional coefficients in the KIDS model, i.e., $t_{3i}$ and $y_{3i}$, allow us to vary more individual EOS parameters, i.e., $Q_0$, $K_{sym}$, and $Q_{sym}$ as shown in Ref.~\cite{PhysRevC.105.044305}.

\subsection{Relativistic mean-field model}

In the present study, we take the following Lagrangian form of the RMF model
\begin{equation}\label{lrmf}
\mathcal{L} =  \mathcal{L}_{nm} + \mathcal{L}_\sigma + \mathcal{L}_\omega + \mathcal{L}_\rho + \mathcal{L}_{\omega\rho},
\end{equation}
with
\begin{eqnarray}
\mathcal{L}_{nm} &=& \bar{\psi} (i \gamma^\mu \partial_\mu - m) \psi + g_\sigma \sigma \bar{\psi} \psi - g_\omega \bar{\psi} \gamma^\mu \omega_\mu \psi, \notag\\
&-& \frac{g_\rho}{2} \bar{\psi}\gamma^\mu \vec{\rho}_\mu \vec{\tau} \psi, \notag\\
\mathcal{L}_\sigma &=& \frac{1}{2} (\partial^\mu \sigma \partial_\mu \sigma -m_\sigma^2\sigma^2) - \frac{A}{3} \sigma^3 - \frac{B}{4} \sigma^4, \notag\\
\mathcal{L}_\omega &=& -\frac{1}{4} F^{\mu\nu} F_{\mu\nu} + \frac{1}{2} m_\omega^2 \omega_\mu \omega^\mu + \frac{C}{4} (g_\omega^2 \omega_\mu \omega^\mu)^2, \notag\\
\mathcal{L}_\rho &=& -\frac{1}{4} \vec{B}^{\mu\nu} \vec{B}_{\mu\nu} + \frac{1}{2} m_\rho^2 \vec{\rho}_\mu \vec{\rho}^\mu,  \notag\\
\mathcal{L}_{\omega\rho} &=& \frac{1}{2} \alpha_3^\prime g_\omega^2 g_\rho^2 \omega_\mu \omega^\mu \vec{\rho}_\mu \vec{\rho}^\mu. \notag
\end{eqnarray}
In the above, $\mathcal{L}_{nm}$ is the contribution from the kinetic part of nucleons as well as its coupling to $\sigma$, $\omega$, and $\rho$ mesons, with $\psi$, $\sigma$, $\omega_\mu$, and $\vec{\rho}_\mu$ being the fields of nucleons and corresponding mesons, $\mathcal{L}_\sigma$, $\mathcal{L}_\omega$, and $\mathcal{L}_\rho$ are free and self-interacting terms of $\sigma$, $\omega$, and $\rho$ mesons, respectively, with $\vec{\tau}$ being the Pauli matrices, and $\mathcal{L}_{\omega\rho}$ represents the cross interaction term between $\omega$ and $\rho$ mesons. The antisymmetric field tensors $F_{\mu\nu}$ and $\vec{B}_{\mu\nu}$ are defined as $F_{\mu\nu}=\partial_\nu \omega_\mu - \partial_\mu \omega_\nu$ and $\vec{B}_{\mu\nu}=\partial_\nu \vec{\rho}_\mu - \partial_\mu \vec{\rho}_\nu - g_\rho (\vec{\rho}_\mu \times \vec{\rho}_\nu)$.

Based on the mean-field approximation, the meson fields are treated as classical fields, and applying the Euler-Lagrange equations leads to the following coupling equations for these fields
\begin{eqnarray}
m^2_\sigma \sigma &=& g_\sigma \rho_s - A\sigma^2 - B\sigma^3, \label{eqsigma} \\
m^2_\omega \omega_0 &=& g_\omega \rho - C g^4_\omega \omega^3_0 - \alpha_3^\prime g^2_\omega g^2_\rho \rho_{0(3)}^2 \omega_0, \label{eqomega}\\
m^2_\rho \rho_{0(3)} &=& \frac{1}{2} g_\rho \rho_3 - \alpha_3^\prime g^2_\omega g^2_\rho \rho_{0(3)} \omega^2_0, \label{eqrho}
\end{eqnarray}
where $\sigma$, $\omega_0$, and $\rho_{0(3)}$ are expectation values of the meson fields at the ground state, with the subscript `0' representing the time component in the Dirac space, and the subscript `3' representing the $z$ component in the Pauli space. $\rho_s=\rho_{sn}+\rho_{sp}$ is the scalar density, with the contribution from nucleons of isospin index $q$ expressed as
\begin{equation}
\rho_{sq} = 2 \int \frac{m_q^\star}{\sqrt{p^2 + {m_q^\star}^2 }} \frac{d^3 p}{(2\pi)^3},
\end{equation}
where $m_q^\star=m-g_\sigma \sigma$ is the Dirac nucleon effective mass, different from the non-relativistic p-mass in the SHF model (see, e.g., Ref.~\cite{LI201829}). The energy density can be expressed as
\begin{eqnarray}\label{ermf}
\epsilon &=& \epsilon^{RMF}_k + \frac{1}{2}m^2_\sigma \sigma^2 + \frac{A}{3} \sigma^3 + \frac{B}{4} \sigma^4  \notag\\
&-& \frac{1}{2} m^2_\omega \omega_0^2 +g_\omega \omega_0 \rho - \frac{C}{4} (g_\omega^2 \omega_0^2)^2 \notag\\
&-& \frac{1}{2} m^2_\rho \rho_{0(3)}^2 + \frac{g_\rho}{2} \rho_{0(3)} \rho_3 - \frac{1}{2} \alpha_3^\prime g_\omega^2 g_\rho^2 \omega_0^2 \rho_{0(3)}^2,
\end{eqnarray}
where
\begin{equation}
\epsilon^{RMF}_k = 2\sum_q \int \left(\sqrt{p^2+{m_q^\star}^2}-m\right) \frac{d^3p}{(2\pi)^3}
\end{equation}
is the kinetic energy contribution. The EOS in the RMF model is determined by $g_\sigma^2/m_\sigma^2$, $g_\omega^2/m_\omega^2$, $g_\rho^2/m_\rho^2$, $A$, $B$, $\alpha_3^\prime$ and $C$. As shown in Ref.~\cite{PhysRevC.90.044305}, the former 6 parameters can be expressed in terms of $\rho_0$, $E_0$, $K_0$, $E_{sym}^0$, $L$, and $m_s^\star$ for a given $C$. In the present study, we are able to vary another independent EOS parameter $Q_0$ by adjusting the value of $C$, so there are totally 7 independent EOS parameters in the RMF model. For arbitrary values of these EOS parameters, the field equations [Eqs.~(\ref{eqsigma})-(\ref{eqrho})] do not necessarily have solutions in asymmetric nuclear matter at high densities, especially for $\omega_0$. However, we noticed that all RMF parameterization sets in Ref.~\cite{PhysRevC.90.055203} have $C>0$ and $\alpha_3^\prime>0$, which guarantee that the field equations have solutions, and this condition is then used to rule out unphysical EOS parameter sets. In addition, the square of the coupling constants ($g_\sigma^2$, $g_\omega^2$, and $g_\rho^2$) calculated from the macroscopic physics quantities should be positive, adding to the limitation of the parameter space. For the quantitative limits of the parameter space from the present EDF of the RMF model, we refer the reader to Appendix~\ref{app}.

\subsection{Neutron star observables}
\label{nstarfor}

In the present study, we assume that the neutron star from the center to the surface contains the liquid core of uniform neutron star matter, the inner crust consists of nuclear pasta phase, and the outer crust is composed of ion lattice and relativistic electron gas. The neutron star matter contains neutrons, protons, electrons, and possibly muons if the charge chemical potential is large enough. The fraction of each component is determined by the $\beta$-equilibrium and the charge-neutrality condition. The total energy density of neutron star matter can be expressed as
\begin{equation}
V = \epsilon + \rho m + \epsilon_l,
\end{equation}
where $\epsilon$ is obtained from the standard SHF [Eq.~(\ref{eshf})], KIDS [Eq.~(\ref{ekids})], or RMF [Eq.~(\ref{ermf})] model, and $\epsilon_l$ is the energy density of electrons and muons by assuming that they are free massive Fermions. The pressure of neutron star matter can be calculated through the relation
\begin{equation}
P = P_{nuc} + P_l
\end{equation}
where
\begin{equation}
P_{nuc} = \sum_q \mu_q\rho_q - \epsilon
\end{equation}
is the pressure from nucleons, with the chemical potential for nucleons of isospin $q$ obtained from $\mu_q = \partial \epsilon/ \partial \rho_q$, and $P_l$ is the pressure from leptons. The thermodynamic consistency relation is satisfied for the global neutron star matter and for each component. The transition density $\rho_t$ between the liquid core and the inner crust is self-consistently determined with a thermodynamical approach as detailed in Refs.~\cite{Xu:2009vi,PhysRevC.79.035802}, i.e., below the transition density the system is unstable and satisfies the relation
\begin{equation}
\frac{\partial \mu_n}{\partial \rho_n}  \frac{\partial \mu_p}{\partial \rho_p} - \left( \frac{\partial \mu_n}{\partial \rho_p} \right)^2 <0.
\end{equation}
The EOS of the inner crust is parameterized as
\begin{equation}\label{inner}
P = a + b V^{\gamma},
\end{equation}
where the default value of $\gamma$ is taken to be $4/3$~\cite{PhysRevLett.83.3362,LATTIMER2000121,2001ApJ...550..426L} while results from other values are compared in order to investigate the effect of the crust EOS on the constraints of the EOS parameters, and $a$ and $b$ are determined by the continuity condition~\cite{Xu:2009vi} of the EOS at $\rho_t$ and at the boundary between the inner crust and the outer crust, with the density in the latter case taken to be $\rho_{out}=2.46 \times 10^{-4}$ fm$^{-3}$. In the outer crust, we use the BPS EOS~\cite{1971ApJ...170..299B,Iida_1997} in the density range $6.93\times10^{-13}$ fm$^{-3} < \rho < \rho_{out}$, and we use the FMT EOS~\cite{1971ApJ...170..299B} in the density range of $4.73\times10^{-15}$ fm$^{-3} < \rho <6.93\times10^{-13}$ fm$^{-3}$. For special parameter sets that the neutron star matter is always stable, there is no core-crust transition, and we use the EOS of neutron star matter in the whole density range. More consistent studies using unified EOSs from core to crust can be found in Refs.~\cite{Tews:2016ofv,Lim:2017luh,Carreau:2019zdy,Newton:2021rni}.

With the EOS from high to low densities constructed above, the mass and radius of a neutron star can be calculated through the TOV equation
\begin{eqnarray}
\frac{dP(r)}{dr}&=&-\frac{M(r)[V(r)+P(r)]}{r^2}\left[1+\frac{4 \pi P(r)r^3}{M(r)}\right]\notag\\
&\times&\left[1-\frac{2M(r)}{r}\right]^{-1},
\end{eqnarray}
where $M(r)$ is the gravitational mass inside the radius $r$ of the compact star and can be obtained from the integral of the following equation
\begin{eqnarray}
\frac{dM(r)}{dr}&=&4 \pi r^2 V(r).
\end{eqnarray}
The tidal deformability $\Lambda$ of compact stars during their merge is related to the love number $k_2$ through the relation $\Lambda = \frac{2}{3}k_2\beta^{-5}$, with the latter given by~\cite{Hinderer:2007mb,Hinderer_2009,PhysRevD.82.024016}
\begin{eqnarray}
k_2 &=&\frac{8}{5}(1-2\beta)^2[2-y_R+2\beta(y_R-1)]
\notag\\
&\times& \{2\beta[6-3y_R+3\beta(5y_R-8)]
\notag\\
&+& 4\beta^3[13-11y_R+\beta(3y_R-2)+2\beta^2(1+y_R)]
\notag\\
&+&3(1-2\beta)^2[2-y_R+2\beta(y_R-1)]\text{ln}(1-2\beta)\}^{-1}.\notag\\
\end{eqnarray}
In the above, $\beta \equiv M/R $ is the compactness of the neutron star, and $y_R \equiv y(R) $ is the solution at the star surface to the first-order differential equation
\begin{eqnarray}
r \frac{dy(r)}{dr}+y(r)^2+y(r)F(r)+r^2Q(r)=0,
\end{eqnarray}
with
\begin{eqnarray}
F(r) &=& \frac{r-4\pi r^3[V(r)-P(r)]}{r-2M(r)},
\notag\\
Q(r)&=&\frac{4\pi r\left[5V(r)+9P(r)+\frac{V(r)+P(r)}{\partial P(r)/\partial V(r)}-\frac{6}{4\pi r^2}\right]}{r-2M(r)}
\notag\\
&-&4\left[\frac{M(r)+4\pi r^3P(r)}{r^2(1-2M(r)/r)}\right]^2.
\end{eqnarray}
For a given central density $\rho(r=0)$, the above equations can be solved from the center ($r=0$) to the surface ($r=R$) where the density is lower than $\rho_{cut} \sim 4.73 \times 10^{-15}$ fm$^{-3}$. For special parameter sets that the neutron star matter is always stable but the pressure becomes negative at low densities, the above equations are solved from the center to where the pressure becomes zero.

\subsection{Bayesian analysis}

To obtain the probability distribution functions (PDFs) of EOS parameters under the constraints of astrophysical observables, we employ the Bayesian approach, and the analysis method can be formally expressed as the Bayes' theorem
\begin{equation}
P(M|D) = \frac{P(D|M)P(M)}{\int P(D|M)P(M)dM},
\end{equation}
where $P(M|D)$ is the posterior probability for the model $M$ given the data set $D$, $P(D|M)$ is the likelihood function or the conditional probability for a given theoretical model $M$ to predict correctly the data $D$, and $P(M)$ denotes the prior probability of the model $M$ before being confronted with the data. The denominator of the right-hand side of the above equation is the normalization constant.

Since the coefficients in the standard SHF, KIDS, and RMF model can now be expressed in terms of physics quantities, we vary the physics quantities as model parameters in the Bayesian analysis. Due to the different numbers of coefficients in different models, the numbers of independent model parameters are also different. Table~\ref{T1} lists the default values of model parameters in each model as well as their prior ranges. In the sensitivity analysis, we will check with the sensitivity of a single model parameter within its prior range to astrophysical observables, with the values of other model parameters fixed at their default values. In the Bayesian analysis, we will vary all independent model parameters within their prior ranges. We try to set the default values of model parameters to be the same so that the model dependence can be investigated on the same basis. In the standard SHF model, we try to get a two-solar mass neutron star by setting the default value of $K_0$ as the upper limit from its prior range obtained from studies on ISGMR~\cite{PhysRevLett.109.092501,PhysRevC.97.025805,Shlomo2006,Colo:2013yta,Garg:2018uam}, while values of other quantities are taken as the default ones in the MSL0 force~\cite{Chen:2010qx}. $Q_0$, $K_{sym}$, and $Q_{sym}$ are not independent quantities in the standard SHF model but are calculated from other quantities. In the KIDS model, $Q_0$, $K_{sym}$, and $Q_{sym}$ can be varied independently, while we set their default values as those calculated from the default parameter set for the standard SHF model. Although $Q_0$ is an independent variable in the RMF model, the parameter space is limited, so the largest available value of $Q_0$ is chosen as the default value. For the Dirac isoscalar effective mass in the RMF model, we set its default value to be $m_s^\star=0.73m$ so that it corresponds effectively to the same non-relativistic isoscalar effective mass~\cite{PhysRevLett.95.022302,LI201829} as in SHF and KIDS models. The Dirac isovector effective mass is then $m_v^\star=0.73m$ from the present RMF Lagrangian without $\delta$-meson coupling.

\begin{table}\small
  \caption{Default values of macroscopic quantities in the standard SHF, KIDS, and RMF models used in the present study. Quantities with asterisk are not independent ones but are calculated from other independent quantities. For independent quantities, they are varied within their prior ranges in the sensitivity analysis and Bayesian analysis.}
    \begin{tabular}{|c|c|c|c|c|}
  \hline
   & SHF & KIDS & RMF & prior range \\
   \hline
    $\rho_0$ (fm$^{-3}$) & 0.16 & 0.16 & 0.16 & -\\
    $E_0$ (MeV) & $-$16 & $-$16 & $-$16 & -\\
    $K_0$ (MeV) & 260 & 260 & 260 & 220 $\sim$ 260~\cite{PhysRevLett.109.092501,PhysRevC.97.025805,Shlomo2006,Colo:2013yta,Garg:2018uam} \\
    $Q_0$ (MeV) & $-$323$^\star$ & $-$323 & $-$389 & $-$800 $\sim$ 400~\cite{Tews_2017,Zhang:2017ncy}\\
    $E_{sym}^0$ (MeV) & 30 & 30 & 30 & 28.5 $\sim$ 34.9~\cite{LI2013276,RevModPhys.89.015007} \\
    $L$ (MeV) & 60 & 60 & 60 & 30 $\sim$ 90~\cite{LI2013276,RevModPhys.89.015007} \\
    $K_{sym}$ (MeV) & $-$105$^\star$  & $-$105 & -127$^\star$  & $-$400 $\sim$ 100~\cite{Tews_2017,Zhang:2017ncy} \\
    $Q_{sym}$ (MeV) & 214$^\star$  & 214 & 474$^\star$  & -200 $\sim$ 800~\cite{Tews_2017,Zhang:2017ncy} \\
    $m_s^\star/m$ & 0.8 & 0.8 & 0.73 & 0.5 $\sim$ 0.9\\
    $m_v^\star/m$ & 0.7 & 0.7 & 0.73$^\star$ & 0.5 $\sim$ 0.9 \\
    $G_S$ (MeVfm$^5$) & 132 & 132 & - &  -\\
    $G_V$ (MeVfm$^5$) & 5 & 5 & - & -\\
   \hline
    \end{tabular}
  \label{T1}
\end{table}

For the standard SHF model, there are totally 10 independent variables, and we choose to vary EOS parameters $p_1=K_0$ uniformly within $220-260$ MeV from ISGMR studies~\cite{PhysRevLett.109.092501,PhysRevC.97.025805,Shlomo2006,Colo:2013yta,Garg:2018uam}, and $p_2=E_{sym}^0$ and $p_3=L$ uniformly within $28.5-34.9$ MeV and $30-90$ MeV, respectively, according to Refs.~\cite{LI2013276,RevModPhys.89.015007}. For the KIDS model, there are totally 13 independent variables, and we choose to vary higher-order EOS parameters $p_4=Q_0$, $p_5=K_{sym}$, and $p_6=Q_{sym}$ uniformly within their prior ranges obtained based on analyses of terrestrial nuclear experiments and EDFs~\cite{Tews_2017,Zhang:2017ncy}, in additional to those in the standard SHF model. For the RMF model, there are totally 7 independent variables, and we choose to vary EOS parameters $p_1=K_0$, $p_2=E_{sym}^0$, $p_3=L$, and $p_4=Q_0$. We also vary $m_s^\star/m$ and $m_v^\star/m$ in the standard SHF and KIDS models within their empirical ranges. In non-relativistic models, especially for KIDS~\cite{Gil:2018yah}, we expect that the effective masses are decoupled from the nuclear matter EOS, to be confirmed by the results. For the Dirac effect mass $m_s^\star/m$ in the RMF model, it is expected to be closely related to the EOS, and we will vary it in the same empirical range.

Results of representative astrophysical observables from a certain model parameter set are compared with data sets, for which we choose the radius $d_1^{exp}=R_{1.4}$ of a canonical neutron star with $M=1.4 M_{\odot}$, the radius $d_2^{exp}=R_{2.08}$ of PSR J0740+6620 with $M=2.08 M_{\odot}$, and the tidal deformability $d_3^{exp}=\Lambda_{1.4}$ of a canonical neutron star. The likelihood function, which describes quantitatively how well the theoretical results $d_{1,2,3,...}^{th}$ reproduces the corresponding observables $d_{1,2,3,...}^{exp}$, is defined as
\begin{eqnarray}
&&P[D(d_1,d_2,d_3,...)|M(p_1,p_2,p_3,...)] \notag\\
&=& \Pi_{i=1} \Bigg \{ \frac{1}{2\pi \sigma_i} \exp\left[-\frac{(d^{th}_i-d^{exp}_i)^2}{2\sigma_i^2}\right] \notag\\
&\times& \Theta(M_{max}-2.08M_{\odot}) \Theta(1-c_s) \Bigg\}, \label{llh}
\end{eqnarray}
where the Heavyside functions put solid constraints that the maximum mass of the neutron star should be larger than $2.08M_{\odot}$ and the speed of sound $c_s = \sqrt{\partial P/\partial \epsilon}$ inside a neutron star should be smaller than the speed of light, otherwise the likelihood function is zero. $\sigma_i$ are estimated from uncertainty ranges for the astrophysical data. In the case of asymmetric uncertainties, we use different values of $\sigma_i$ for $d^{th}_i>d^{exp}_i$ and $d^{th}_i<d^{exp}_i$. Table~\ref{T2} lists the values of $d^{exp}_i$ as well as the corresponding uncertainty ranges to be used in the Bayesian analysis.

\begin{table}\small
  \caption{Most probable values and uncertainties of adopted astrophysical observables for the Bayesian analysis.}
    \begin{tabular}{|c|c|}
   \hline
    $R_{1.4}$ (km) & $11.725 \pm 1.105$~\cite{Lattimer:2014sga} \\
    $R_{2.08}$ (km) & $13.7^{+2.6}_{-1.5}$~\cite{Miller_2021} and $12.39^{+1.30}_{-0.98}$~\cite{Riley_2021} \\
    $\Lambda_{1.4}$ & $190^{+390}_{-120}$~\cite{PhysRevLett.121.161101} \\
    $M_{max}$  & $>2.08M_{\odot}$~\cite{Fonseca_2021}\\
    $c_s$ & $<1$ \\
   \hline
    \end{tabular}
  \label{T2}
\end{table}

\section{Results and discussions}
\label{sec:results}

We first do sensitivity analysis by changing each individual EOS parameter within its prior range and thus get a global picture how the resulting astrophysical observables change with these EOS parameters. Then, we vary all EOS parameters within their prior ranges and obtain the constraints on these EOS parameters as well as their correlations from the astrophysical data based on the Bayesian approach. We will also discuss the posterior EOS from the resulting constrained EOS parameters.

\subsection{Sensitivity analysis}

\begin{figure}[!h]
\includegraphics[width=1\linewidth]{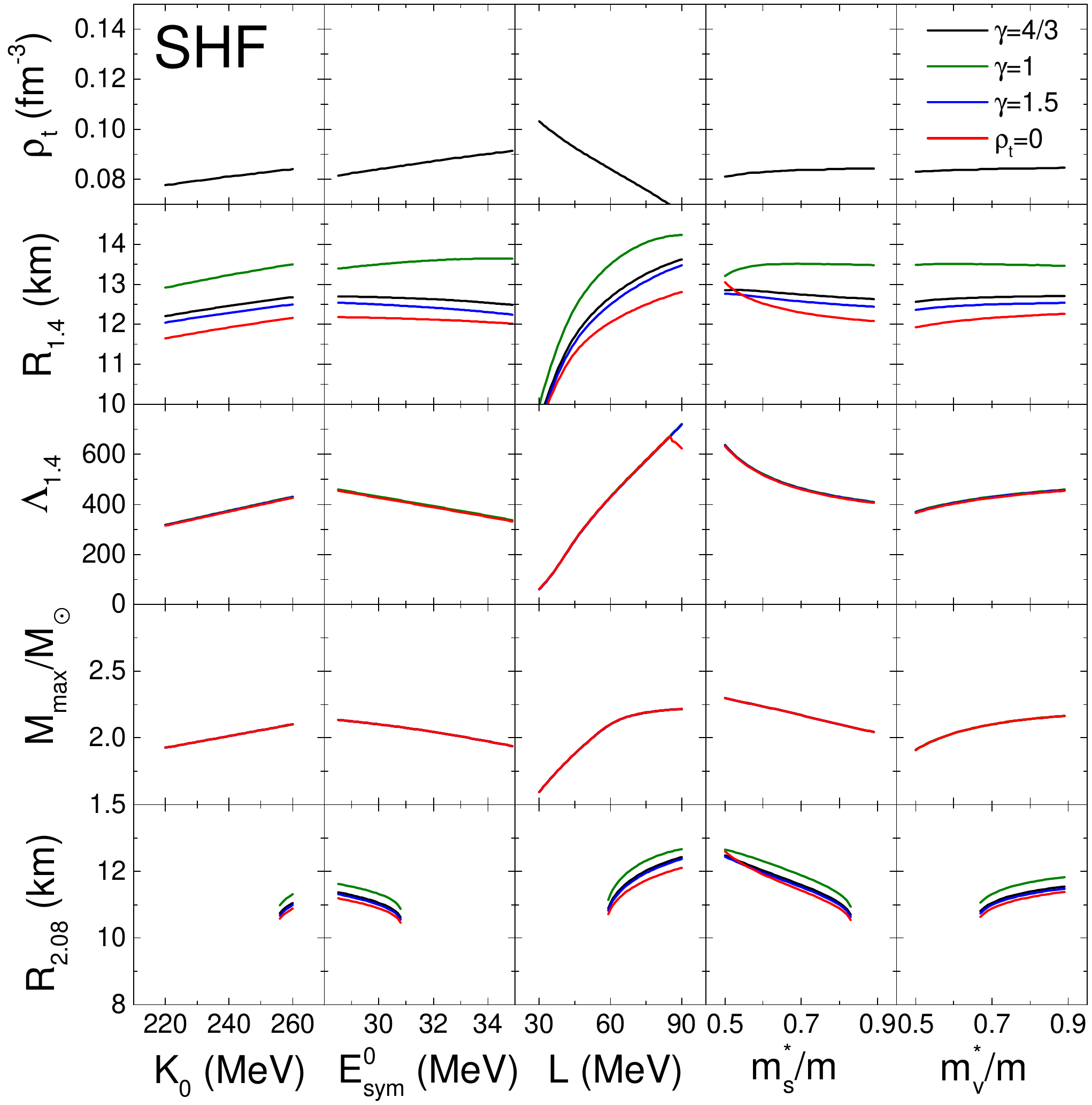}
\caption{\label{sen_SHF} Dependence of the core-crust transition density $\rho_t$ (first row), the radius $R_{1.4}$ (second row) and the tidal deformability $\Lambda_{1.4}$ (third row) of a canonical neutron star, the maximum mass of the neutron star $M_{max}/M_{\odot}$ (fourth row), and the radius $R_{2.08}$ (fifth row) of a neutron star with mass $M=2.08M_{\odot}$ individually on $K_0$, $E_{sym}^0$, $L$, $m_s^\star/m$, and $m_v^\star/m$ within their prior ranges, with the values of other parameters fixed at their default values as in Table~\ref{T1}, based on the standard SHF model. Results from different values of EOS coefficients $\gamma$ for the inner crust are compared, together with those without considering the crust ($\rho_t=0$).}
\end{figure}

\begin{figure*}[!h]
\includegraphics[width=0.8\linewidth]{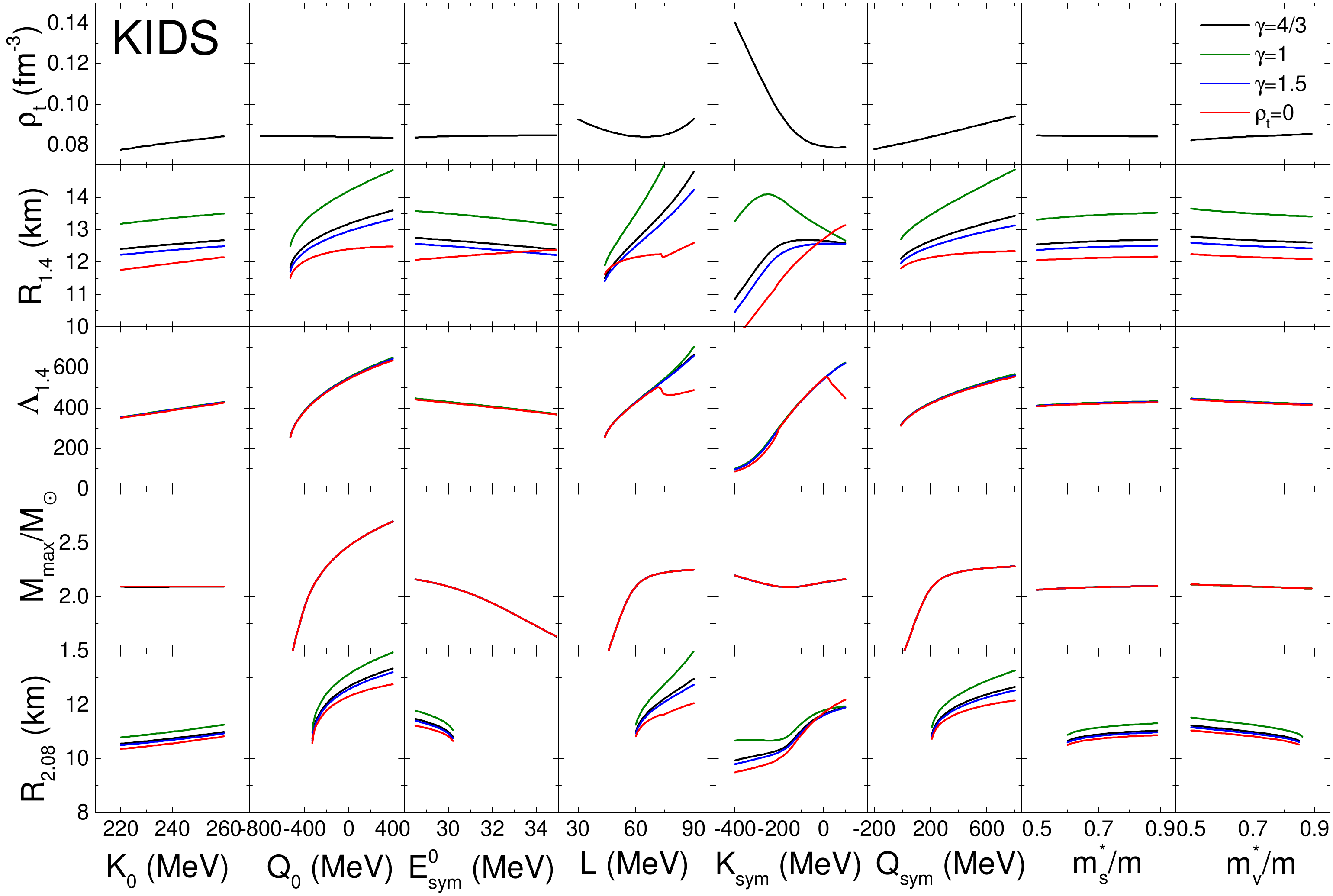}
\caption{\label{sen_KIDS} Similar to Fig.~\ref{sen_SHF} but for the KIDS model showing dependence of observables individually on $K_0$, $Q_0$, $E_{sym}^0$, $L$, $K_{sym}$, $Q_{sym}$, $m_s^\star/m$, and $m_v^\star/m$ within their prior ranges.}
\end{figure*}

\begin{figure}[!h]
\includegraphics[width=1\linewidth]{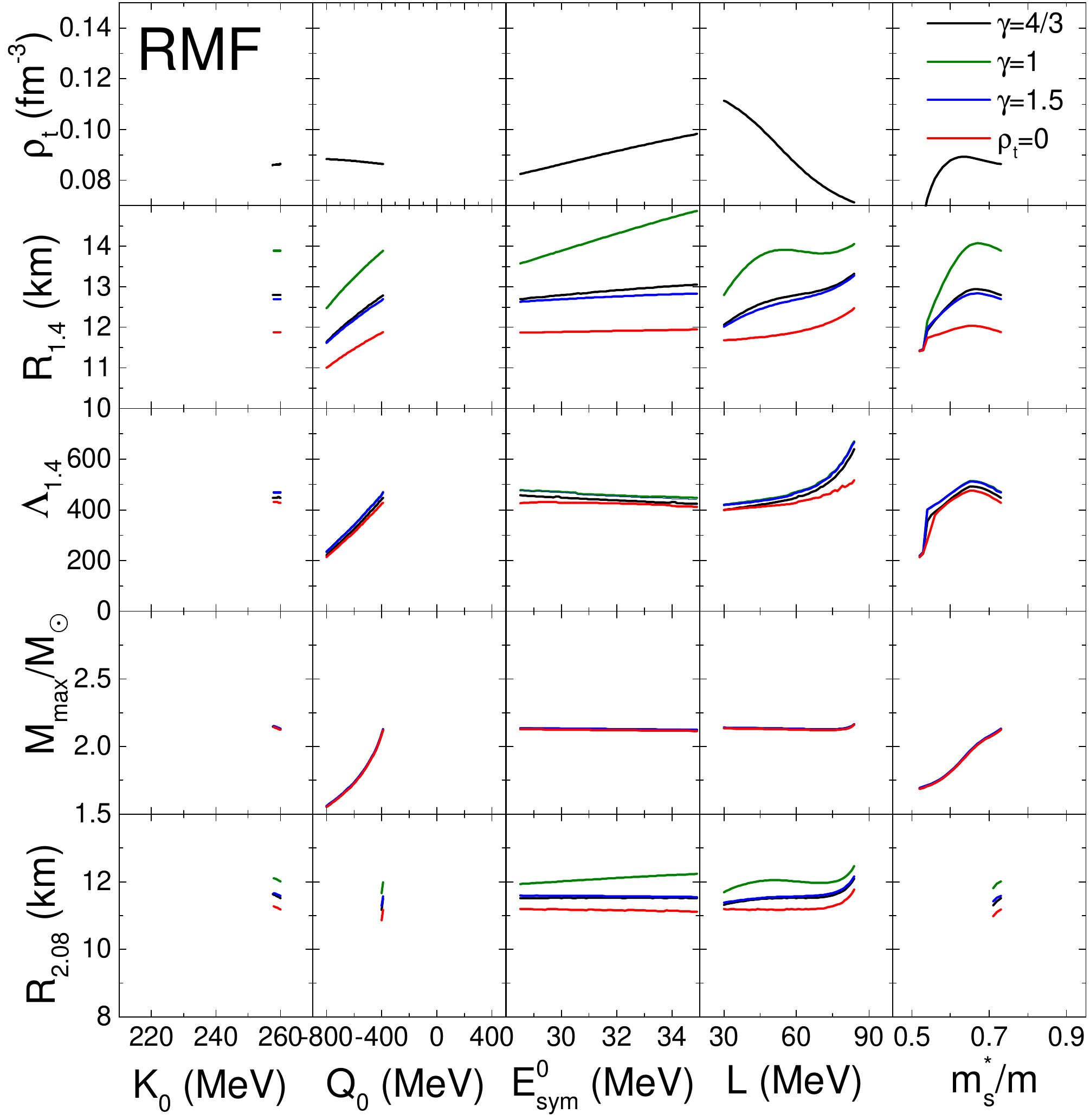}
\caption{\label{sen_RMF} Similar to Fig.~\ref{sen_SHF} but for the RMF model showing dependence of observables individually on $K_0$, $Q_0$, $E_{sym}^0$, $L$, $m_s^\star/m$, and $m_v^\star/m$ within their prior ranges.}
\end{figure}

Figure~\ref{sen_SHF} displays extensively how the core-crust transition density and relevant astrophysical observables change with each individual EOS parameters in the standard SHF model. The transition density is seen to decrease almost linearly with increasing $L$, as already observed in Refs.~\cite{Xu:2009vi,PhysRevC.79.035802}, while it is not very sensitive to other EOS parameters. Both the radius and the tidal deformability of a canonical neutron star increase with increasing $L$ but are not very sensitive to other EOS parameters. The maximum mass of a neutron star is found to be moderately sensitive to $L$ but is not very sensitive to other EOS parameters. We note that the weak sensitivity of the maximum mass of a neutron star to $K_0$ is due to the small prior range of $K_0$ constrained by ISGMR, and the resulting $Q_0$, which increases with increasing $K_0$ according to Eq.~(3) and Fig.~16 of Ref.~\cite{PhysRevC.105.044305}, also has a small range. If a neutron star with mass $M=2.08M_{\odot}$ can be achieved, its radius becomes sensitive to most EOS parameters. We have also compared results with different EOSs for crust. Using a soft EOS ($\gamma=1$) for inner crust increases the radius of a canonical neutron star by $1-2$ km compared with a stiff EOS ($\gamma=1.5$). Using the EOS of the neutron star matter as that for the crust, or identically by setting $\rho_t=0$, the radius of a canonical neutron star becomes even smaller. The EOS of the crust has a smaller effect on the radius of heavier neutron stars, and has a minor effect on the tidal deformability and the maximum mass of a neutron star.

Figures~\ref{sen_KIDS} displays similar content as Fig.~\ref{sen_SHF} but for the KIDS model, and additional dependencies on $Q_0$, $K_{sym}$, and $Q_{sym}$ are shown. It is seen that the core-crust transition density is most sensitivity to $K_{sym}$ rather than $L$. We note that $K_{sym}$ increases linearly with increasing $L$ in the standard SHF model as shown by Eq.~(4) of Ref.~\cite{PhysRevC.105.044305}. Both the radius and the tidal deformability of a canonical neutron star is sensitive to $L$, $K_{sym}$, as well as higher-order EOS parameters $Q_0$ and $Q_{sym}$. The maximum mass is again insensitive to $K_0$, but most sensitive to $Q_0$, and moderately sensitive to $E_{sym}^0$, $L$, and $Q_{sym}$, within their prior ranges. Again, if a neutron star with mass $M=2.08M_{\odot}$ can be achieved, its radius becomes sensitive to most EOS parameters. The crust EOS has larger effects on the radius of a canonical neutron star, smaller effects on the radius of a heavy neutron star, and minor effects on the tidal deformability and the maximum mass of a neutron star. In the case of $\rho_t=0$, the kinks for $R_{1.4}$ and $\Lambda_{1.4}$ at larger $L$ are from the negative pressure at low densities of neutron star matter, so the TOV equation is solved until the pressure is zero rather than a density cut, as mentioned in Sec.~\ref{nstarfor}.

Figures~\ref{sen_RMF} displays similar content as Fig.~\ref{sen_SHF} but for the RMF model, and the individual variables $K_0$, $Q_0$, $E_{sym}^0$, $L$, and $m_s^\star/m$ are varied independently within their limited parameter space with other parameters fixed at their default values. For example, the available ranges of $K_0$, $Q_0$, and $m_s^\star/m$ are much smaller than the prior ranges as shown in Appendix~\ref{app}. Here $K_{sym}$ is not an independent variable, and the transition density decreases almost linearly with increasing $L$, similar to the standard SHF model. Incorporating $Q_0$ as an independent variable, the radius, the tidal deformability, and the maximum mass of a neutron star become sensitive to $Q_0$, similar to the KIDS model. The moderate sensitivities of most astrophysical observables to the Dirac effective mass $m_s^\star/m$ is a special feature in the RMF model compared to non-relativistic models. The considerable sensitivity of the neutron-star radius and the less sensitivity of the tidal deformability and maximum mass to the crust EOS are also observed in Fig.~\ref{sen_RMF}.

\subsection{Constraints on EOS parameters}

\begin{figure*}[!h]
\includegraphics[width=0.22\linewidth]{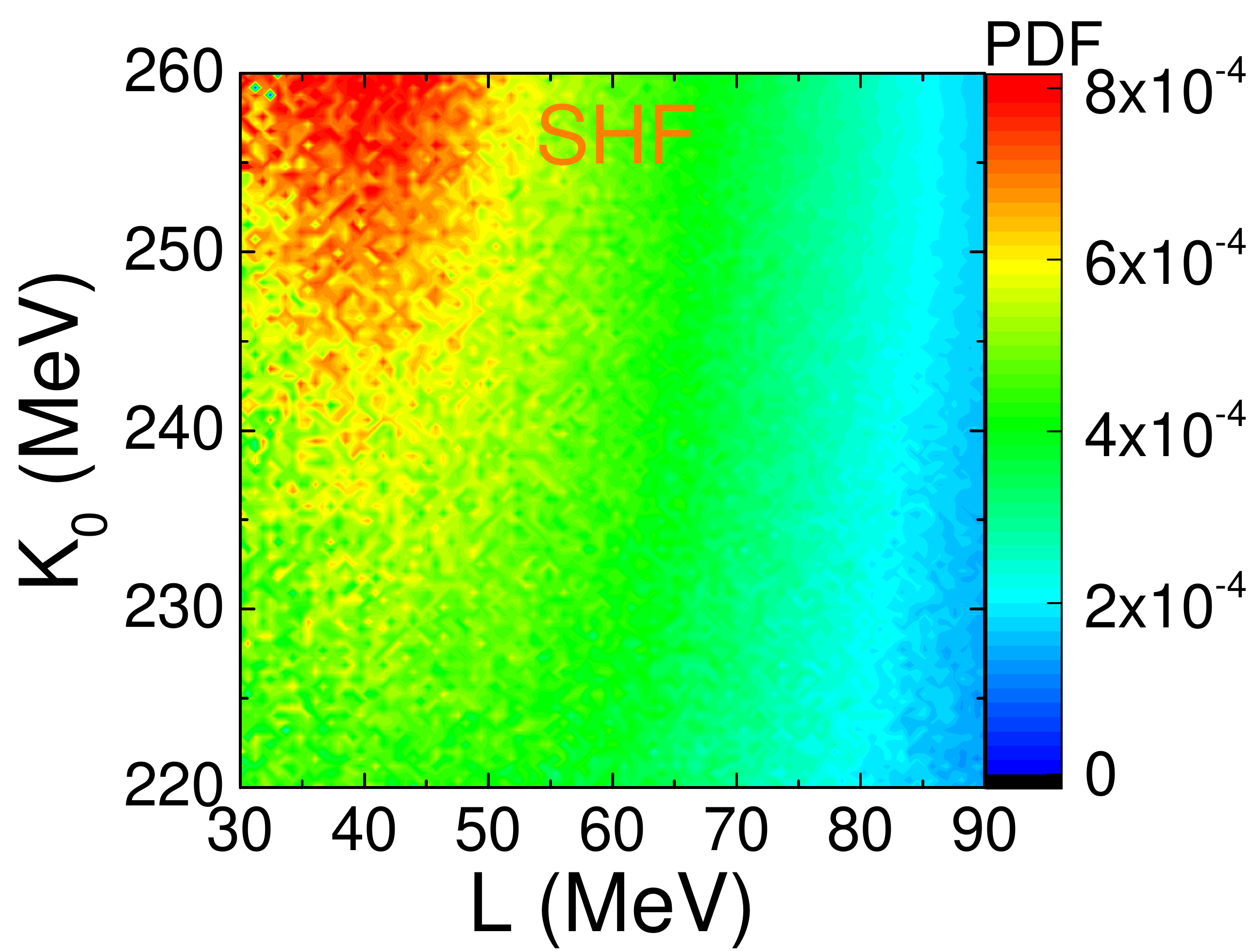}
\includegraphics[width=0.22\linewidth]{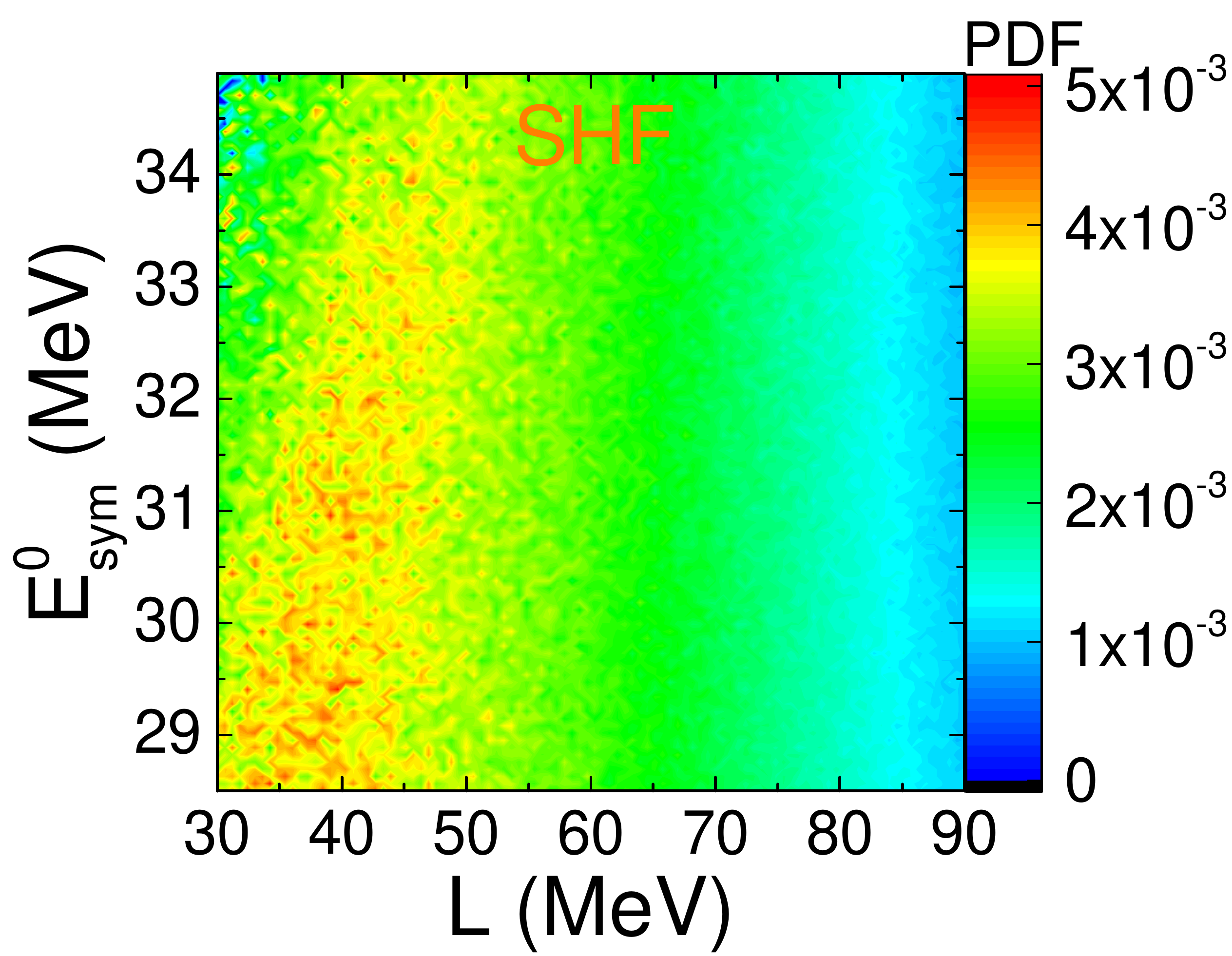}\\
\includegraphics[width=0.22\linewidth]{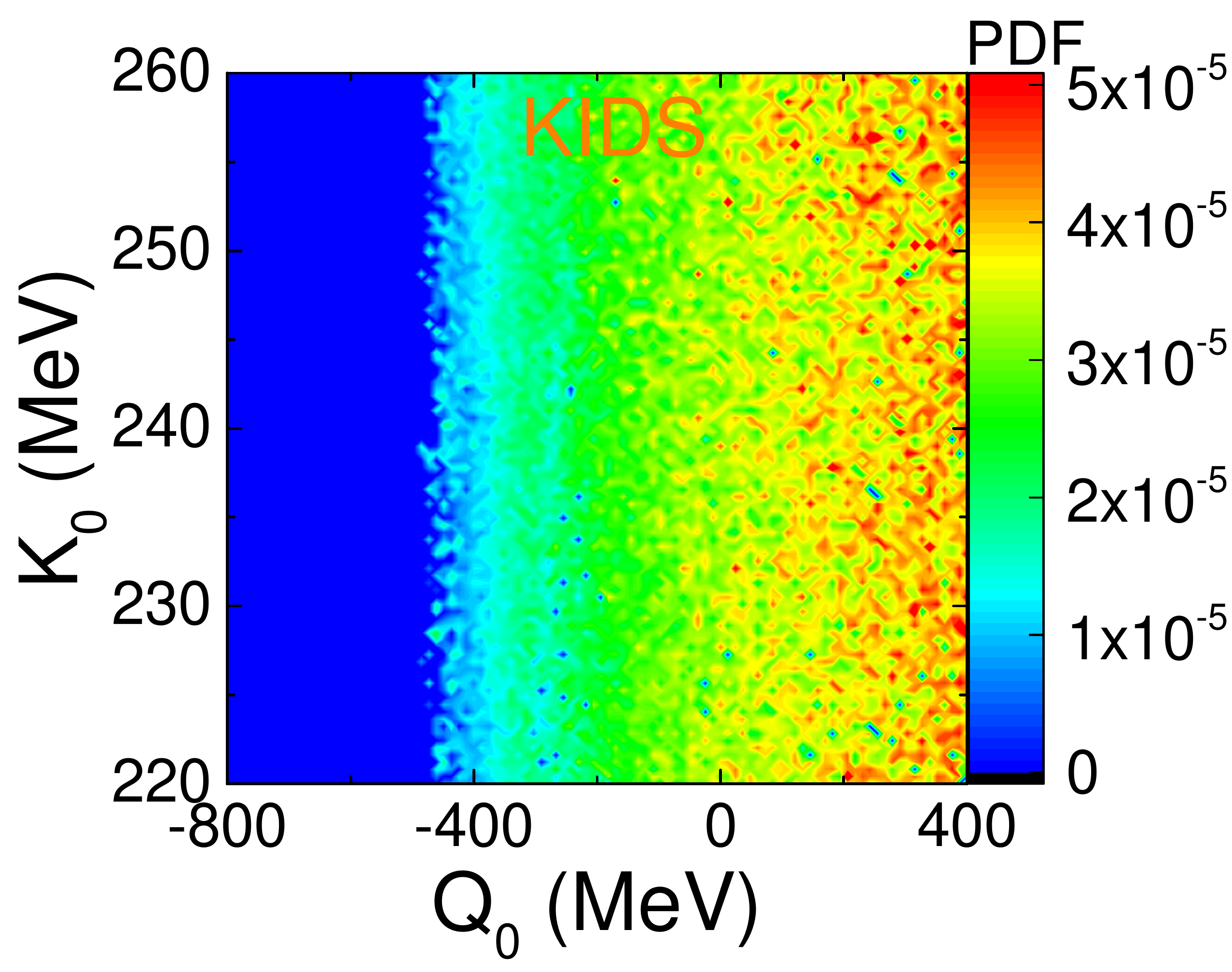}
\includegraphics[width=0.22\linewidth]{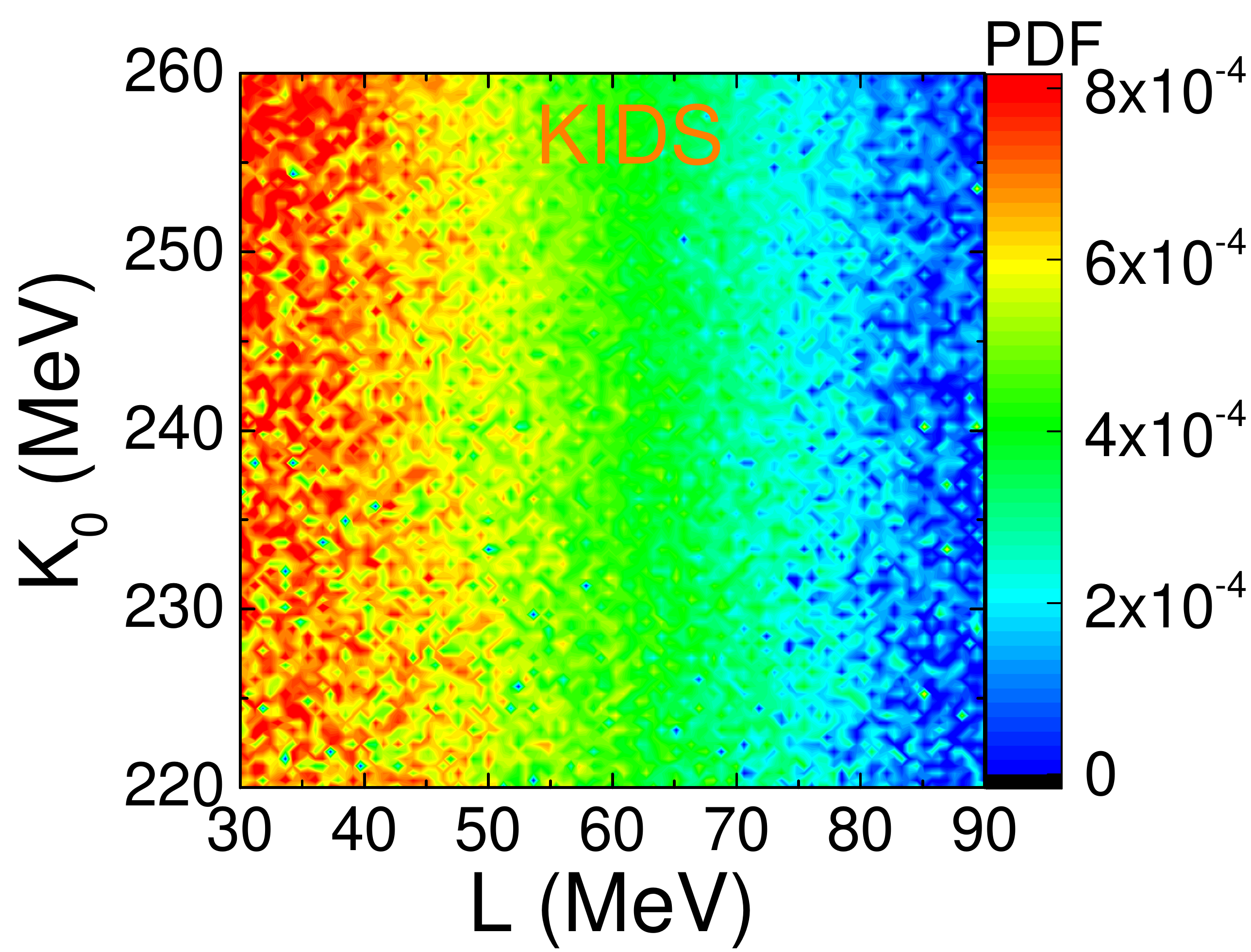}
\includegraphics[width=0.22\linewidth]{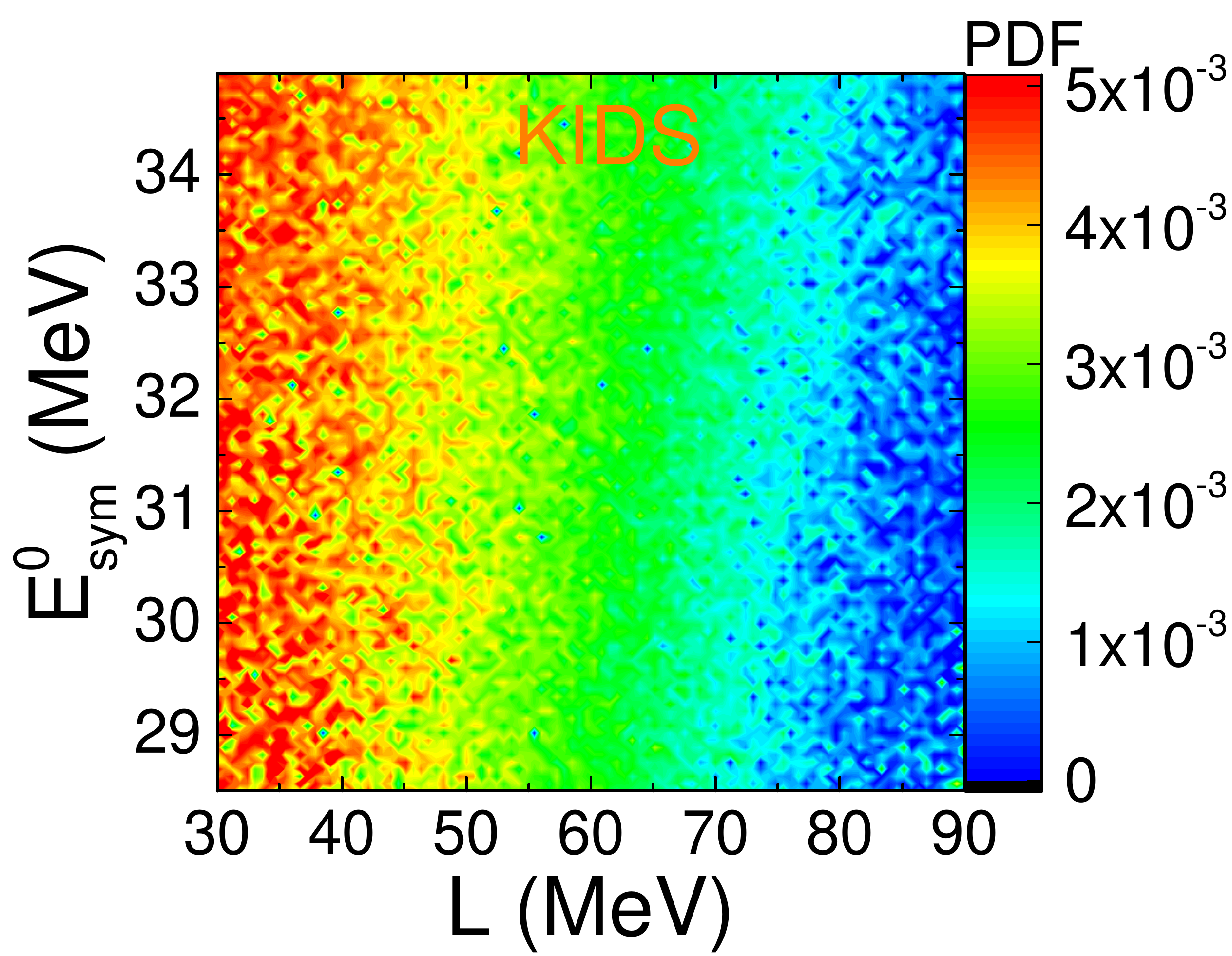}
\includegraphics[width=0.24\linewidth]{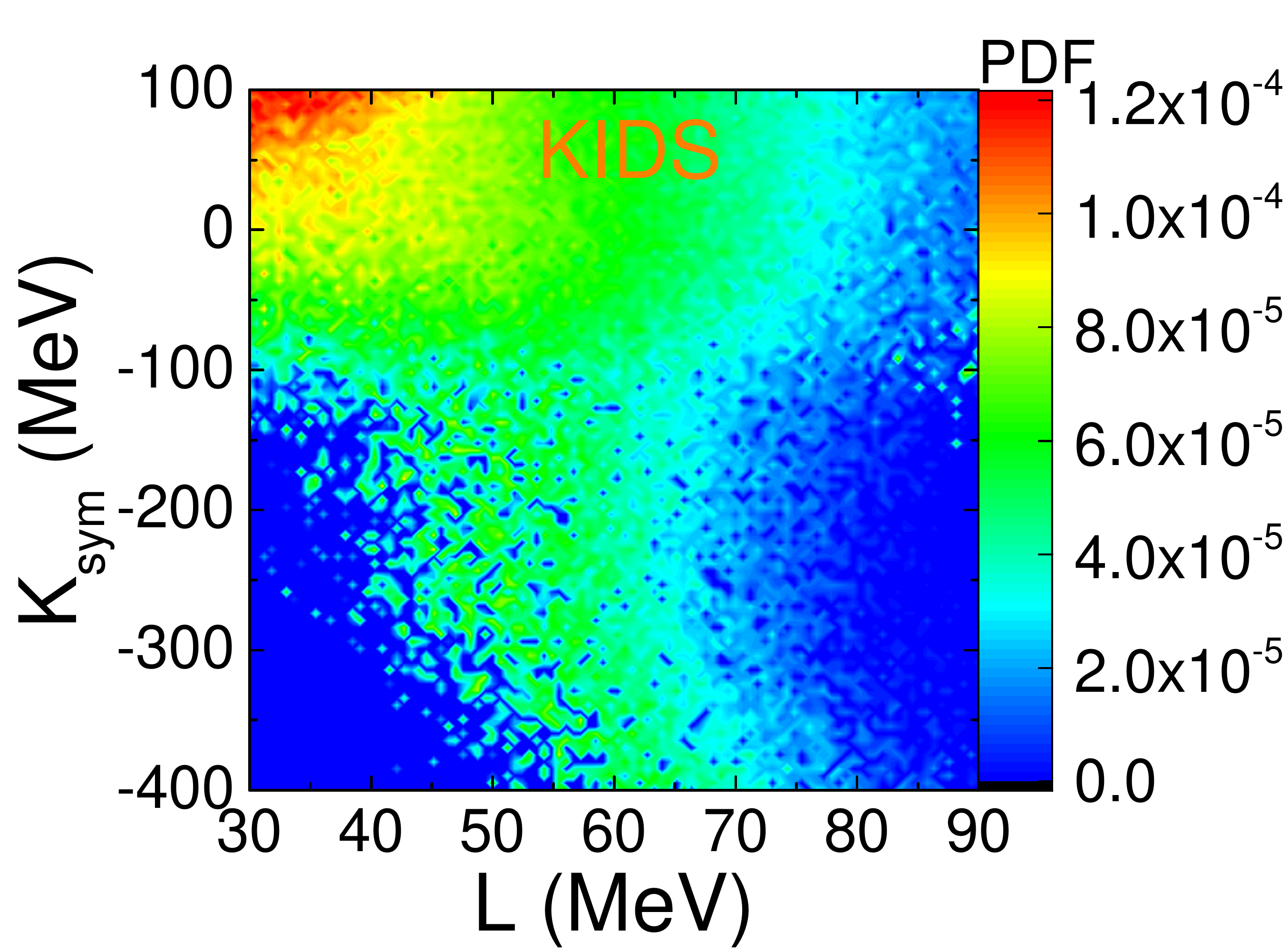}\\
\includegraphics[width=0.22\linewidth]{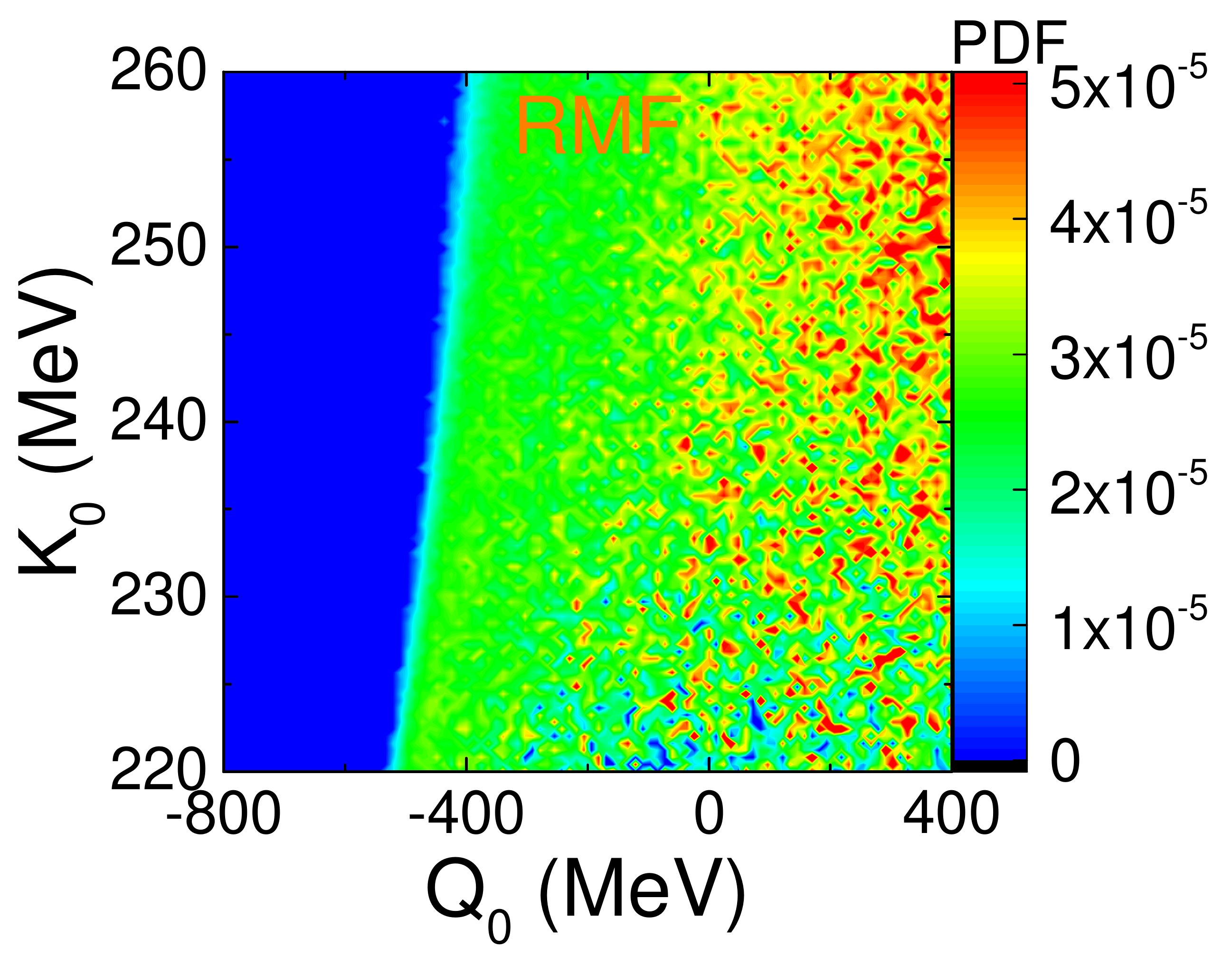}
\includegraphics[width=0.22\linewidth]{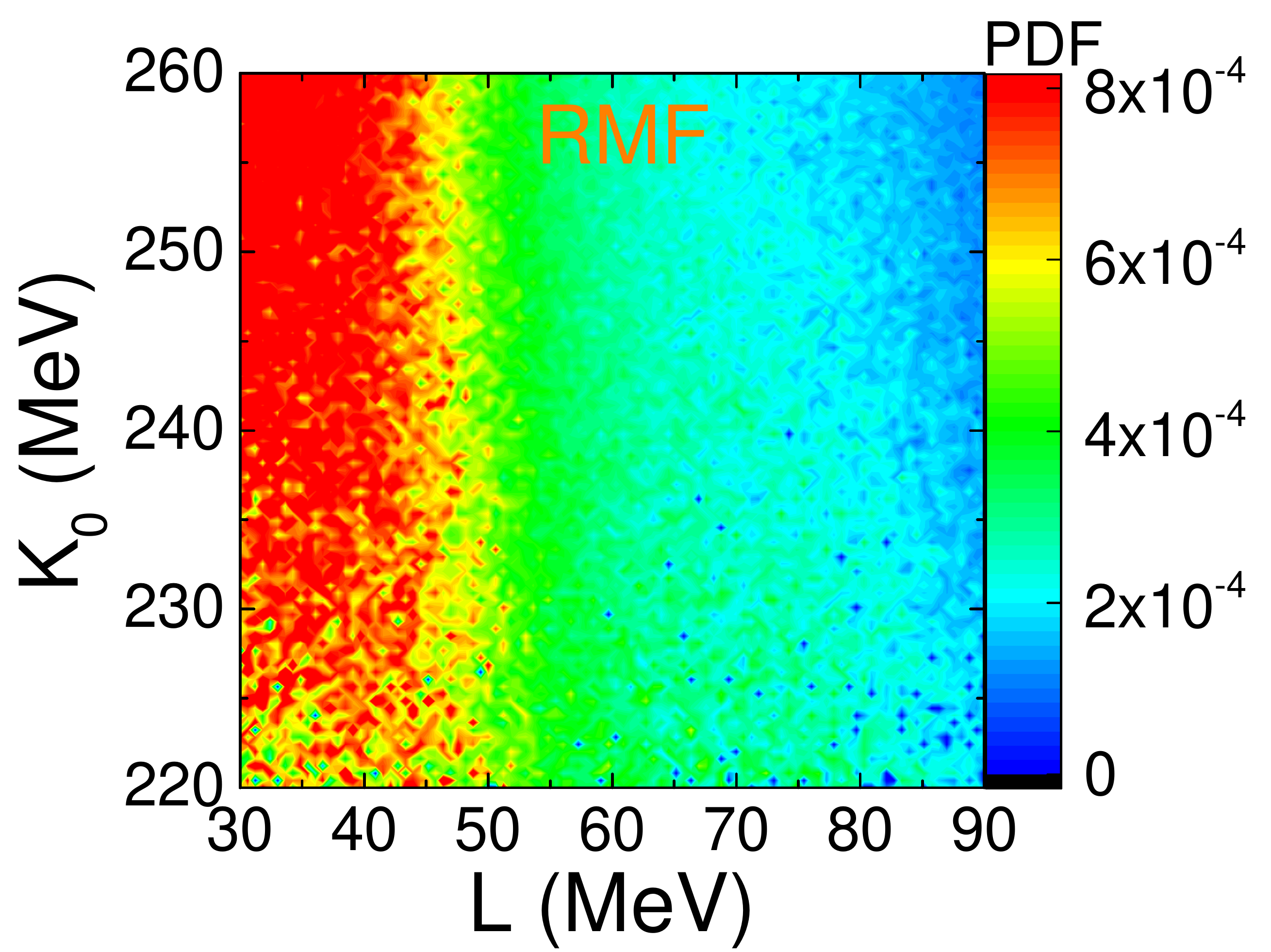}
\includegraphics[width=0.22\linewidth]{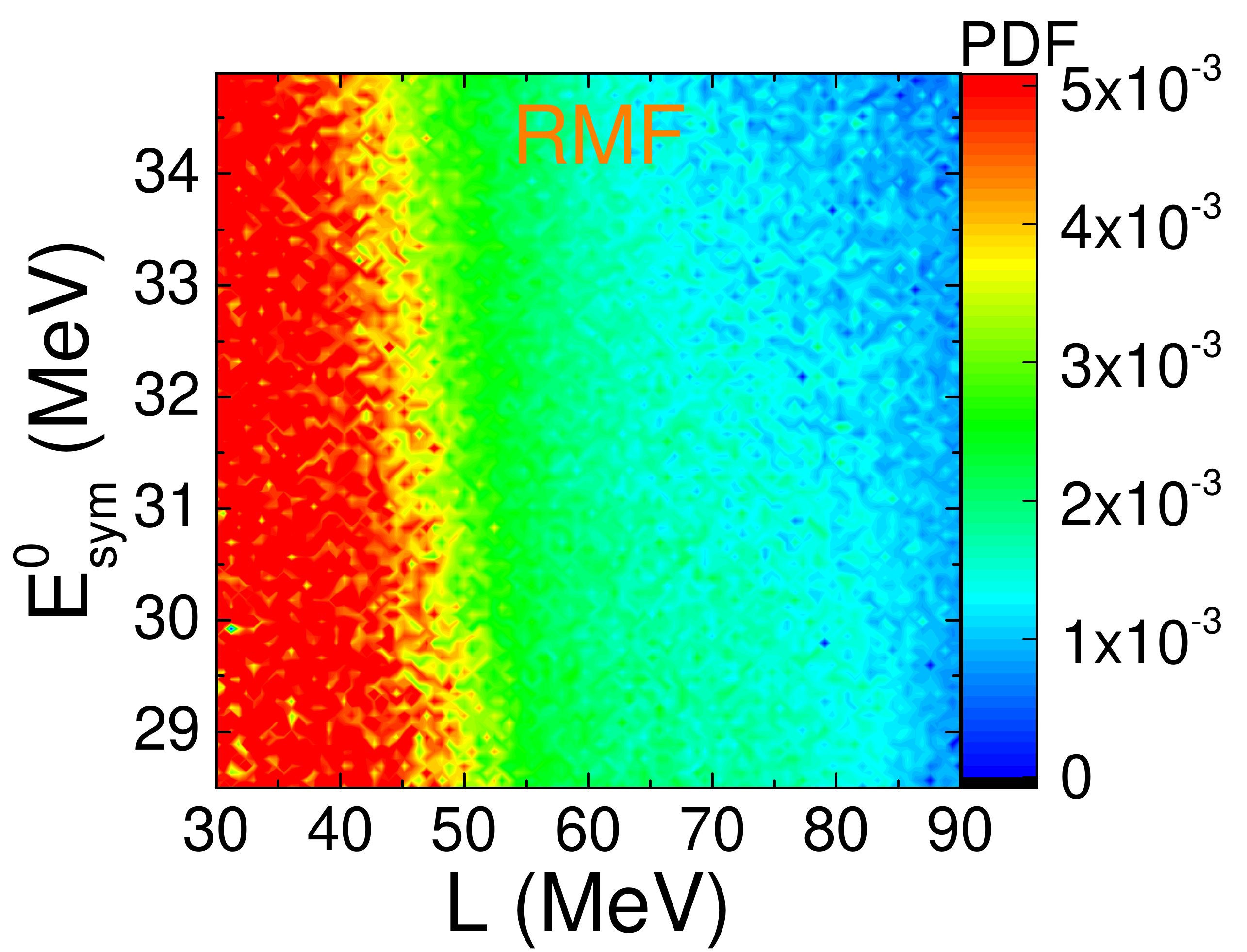}
\caption{\label{corr} Posterior correlated PDFs in the ($L$, $K_0$) and ($L$, $E_{sym}^0$) planes in the standard SHF model (top row), in the ($Q_0$, $K_0$), ($L$, $K_0$), ($L$, $E_{sym}^0$), and ($L$, $K_{sym}$) planes in the KIDS model (middle row), and in the ($Q_0$, $K_0$), ($L$, $K_0$), and ($L$, $E_{sym}^0$) planes in the RMF model (bottom row) from the constraints of astrophysical observables.}
\end{figure*}

\begin{figure*}[!h]
\includegraphics[width=1.0\linewidth]{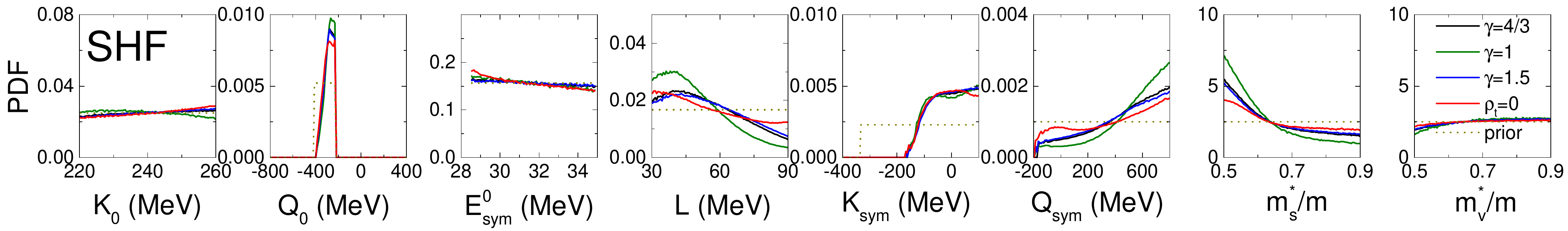}\\
\includegraphics[width=1.0\linewidth]{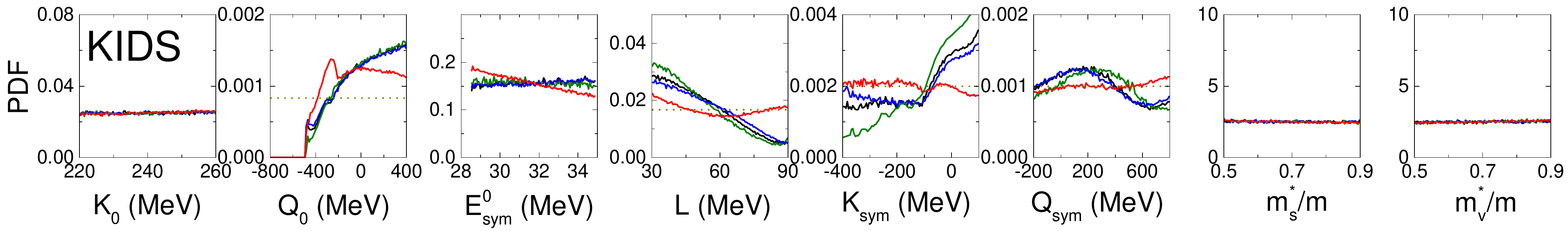}\\
\includegraphics[width=1.0\linewidth]{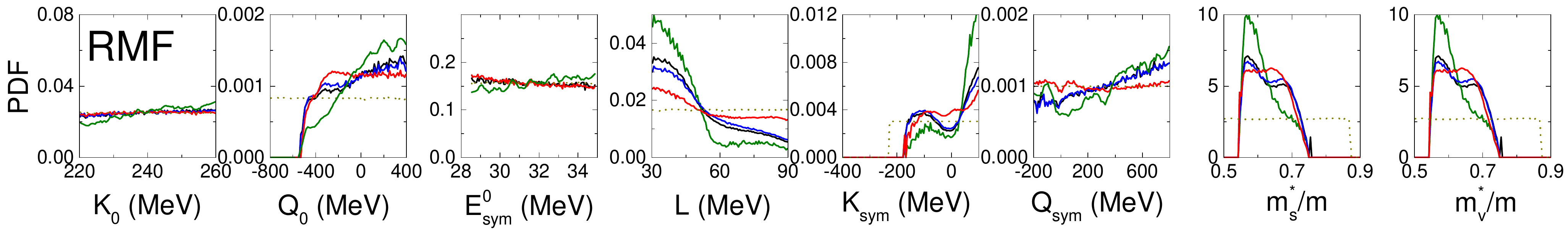}
\caption{\label{pdf} Comparison of the posterior PDFs of $K_0$, $Q_0$, $E_{sym}^0$, $L$, $K_{sym}$, $Q_{sym}$, $m_s^\star$, and $m_v^\star$ in the standard SHF model (top row), the KIDS model (middle row), and the RMF model (bottom row) from the constraints of astrophysical observables with different crust EOSs. Results from different crust EOSs with different coefficients $\gamma$ are compared, and the prior PDF of each individual quantity is also displayed.}
\end{figure*}

For the default scenario by considering $\gamma=4/3$ for the EOS of the inner crust, we now display in Fig.~\ref{corr} the posterior correlated PDFs between lower-order and higher-order EOS parameters in both the isoscalar and isovector channels, as well as the correlated PDFs between the isoscalar and isovector EOS parameters. In the standard SHF model, it is seen that a smaller $L$ is generally associated with a larger $K_0$, due to the constraint of a large $M_{max}$ but a small $R_{1.4}$. A slightly positive correlation between $L$ and $E_{sym}^0$ is observed in the same model, likely due to the opposite dependence of neutron star radii, $\Lambda_{1.4}$, and $M_{max}$ on $L$ and $E_{sym}^0$, as shown in Fig.~\ref{sen_SHF}. In the KIDS model, where both lower-order and higher-order EOS parameters can be varied independently, there are no non-trivial correlations, except for the slightly negative correlation between $L$ and $K_{sym}$, likely due to the similar dependence of neutron star radii and $\Lambda_{1.4}$ on $L$ and $K_{sym}$, as shown in Fig.~\ref{sen_KIDS}. In both the KIDS and RMF models, the sharp cut on the correlated PDF in the ($Q_0$, $K_0$) plane is from the criterion $M_{max}>2.08M_{\odot}$ in the definition of the likelihood function [Eq.~(\ref{llh})]. Similar to the situation in the KIDS model, there are no significant correlations between $K_0$ and $Q_0$, $K_0$ and $L$, or $E_{sym}^0$ and $L$ in the RMF model.

Integrating over all the other variables leads to the one-dimensional PDF of each individual physics quantity. We compare in Fig.~\ref{pdf} the posterior PDFs of EOS parameters and nucleon effective masses in the three models from the constraints of astrophysical observables using different crust EOSs. $Q_0$, $K_{sym}$, and $Q_{sym}$ in the standard SHF model are not independent variables, but are constrained through the posterior PDFs of other EOS parameters. Comparing the default scenario ($\gamma=4/3$), without considering crust ($\rho_t=0$) may lead to significantly different constraints on most EOS parameters, depending on the chosen EDFs. A too soft EOS for the inner crust ($\gamma=1$) may lead to a smaller $L$ and/or a larger $K_{sym}$, compared with results from $\gamma=4/3$ and 1.5. Basically, the astrophysical observables do not put much constraint on $K_0$ and $E_{sym}^0$. On the other hand, a small $L$ is favored by the small $R_{1.4}$, while a large $K_{sym}$ is favored by the large $R_{2.08}$, in all three models for the default case of $\gamma=4/3$. The constraint on $K_{sym}$ is roughly consistent with $-200<K_{sym}<0$ MeV extracted in Refs.~\cite{Gil:2020wqs,Gil:2021ols} based on the KIDS EDF. A large $Q_{sym}$ is favored by the neutron star radii in the standard SHF model, while $Q_{sym}$ is not much constrained in the KIDS and RMF model. The constraint of $M_{max}$ favors a large $Q_0$ in the KIDS and RMF model. Due to the limited prior ranges of $Q_0$ and $K_{sym}$ in the standard SHF model and $K_{sym}$ in the RMF model, which are calculated from other variables, the corresponding posterior PDFs of these higher-order EOS parameters are narrower compared to those in the KIDS model. In the standard SHF model there are some constraints on the non-relativistic p-mass of nucleons, since these effective masses are related to higher-order EOS parameters, e.g., $m_s^\star$ is related to $Q_0$ and $K_{sym}$ according to Eqs.~(3-5) in Ref.~\cite{PhysRevC.105.044305}. In the KIDS model, where higher-order EOS parameters can be varied independently, there are almost no constraints on these non-relativistic effective masses. In the RMF model, where the Dirac mass is closely related to the EOS, the constraint of $M_{max}$ favors a smaller Dirac effective mass of nucleons, corresponding to a stiffer $E_{SNM}$.

\subsection{Constraints on EOS}

\begin{figure*}[!h]
\includegraphics[width=0.22\linewidth]{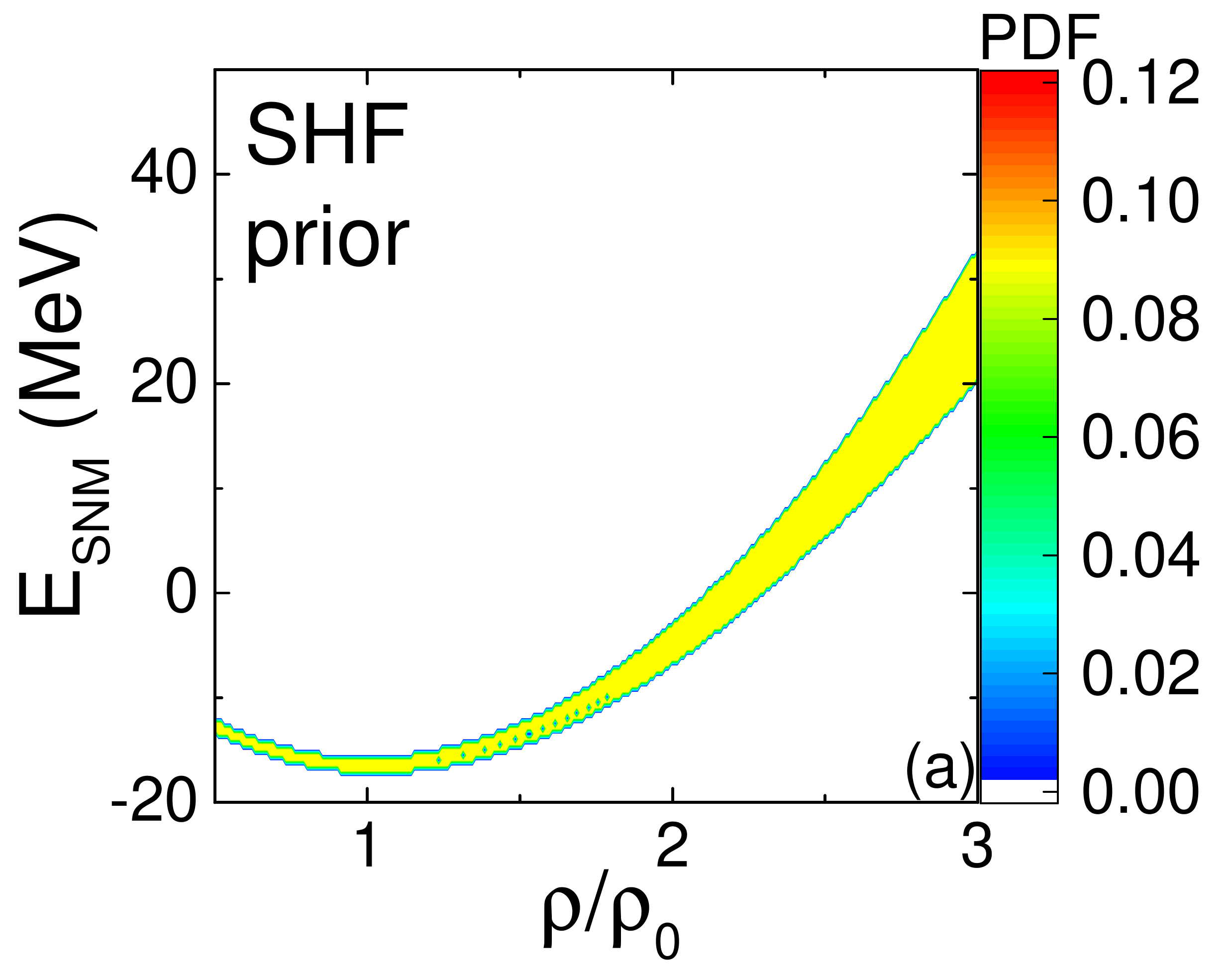}
\includegraphics[width=0.22\linewidth]{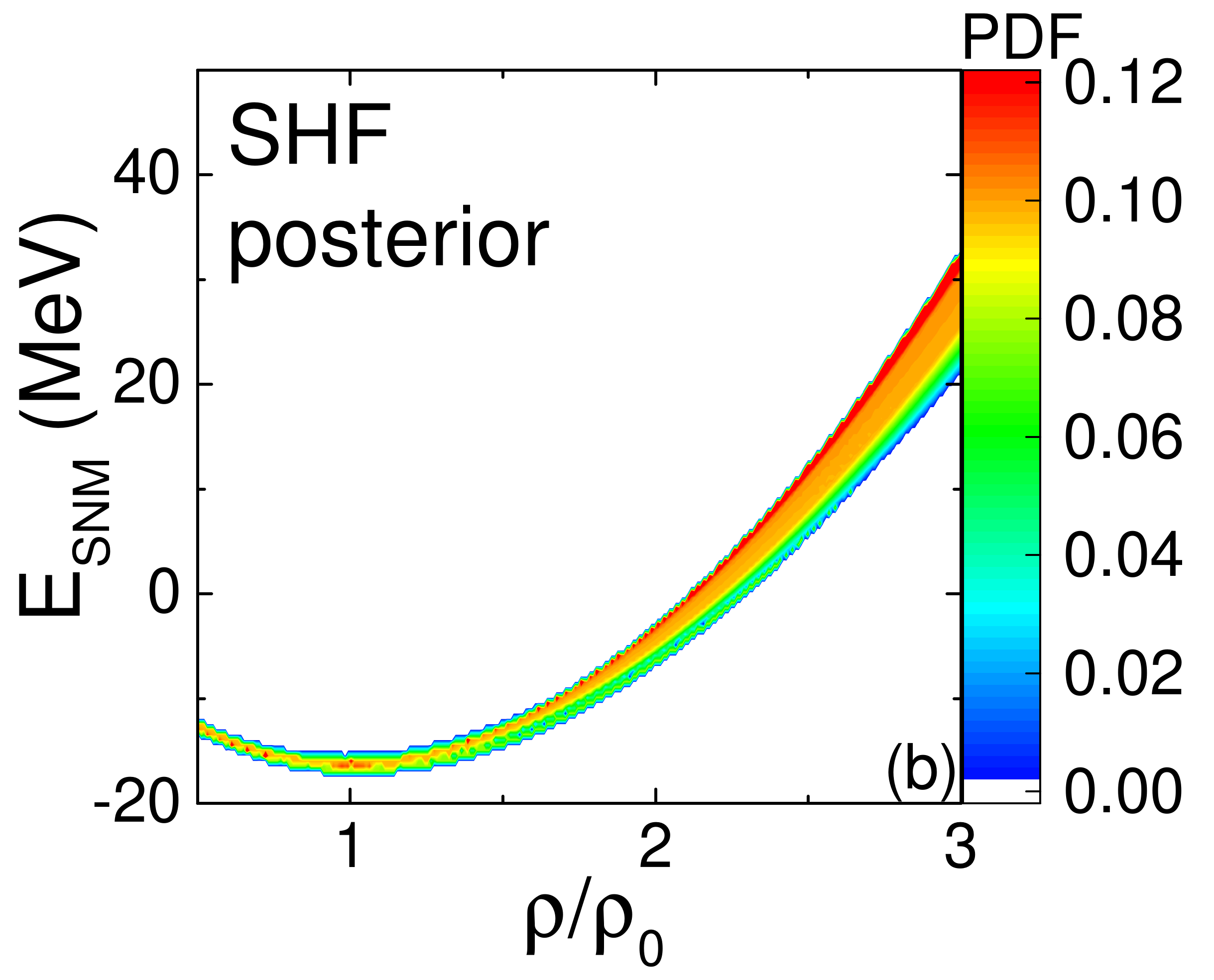}
\includegraphics[width=0.22\linewidth]{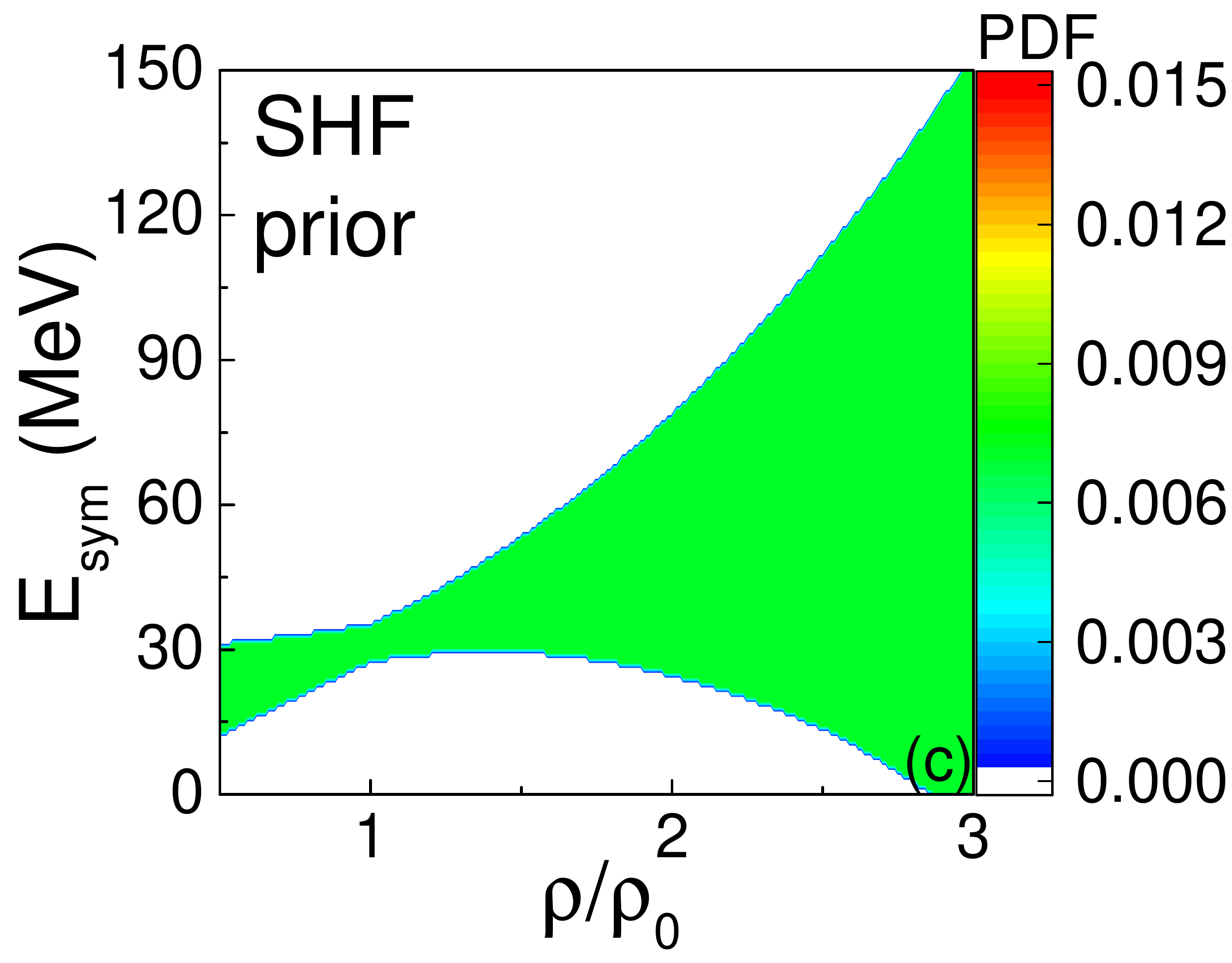}
\includegraphics[width=0.22\linewidth]{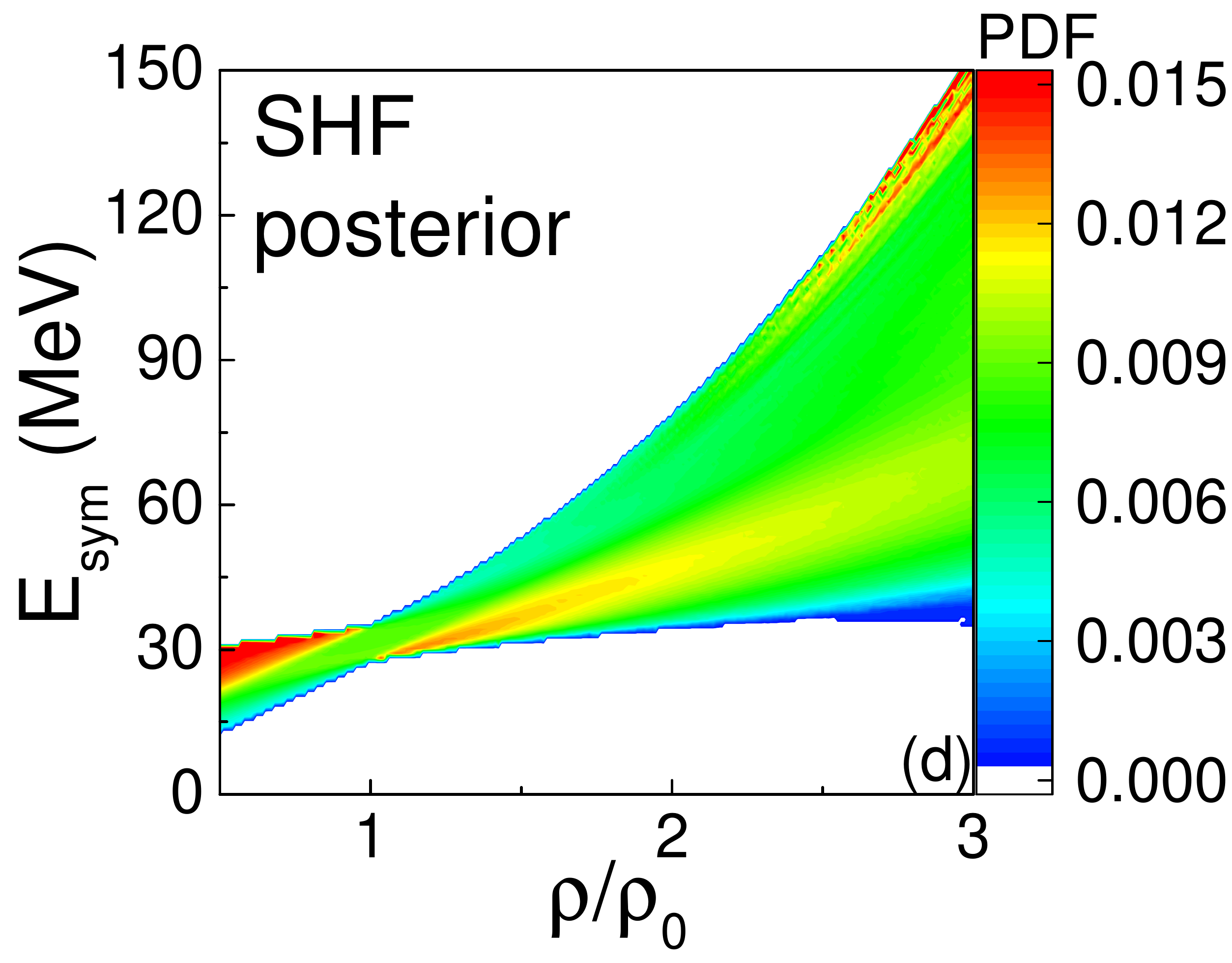}\\
\includegraphics[width=0.22\linewidth]{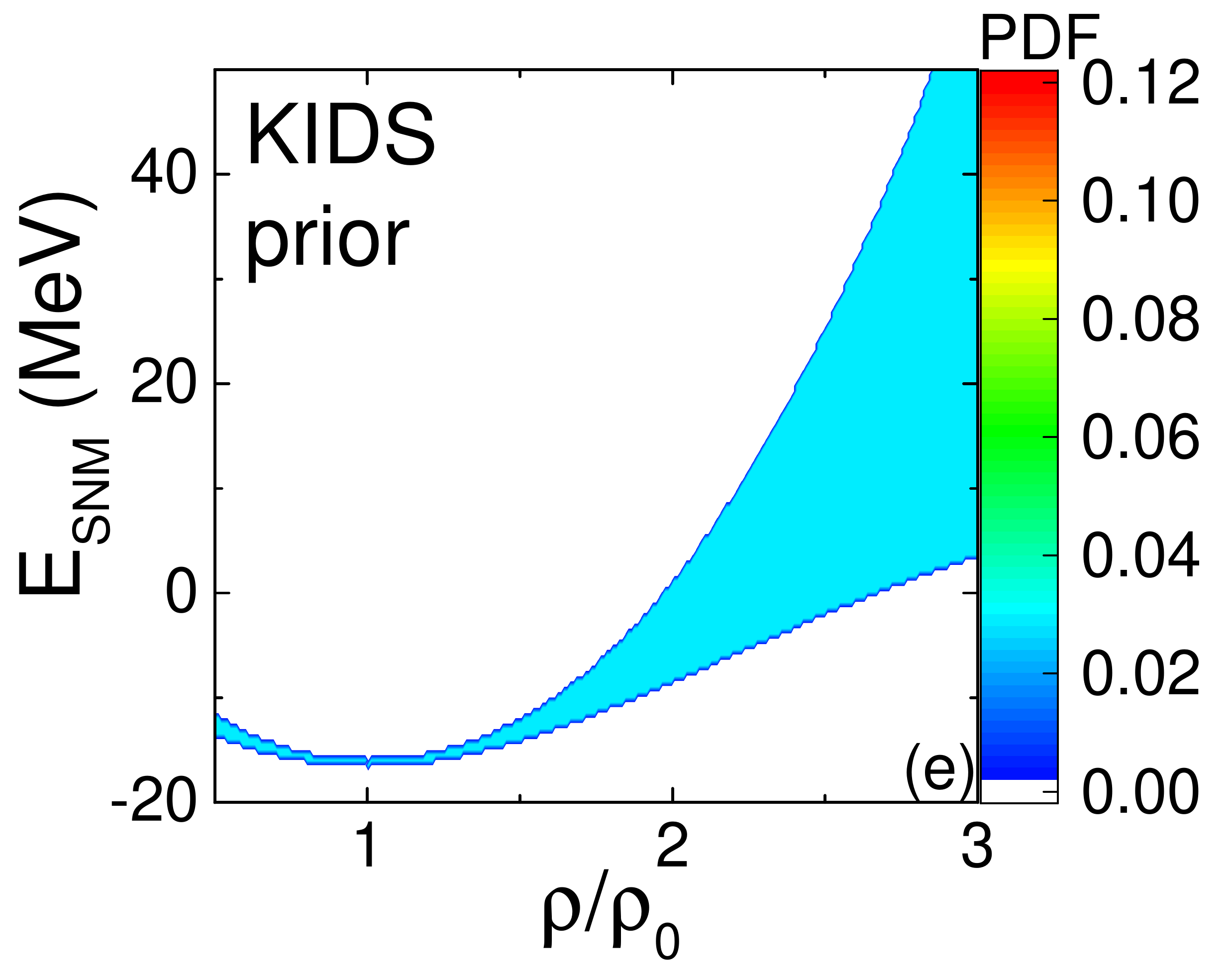}
\includegraphics[width=0.22\linewidth]{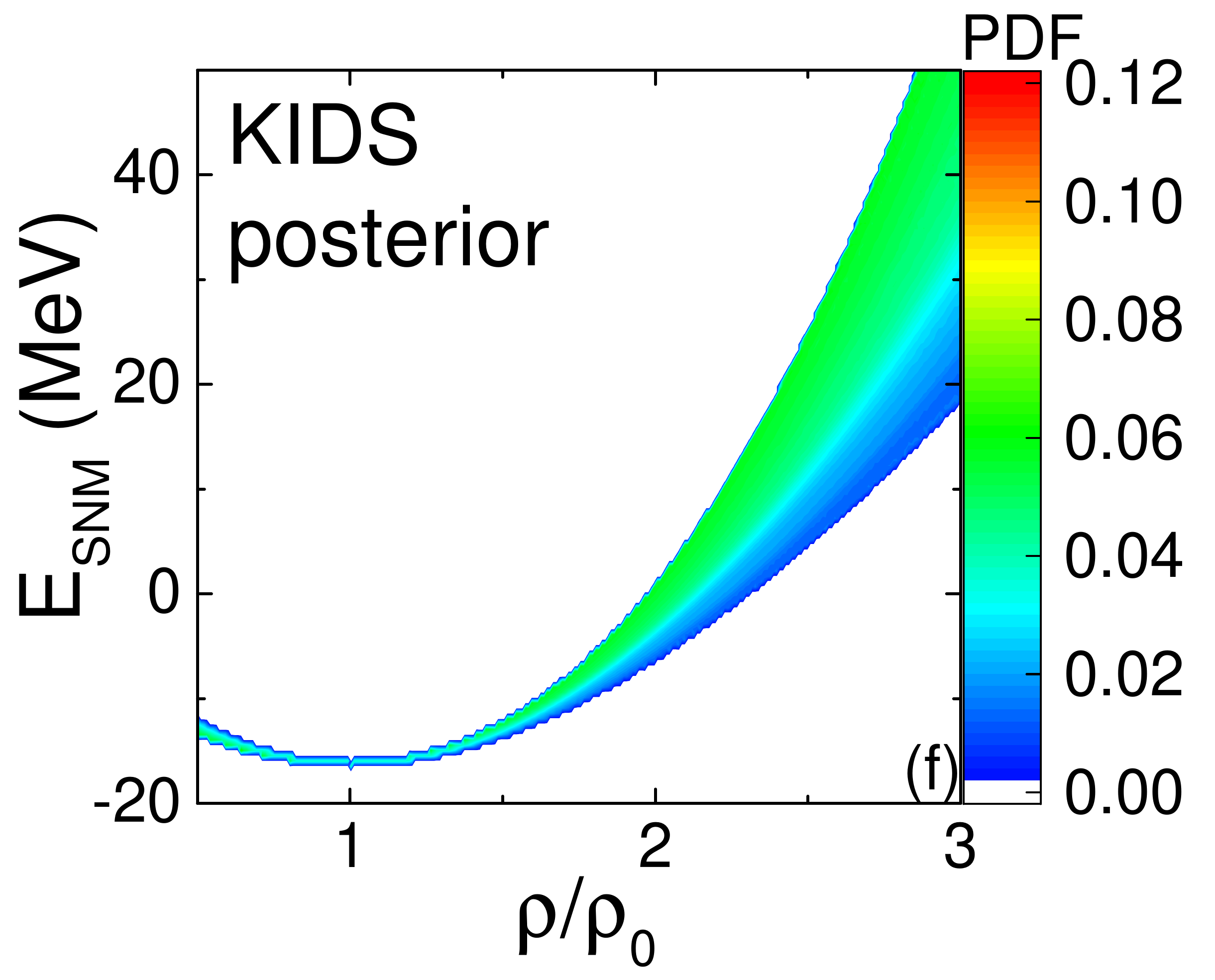}
\includegraphics[width=0.22\linewidth]{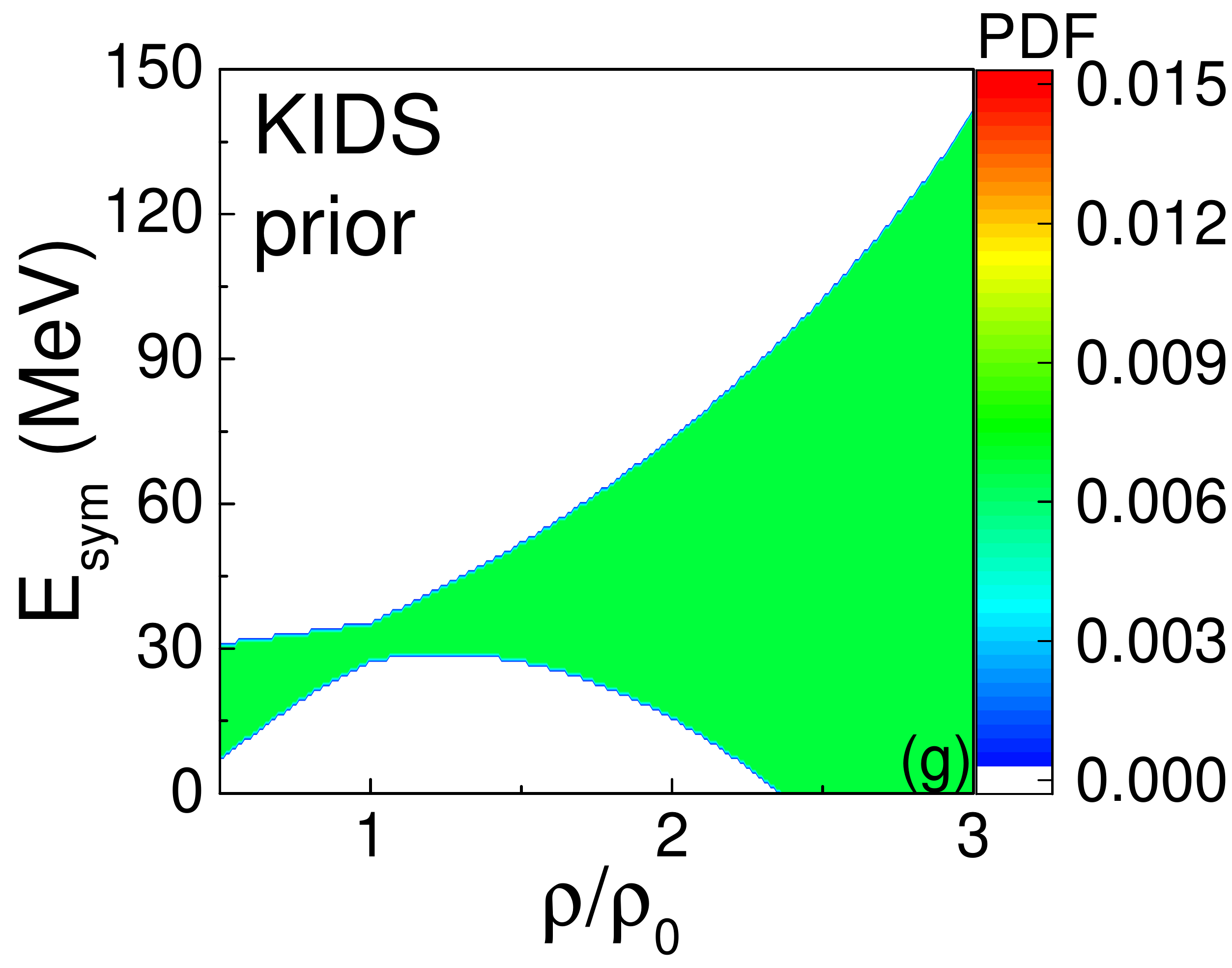}
\includegraphics[width=0.22\linewidth]{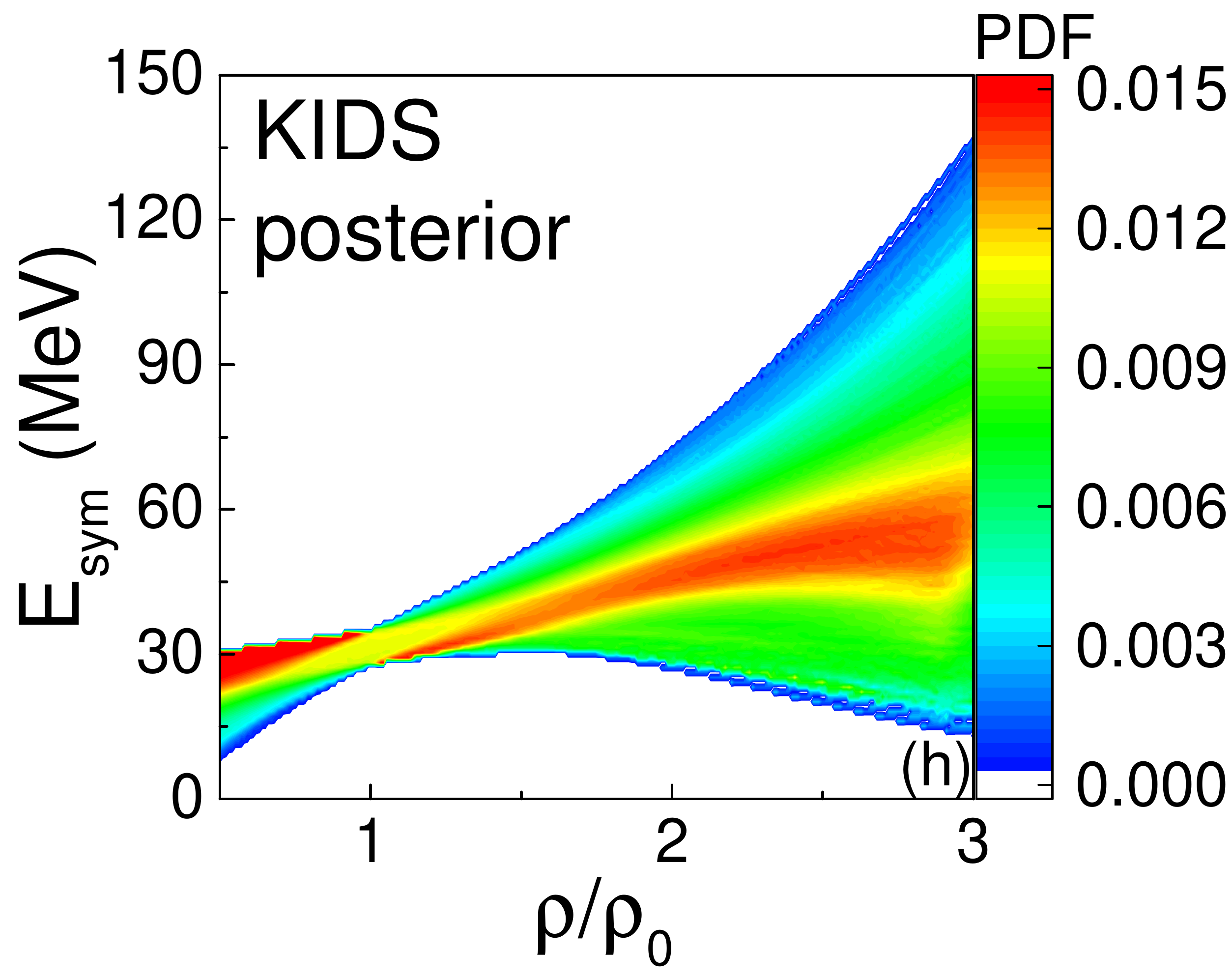}\\
\includegraphics[width=0.22\linewidth]{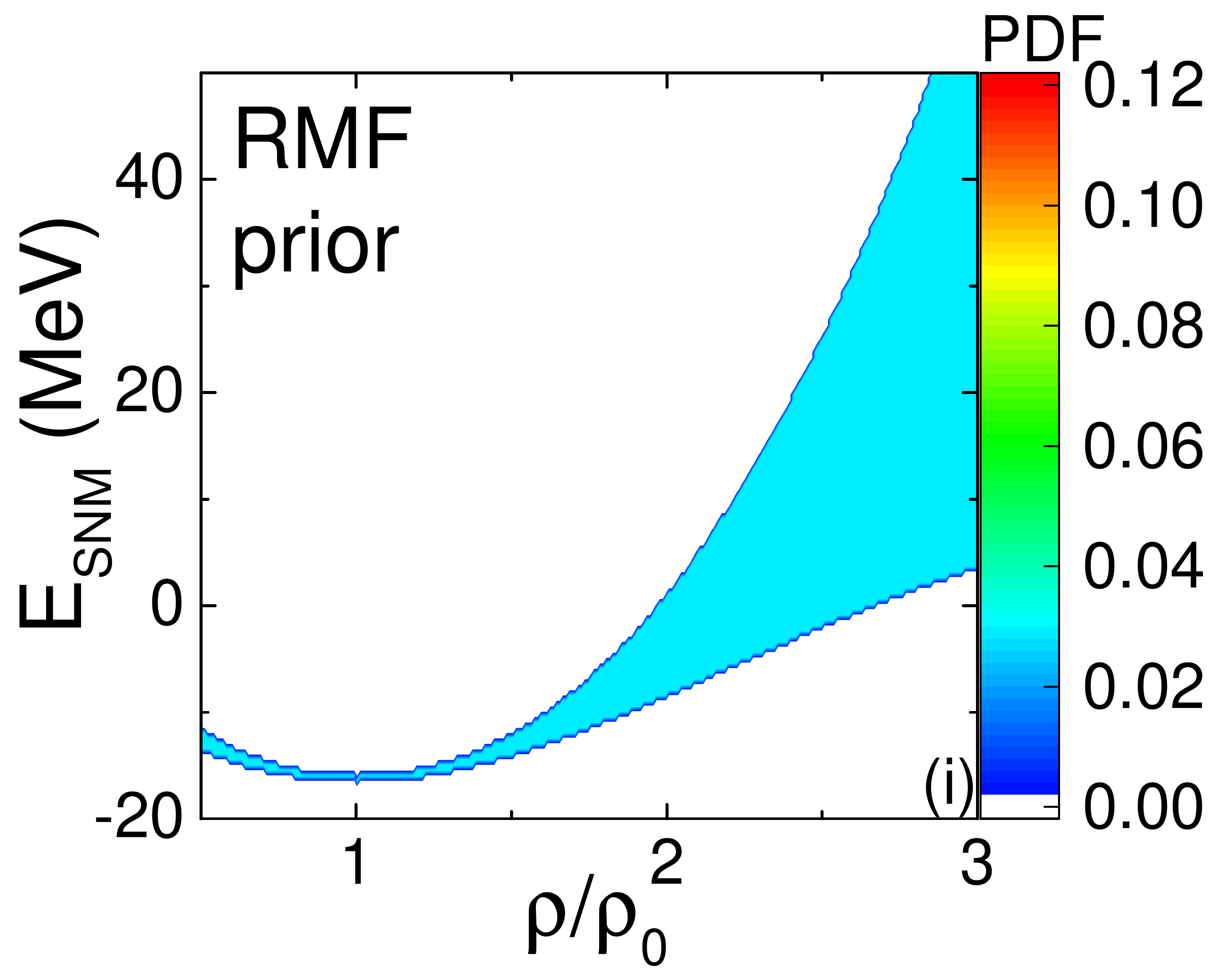}
\includegraphics[width=0.22\linewidth]{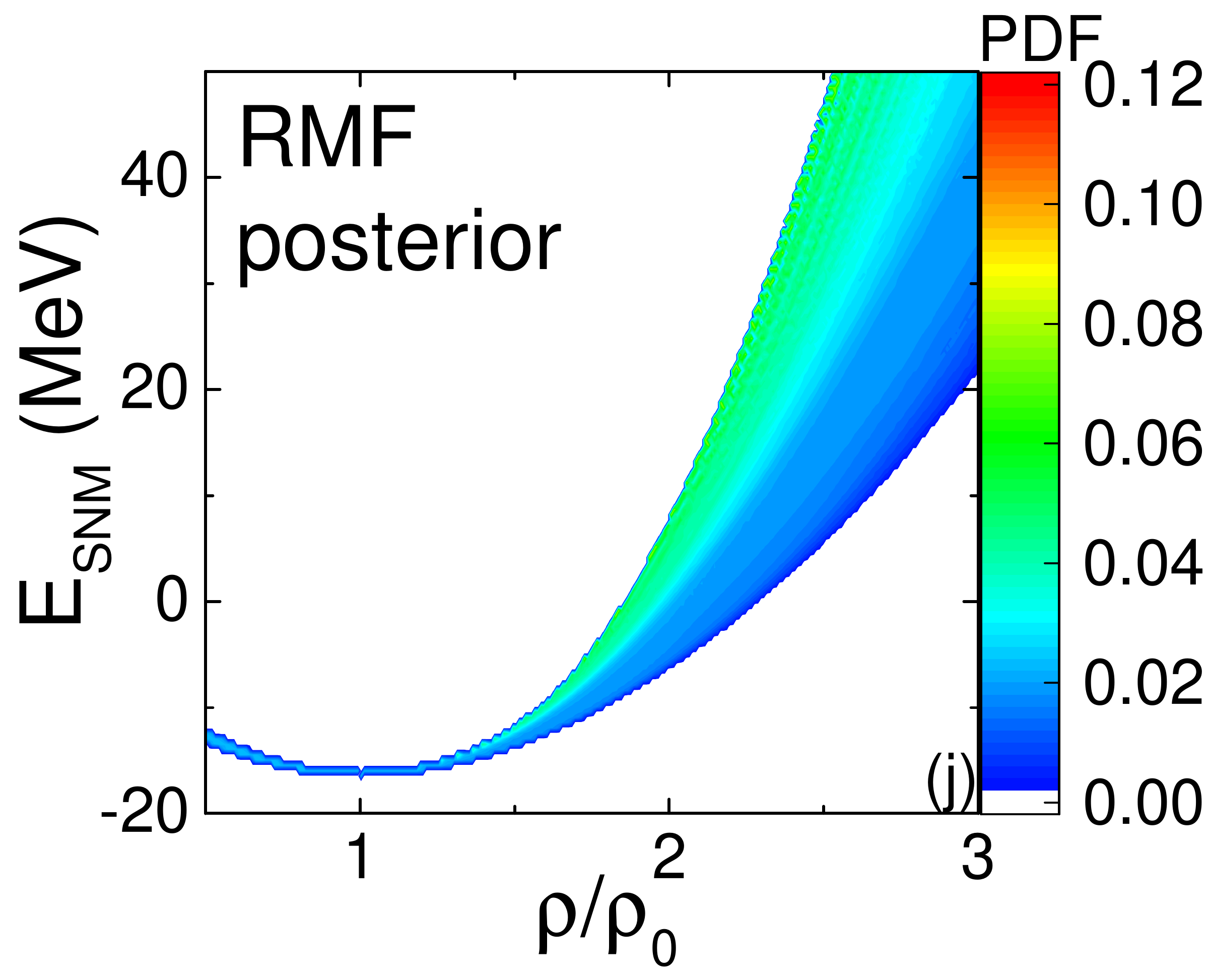}
\includegraphics[width=0.22\linewidth]{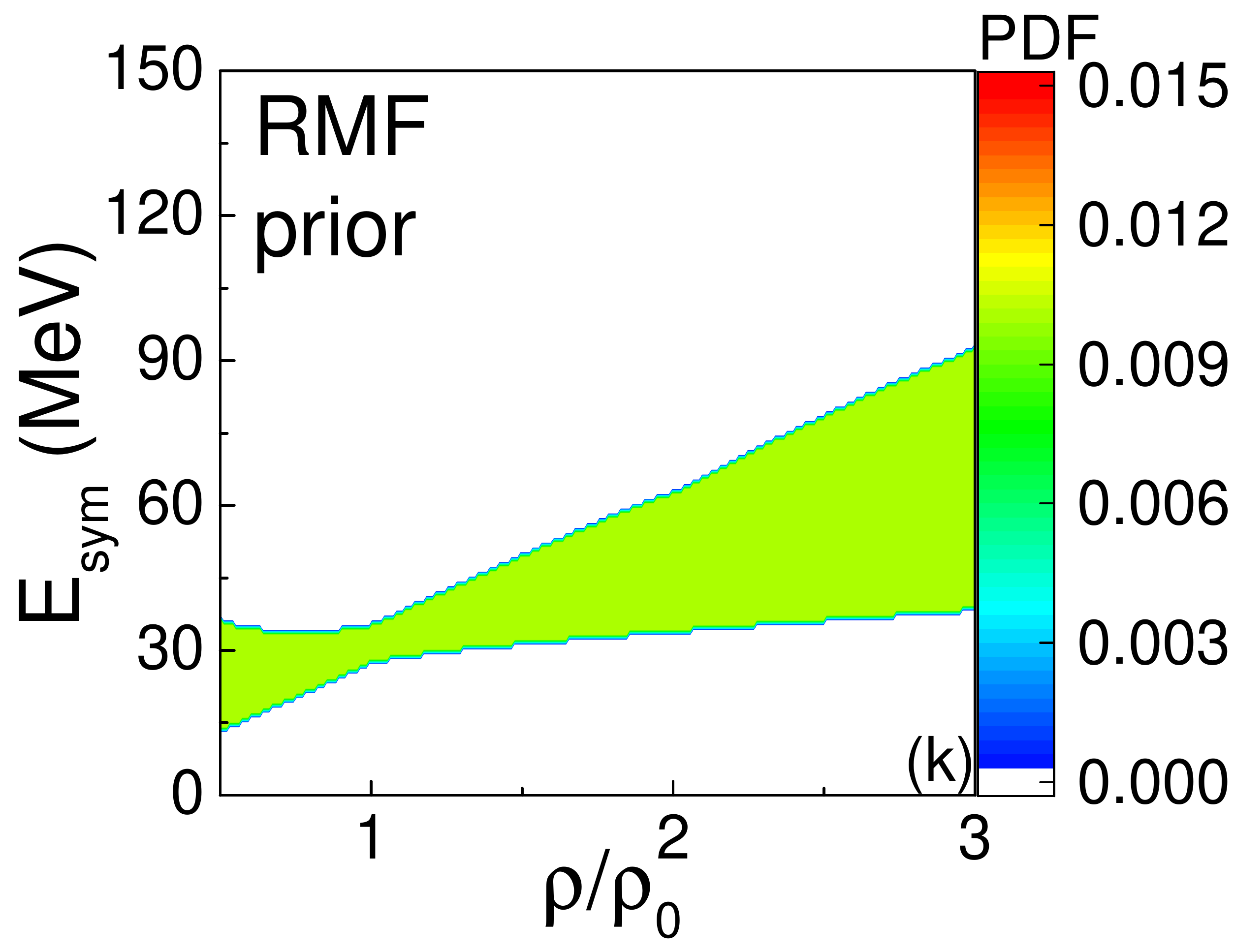}
\includegraphics[width=0.22\linewidth]{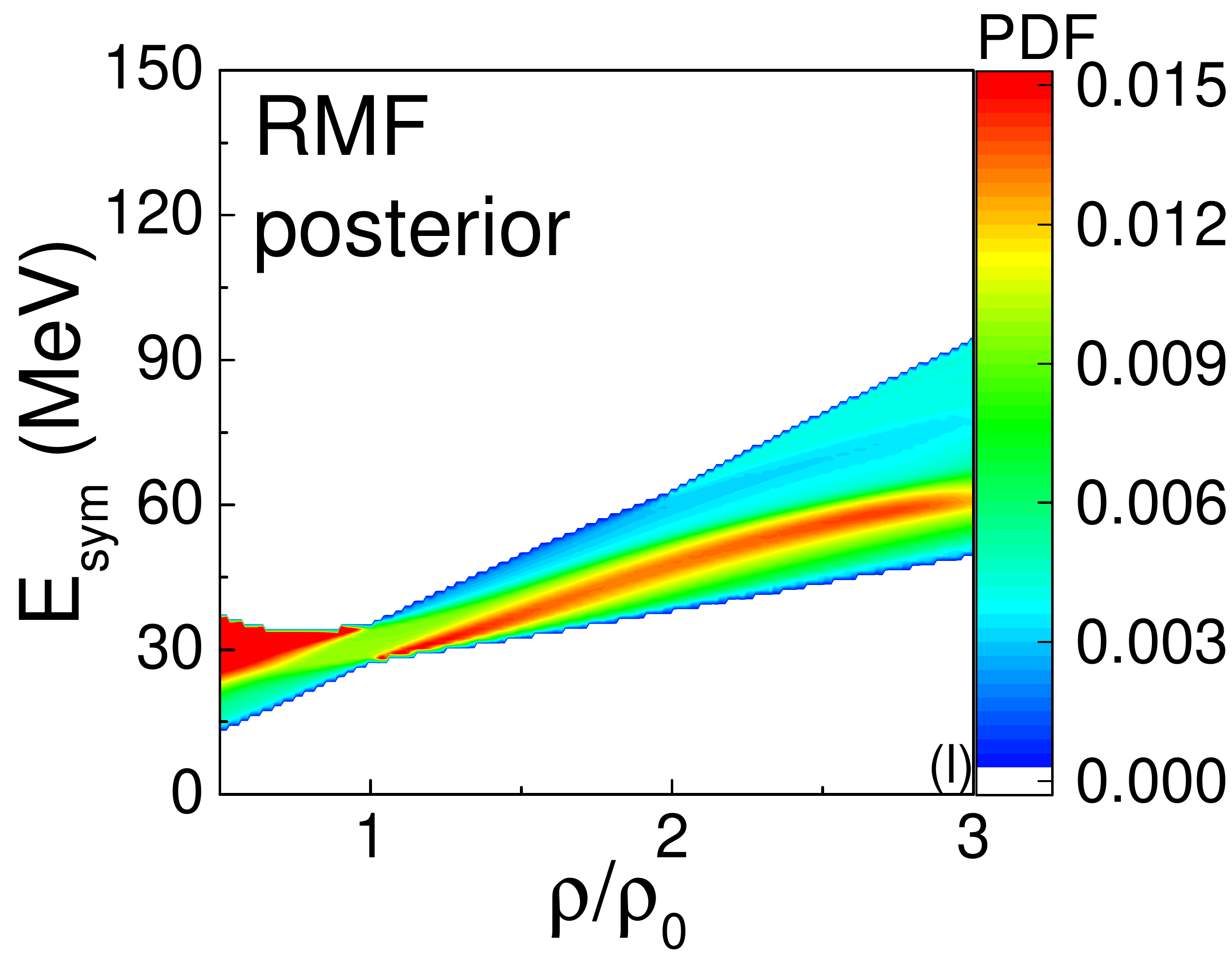}
\caption{\label{EOS} Probability distributions of $E_{SNM}(\rho)$ and $E_{sym}(\rho)$ from parameter ranges in Table~\ref{T1} (prior) and under the constraints of astrophysical observables (posterior) in the standard SHF model (top), the KIDS model (middle), and the RMF model (bottom).}
\end{figure*}

\begin{figure}[!h]
\includegraphics[width=0.8\linewidth]{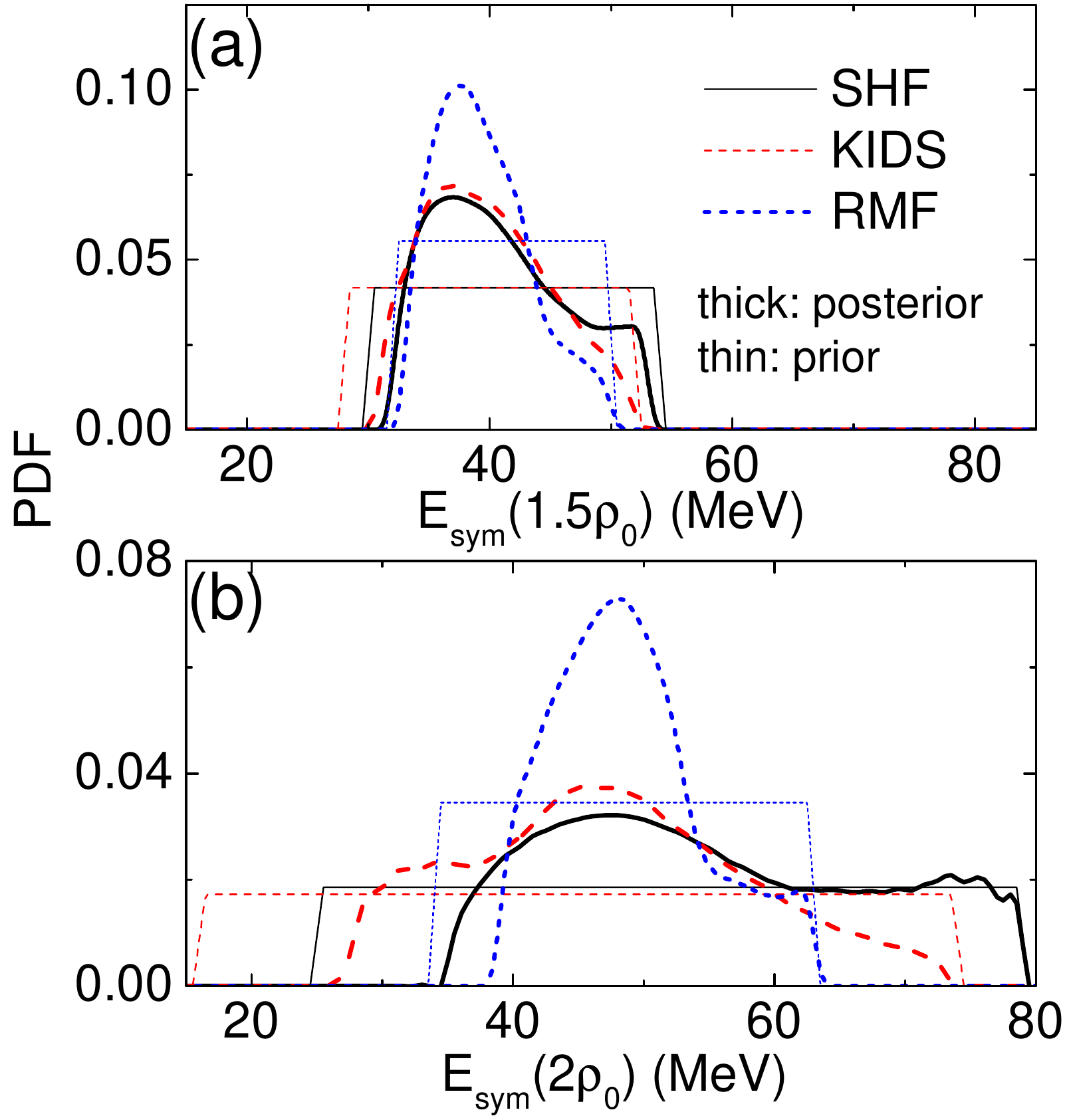}
\caption{\label{Esym} Posterior and prior PDFs of $E_{sym}(\rho)$ at $\rho=1.5\rho_0$ (a) and $2\rho_0$ (b) in the standard SHF model, the KIDS model, and the RMF model.}
\end{figure}

The parameters of effective models are constrained by the astrophysical observables through the Bayesian analysis, resulting in the constraints on the EOS of nuclear matter characterized by $E_{SNM}(\rho)$ and $E_{sym}(\rho)$ according to the EDF. Based on the three effective models, we compare the prior and posterior probability distributions of $E_{SNM}(\rho)$ and $E_{sym}(\rho)$ in Fig.~\ref{EOS}, with the prior distribution obtained based on parameter ranges in Table~\ref{T1}, and the posterior distribution from the Bayesian analysis under the constraints of astrophysical observables. Since the major constraints from the adopted astrophysical observables are on the EOS around and above the saturation density, these figures are plotted in the density range from $0.5\rho_0$ to $3\rho_0$. One sees that the prior distributions of both $E_{SNM}(\rho)$ and $E_{sym}(\rho)$ are broader in the KIDS model than in the standard SHF model, due to the larger parameter space in the KIDS model. While a broad prior distribution of $E_{SNM}(\rho)$ is seen in the RMF model, that of $E_{sym}(\rho)$ is very different from the other two models. A large neutron star mass favors a stiffer $E_{SNM}(\rho)$, especially for the KIDS model and the RMF model, where $Q_0$ is incorporated as an independent variable. The posterior $E_{SNM}(\rho)$ is even stiffer in the RMF model than in the KIDS model, since in the former case the resulting smaller Dirac effective mass also stiffens the EOS. While a very stiff $E_{sym}(\rho)$ is still favored to support a heavy neutron star in the standard SHF model, the radius data mostly favors a moderately soft $E_{sym}(\rho)$ at suprasaturation densities, corresponding to a small $L$ and a large $K_{sym}$ from Fig.~\ref{pdf}, based on all three models. The resulting soft $E_{sym}(\rho)$ is qualitatively consistent with results from other studies based on nucleonic models~\cite{Margueron:2017eqc,Lim:2018bkq,Yue:2021yfx,Newton:2021yru,Zhu:2022ibs}, where different astrophysical observables are adopted.

The symmetry energy at suprasaturation densities is of special interest for the nuclear physics community, and we compare its values at $\rho=1.5\rho_0$ and $2\rho_0$ from the constraints of astrophysical observables based on the three effective models in Fig.~\ref{Esym}. As expected, the constraint on $E_{sym}(1.5\rho_0)$ is stronger than that on $E_{sym}(2\rho_0)$. One sees that the RMF model gives the most stringent constraint of the symmetry energy at suprasaturation densities, mostly due to the narrow prior range of $E_{sym}$, compared to the other two models. Interestingly, despite of the different widths of the posterior PDFs, $E_{sym}(1.5\rho_0)$ peak around 38 MeV and $E_{sym}(2\rho_0)$ peak around 48 MeV for all three models. Within $68\%$ confidence intervals, we obtain $E_{sym}(1.5\rho_0)=38^{+6}_{-5}$ MeV in the standard SHF model, $E_{sym}(1.5\rho_0)=38^{+6}_{-5}$ MeV in the KIDS model, and $E_{sym}(1.5\rho_0)=38^{+4}_{-4}$ MeV in the RMF model, and we obtain $E_{sym}(2\rho_0)=48^{+15}_{-11}$ MeV in the standard SHF model, $E_{sym}(2\rho_0)=48^{+8}_{-15}$ MeV in the KIDS model, and $E_{sym}(2\rho_0)=48^{+5}_{-6}$ MeV in the RMF model. Our constraints of $E_{sym}(2\rho_0)$ are in good agreement with the fiducial value of about 47 MeV (see Fig. 1 of Ref.~\cite{Xie:2020tdo} and corresponding discussions).

\section{Summary and outlook}
\label{sec:summary}

Based on three effective nuclear interactions, we have studied the constraints on the EOS of both isoscalar and isovector channels from adopted astrophysical observables using the Bayesian approach. In all three models, i.e., the standard SHF model, the KIDS model, and the RMF model, a stiff isoscalar EOS is favored by the heavy mass of PSR J0740+6620. While a soft symmetry energy with a small $L$ is favored by the empirical radii of canonical neutron stars, $K_{sym}>-200$ MeV is favored by the radius of PSR J0740+6620. Due to the limit number of independent parameters in the SHF model, higher-order EOS parameters are related to lower-order ones, and correlation between EOS parameters are observed under the astrophysical constraints. With higher-order EOS parameters incorporated as independent variables, there are almost no such correlations between different EOS parameters in the KIDS and RMF models. The resulting smaller Dirac effective mass in the relativistic model further stiffens the isoscalar EOS compared to the non-relativistic models. In the RMF model, the parameter space is intrinsically limited in order to get physical solutions of model coefficients, and this leads to a different and actually more narrow constraint on the symmetry energy at suprasaturation densities. The symmetry energy at twice saturation density is constrained to be $48^{+15}_{-11}$ MeV in the standard SHF model, $48^{+8}_{-15}$ MeV in the KIDS model, and $48^{+5}_{-6}$ MeV in the RMF model, within their $68\%$ confidence intervals, and these values are in good agreement with those from state-of-art studies.

In the present study, three models with different numbers of free parameters and EDF forms are compared. While there are some model dependencies, the constraints from the adopted astrophysical observables on the EOS, especially on the $E_{sym}(\rho)$ at $\rho=1-2 \rho_0$, are robust and less sensitive to model details. Generally, increasing the number of parameters enhances the flexibility of the model and further enables it to explain better the data. On the other hand, a model with less number of free parameters but has a stronger prediction power is always favored. With the limited astrophysical observables adopted in the present study, although we are unable to judge the effectiveness of the three models, some lessons have been learnt. On the other hand, while all three models are nucleonic models, one can consider them as effective models to mimic the high-density EOS with hyperon or quark degrees of freedom. Non-parameterized models, e.g., studies using Gaussian processes~\cite{Essick:2021kjb}, are free from the possible hadron-quark phase transition at high densities. Furthermore, it will be of great interest to constrain the EOS parameters from not only astrophysical observables but also nuclear structure data, e.g., neutron-skin thickness and nucleus resonances, based on different models. In that case, we can constrain the EOS from high to low densities, and have a deeper understanding on the performance of EDFs from effective nuclear interactions.

\appendix

\section{Limited parameter space for the RMF model}
\label{app}

\begin{figure*}[!h]
\includegraphics[width=0.3\linewidth]{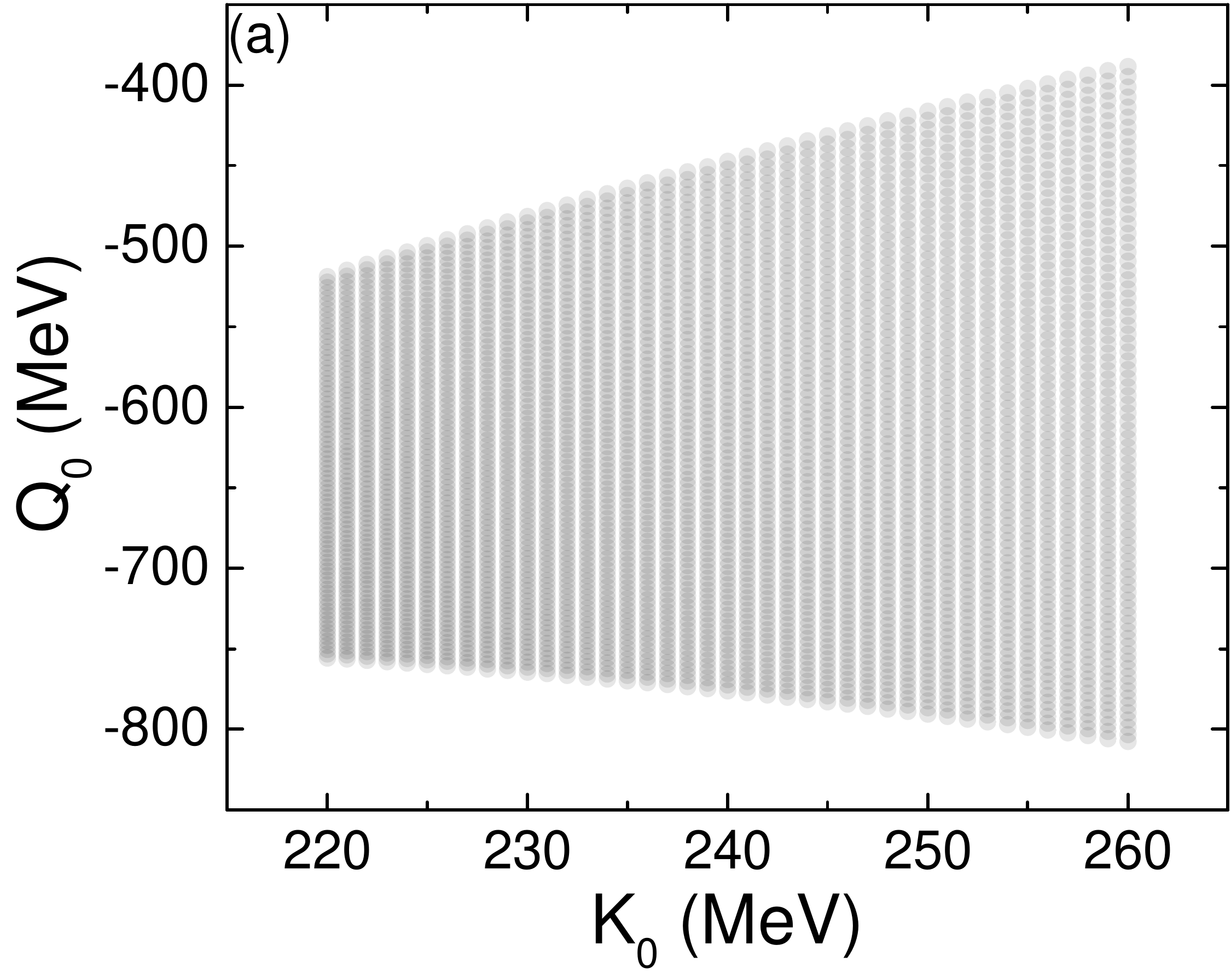}
\includegraphics[width=0.31\linewidth]{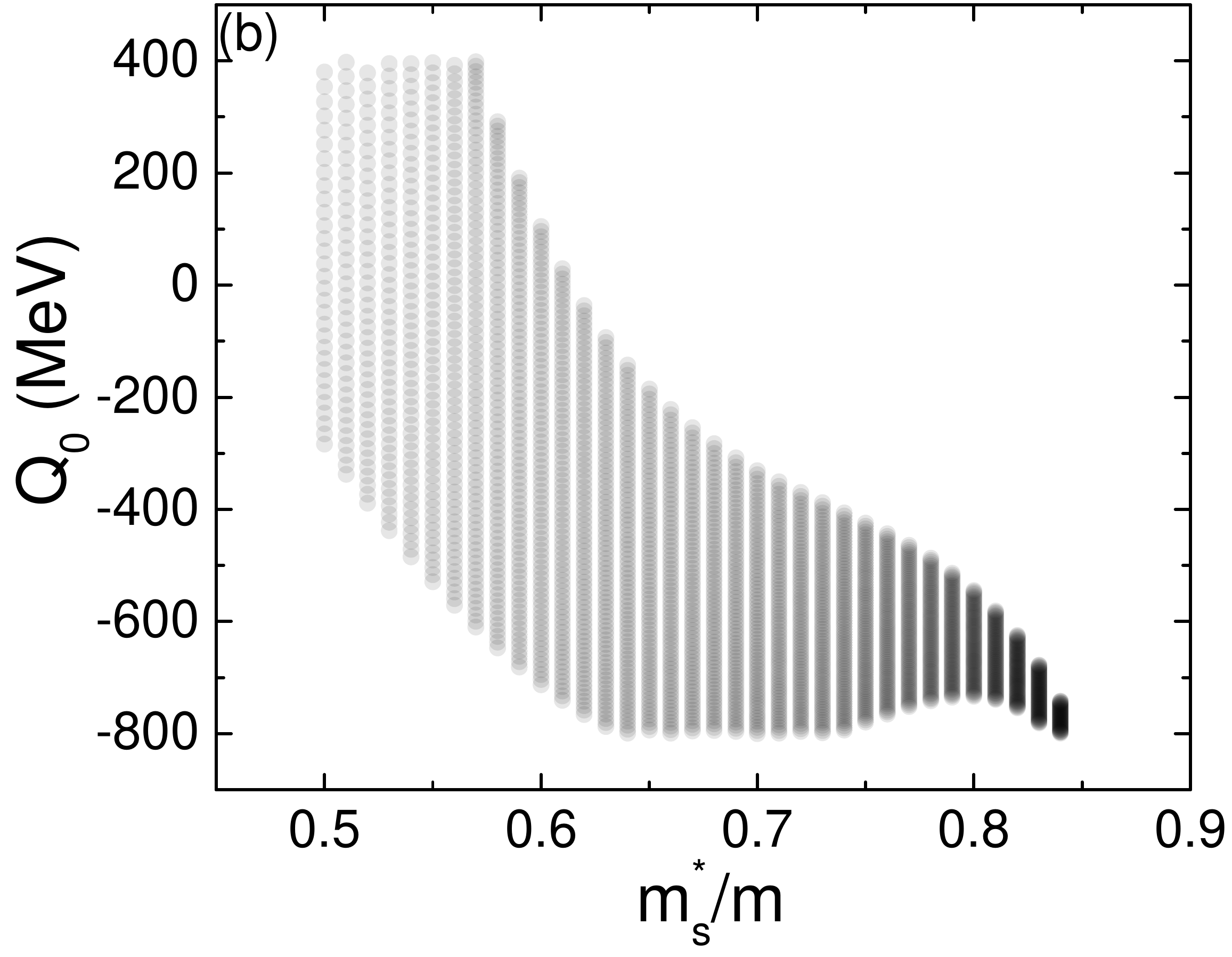}
\includegraphics[width=0.28\linewidth]{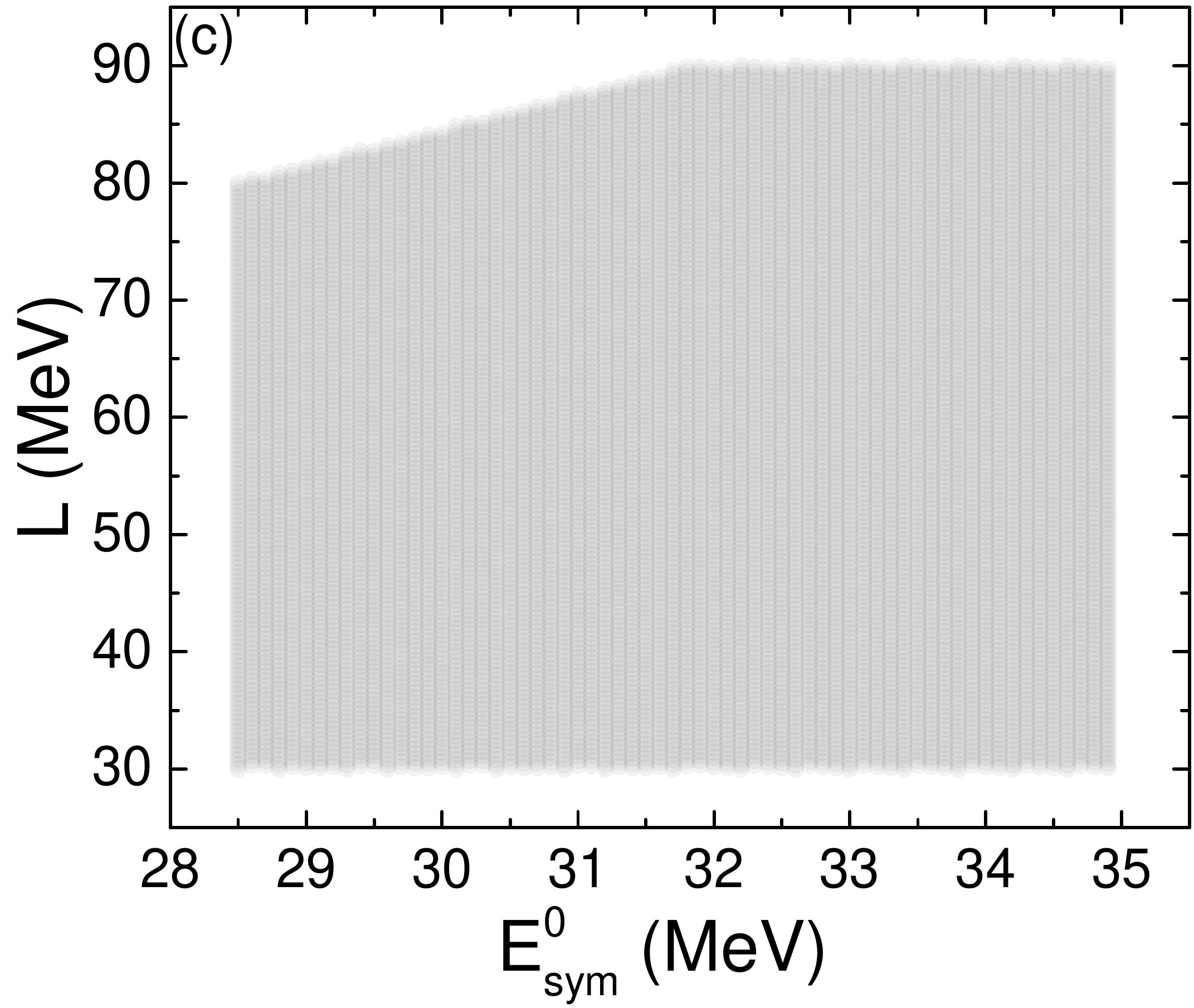}
\caption{\label{rmf_limit} Illustration of the limited parameter space in the $K_0-Q_0$ plane (a), the $m_s^\star/m-Q_0$ plane (b), and the $E_{sym}^0-L$ plane (c) with other parameters set as their default values in Table~\ref{T1} in the RMF model.}
\end{figure*}

In the present study on neutron stars using the Lagrangian form as Eq.~(\ref{lrmf}) in the RMF model, we set additional constraints of $C>0$ and $\alpha_3^\prime>0$, otherwise the field equations [Eqs.~(\ref{eqsigma})-(\ref{eqrho})] do not necessarily have solutions in asymmetric nuclear matter at high densities for an arbitrary parameter set, especially for $\omega_0$. In addition, the square of the coupling constants ($g_\sigma^2$, $g_\omega^2$, and $g_\rho^2$) calculated inversely from macroscopic physics quantities must be positive. These lead to certain intrinsic limits of the parameter space for the present EDF of the RMF model. For instance, with other parameters set as their default values as in Table.~\ref{T1}, the value of $Q_0$ can only be varied within about $-400$ to $-800$ MeV by changing the value of $K_0$, mapping out a much smaller space compared to its prior range, as shown in Fig.~\ref{rmf_limit} (a).  We have also observed that the available values in the $m_s^\star/m-Q_0$ plane are quite limited as shown in Fig.~\ref{rmf_limit} (b). In addition, a large $L$ can't be achieved for a small $E_{sym}^0$ MeV as shown in Fig.~\ref{rmf_limit} (c).

\begin{acknowledgments}
We acknowledge helpful discussions with Bao-Jun Cai and Lu-Meng Liu. JX is supported by the Strategic Priority Research Program of the Chinese Academy of Sciences under Grant No. XDB34030000, the National Natural Science Foundation of China under Grant No. 11922514, and the Fundamental Research Funds for the Central Universities. PP is supported by the Rare Isotope Science Project of the Institute for Basic Science funded by the Ministry of Science, ICT and Future Planning and the National Research Foundation (NRF) of Korea (2013M7A1A1075764).
\end{acknowledgments}

\bibliography{Nstar_bayes}

\begin{thebibliography}{63}%
\makeatletter
\providecommand \@ifxundefined [1]{%
 \@ifx{#1\undefined}
}%
\providecommand \@ifnum [1]{%
 \ifnum #1\expandafter \@firstoftwo
 \else \expandafter \@secondoftwo
 \fi
}%
\providecommand \@ifx [1]{%
 \ifx #1\expandafter \@firstoftwo
 \else \expandafter \@secondoftwo
 \fi
}%
\providecommand \natexlab [1]{#1}%
\providecommand \enquote  [1]{``#1''}%
\providecommand \bibnamefont  [1]{#1}%
\providecommand \bibfnamefont [1]{#1}%
\providecommand \citenamefont [1]{#1}%
\providecommand \href@noop [0]{\@secondoftwo}%
\providecommand \href [0]{\begingroup \@sanitize@url \@href}%
\providecommand \@href[1]{\@@startlink{#1}\@@href}%
\providecommand \@@href[1]{\endgroup#1\@@endlink}%
\providecommand \@sanitize@url [0]{\catcode `\\12\catcode `\$12\catcode
  `\&12\catcode `\#12\catcode `\^12\catcode `\_12\catcode `\%12\relax}%
\providecommand \@@startlink[1]{}%
\providecommand \@@endlink[0]{}%
\providecommand \url  [0]{\begingroup\@sanitize@url \@url }%
\providecommand \@url [1]{\endgroup\@href {#1}{\urlprefix }}%
\providecommand \urlprefix  [0]{URL }%
\providecommand \Eprint [0]{\href }%
\providecommand \doibase [0]{http://dx.doi.org/}%
\providecommand \selectlanguage [0]{\@gobble}%
\providecommand \bibinfo  [0]{\@secondoftwo}%
\providecommand \bibfield  [0]{\@secondoftwo}%
\providecommand \translation [1]{[#1]}%
\providecommand \BibitemOpen [0]{}%
\providecommand \bibitemStop [0]{}%
\providecommand \bibitemNoStop [0]{.\EOS\space}%
\providecommand \EOS [0]{\spacefactor3000\relax}%
\providecommand \BibitemShut  [1]{\csname bibitem#1\endcsname}%
\let\auto@bib@innerbib\@empty
\bibitem [{\citenamefont {Li}\ \emph {et~al.}(2021)\citenamefont {Li},
  \citenamefont {Cai}, \citenamefont {Xie},\ and\ \citenamefont
  {Zhang}}]{Li:2021thg}%
  \BibitemOpen
  \bibfield  {author} {\bibinfo {author} {\bibfnamefont {Bao-An}\ \bibnamefont
  {Li}}, \bibinfo {author} {\bibfnamefont {Bao-Jun}\ \bibnamefont {Cai}},
  \bibinfo {author} {\bibfnamefont {Wen-Jie}\ \bibnamefont {Xie}}, \ and\
  \bibinfo {author} {\bibfnamefont {Nai-Bo}\ \bibnamefont {Zhang}},\ }\bibfield
   {title} {\enquote {\bibinfo {title} {{Progress in Constraining Nuclear
  Symmetry Energy Using Neutron Star Observables Since GW170817}},}\ }\href
  {\doibase 10.3390/universe7060182} {\bibfield  {journal} {\bibinfo  {journal}
  {Universe}\ }\textbf {\bibinfo {volume} {7}},\ \bibinfo {pages} {182}
  (\bibinfo {year} {2021})},\ \Eprint {http://arxiv.org/abs/2105.04629}
  {arXiv:2105.04629 [nucl-th]} \BibitemShut {NoStop}%
\bibitem [{\citenamefont {Lattimer}(2021)}]{Lattimer:2021emm}%
  \BibitemOpen
  \bibfield  {author} {\bibinfo {author} {\bibfnamefont {J.~M.}\ \bibnamefont
  {Lattimer}},\ }\bibfield  {title} {\enquote {\bibinfo {title} {{Neutron Stars
  and the Nuclear Matter Equation of State}},}\ }\href {\doibase
  10.1146/annurev-nucl-102419-124827} {\bibfield  {journal} {\bibinfo
  {journal} {Ann. Rev. Nucl. Part. Sci.}\ }\textbf {\bibinfo {volume} {71}},\
  \bibinfo {pages} {433--464} (\bibinfo {year} {2021})}\BibitemShut {NoStop}%
\bibitem [{\citenamefont {Lattimer}\ and\ \citenamefont
  {Prakash}(2007)}]{Lattimer:2006xb}%
  \BibitemOpen
  \bibfield  {author} {\bibinfo {author} {\bibfnamefont {James~M.}\
  \bibnamefont {Lattimer}}\ and\ \bibinfo {author} {\bibfnamefont {Maddapa}\
  \bibnamefont {Prakash}},\ }\bibfield  {title} {\enquote {\bibinfo {title}
  {{Neutron Star Observations: Prognosis for Equation of State Constraints}},}\
  }\href {\doibase 10.1016/j.physrep.2007.02.003} {\bibfield  {journal}
  {\bibinfo  {journal} {Phys. Rept.}\ }\textbf {\bibinfo {volume} {442}},\
  \bibinfo {pages} {109--165} (\bibinfo {year} {2007})},\ \Eprint
  {http://arxiv.org/abs/astro-ph/0612440} {arXiv:astro-ph/0612440} \BibitemShut
  {NoStop}%
\bibitem [{\citenamefont {Khan}\ \emph {et~al.}(2012)\citenamefont {Khan},
  \citenamefont {Margueron},\ and\ \citenamefont
  {Vida\~na}}]{PhysRevLett.109.092501}%
  \BibitemOpen
  \bibfield  {author} {\bibinfo {author} {\bibfnamefont {E.}~\bibnamefont
  {Khan}}, \bibinfo {author} {\bibfnamefont {J.}~\bibnamefont {Margueron}}, \
  and\ \bibinfo {author} {\bibfnamefont {I.}~\bibnamefont {Vida\~na}},\
  }\bibfield  {title} {\enquote {\bibinfo {title} {Constraining the nuclear
  equation of state at subsaturation densities},}\ }\href {\doibase
  10.1103/PhysRevLett.109.092501} {\bibfield  {journal} {\bibinfo  {journal}
  {Phys. Rev. Lett.}\ }\textbf {\bibinfo {volume} {109}},\ \bibinfo {pages}
  {092501} (\bibinfo {year} {2012})}\BibitemShut {NoStop}%
\bibitem [{\citenamefont {Margueron}\ \emph
  {et~al.}(2018{\natexlab{a}})\citenamefont {Margueron}, \citenamefont
  {Hoffmann~Casali},\ and\ \citenamefont {Gulminelli}}]{PhysRevC.97.025805}%
  \BibitemOpen
  \bibfield  {author} {\bibinfo {author} {\bibfnamefont {J\'er\^ome}\
  \bibnamefont {Margueron}}, \bibinfo {author} {\bibfnamefont {Rudiney}\
  \bibnamefont {Hoffmann~Casali}}, \ and\ \bibinfo {author} {\bibfnamefont
  {Francesca}\ \bibnamefont {Gulminelli}},\ }\bibfield  {title} {\enquote
  {\bibinfo {title} {Equation of state for dense nucleonic matter from
  metamodeling. i. foundational aspects},}\ }\href {\doibase
  10.1103/PhysRevC.97.025805} {\bibfield  {journal} {\bibinfo  {journal} {Phys.
  Rev. C}\ }\textbf {\bibinfo {volume} {97}},\ \bibinfo {pages} {025805}
  (\bibinfo {year} {2018}{\natexlab{a}})}\BibitemShut {NoStop}%
\bibitem [{\citenamefont {Shlomo}\ \emph {et~al.}(2006)\citenamefont {Shlomo},
  \citenamefont {Kolomietz},\ and\ \citenamefont {Colo}}]{Shlomo2006}%
  \BibitemOpen
  \bibfield  {author} {\bibinfo {author} {\bibfnamefont {S.}~\bibnamefont
  {Shlomo}}, \bibinfo {author} {\bibfnamefont {V.~M.}\ \bibnamefont
  {Kolomietz}}, \ and\ \bibinfo {author} {\bibfnamefont {G.}~\bibnamefont
  {Colo}},\ }\bibfield  {title} {\enquote {\bibinfo {title} {{Deducing the
  nuclear-matter incompressibility coefficient from data on isoscalar
  compression modes}},}\ }\href {\doibase 10.1140/epja/i2006-10100-3}
  {\bibfield  {journal} {\bibinfo  {journal} {Eur. Phys. J. A}\ }\textbf
  {\bibinfo {volume} {30}},\ \bibinfo {pages} {23} (\bibinfo {year}
  {2006})}\BibitemShut {NoStop}%
\bibitem [{\citenamefont {Colo}\ \emph {et~al.}(2014)\citenamefont {Colo},
  \citenamefont {Garg},\ and\ \citenamefont {Sagawa}}]{Colo:2013yta}%
  \BibitemOpen
  \bibfield  {author} {\bibinfo {author} {\bibfnamefont {G.}~\bibnamefont
  {Colo}}, \bibinfo {author} {\bibfnamefont {U.}~\bibnamefont {Garg}}, \ and\
  \bibinfo {author} {\bibfnamefont {H.}~\bibnamefont {Sagawa}},\ }\bibfield
  {title} {\enquote {\bibinfo {title} {{Symmetry energy from the nuclear
  collective motion: constraints from dipole, quadrupole, monopole and
  spin-dipole resonances}},}\ }\href {\doibase 10.1140/epja/i2014-14026-9}
  {\bibfield  {journal} {\bibinfo  {journal} {Eur. Phys. J. A}\ }\textbf
  {\bibinfo {volume} {50}},\ \bibinfo {pages} {26} (\bibinfo {year} {2014})},\
  \Eprint {http://arxiv.org/abs/1309.1572} {arXiv:1309.1572 [nucl-th]}
  \BibitemShut {NoStop}%
\bibitem [{\citenamefont {Garg}\ and\ \citenamefont
  {Col\`o}(2018)}]{Garg:2018uam}%
  \BibitemOpen
  \bibfield  {author} {\bibinfo {author} {\bibfnamefont {Umesh}\ \bibnamefont
  {Garg}}\ and\ \bibinfo {author} {\bibfnamefont {Gianluca}\ \bibnamefont
  {Col\`o}},\ }\bibfield  {title} {\enquote {\bibinfo {title} {{The
  compression-mode giant resonances and nuclear incompressibility}},}\ }\href
  {\doibase 10.1016/j.ppnp.2018.03.001} {\bibfield  {journal} {\bibinfo
  {journal} {Prog. Part. Nucl. Phys.}\ }\textbf {\bibinfo {volume} {101}},\
  \bibinfo {pages} {55--95} (\bibinfo {year} {2018})},\ \Eprint
  {http://arxiv.org/abs/1801.03672} {arXiv:1801.03672 [nucl-ex]} \BibitemShut
  {NoStop}%
\bibitem [{\citenamefont {Li}\ and\ \citenamefont {Han}(2013)}]{LI2013276}%
  \BibitemOpen
  \bibfield  {author} {\bibinfo {author} {\bibfnamefont {Bao-An}\ \bibnamefont
  {Li}}\ and\ \bibinfo {author} {\bibfnamefont {Xiao}\ \bibnamefont {Han}},\
  }\bibfield  {title} {\enquote {\bibinfo {title} {Constraining the
  neutron-proton effective mass splitting using empirical constraints on the
  density dependence of nuclear symmetry energy around normal density},}\
  }\href {\doibase https://doi.org/10.1016/j.physletb.2013.10.006} {\bibfield
  {journal} {\bibinfo  {journal} {Physics Letters B}\ }\textbf {\bibinfo
  {volume} {727}},\ \bibinfo {pages} {276--281} (\bibinfo {year}
  {2013})}\BibitemShut {NoStop}%
\bibitem [{\citenamefont {Oertel}\ \emph {et~al.}(2017)\citenamefont {Oertel},
  \citenamefont {Hempel}, \citenamefont {Kl\"ahn},\ and\ \citenamefont
  {Typel}}]{RevModPhys.89.015007}%
  \BibitemOpen
  \bibfield  {author} {\bibinfo {author} {\bibfnamefont {M.}~\bibnamefont
  {Oertel}}, \bibinfo {author} {\bibfnamefont {M.}~\bibnamefont {Hempel}},
  \bibinfo {author} {\bibfnamefont {T.}~\bibnamefont {Kl\"ahn}}, \ and\
  \bibinfo {author} {\bibfnamefont {S.}~\bibnamefont {Typel}},\ }\bibfield
  {title} {\enquote {\bibinfo {title} {Equations of state for supernovae and
  compact stars},}\ }\href {\doibase 10.1103/RevModPhys.89.015007} {\bibfield
  {journal} {\bibinfo  {journal} {Rev. Mod. Phys.}\ }\textbf {\bibinfo {volume}
  {89}},\ \bibinfo {pages} {015007} (\bibinfo {year} {2017})}\BibitemShut
  {NoStop}%
\bibitem [{\citenamefont {Cromartie}\ \emph {et~al.}(2019)\citenamefont
  {Cromartie}, \citenamefont {Fonseca}, \citenamefont {Ransom}, \citenamefont
  {Demorest}, \citenamefont {Arzoumanian}, \citenamefont {Blumer},
  \citenamefont {Brook}, \citenamefont {DeCesar}, \citenamefont {Dolch},
  \citenamefont {Ellis}, \citenamefont {Ferdman}, \citenamefont {Ferrara},
  \citenamefont {Garver-Daniels}, \citenamefont {Gentile}, \citenamefont
  {Jones}, \citenamefont {Lam}, \citenamefont {Lorimer}, \citenamefont {Lynch},
  \citenamefont {McLaughlin}, \citenamefont {Ng}, \citenamefont {Nice},
  \citenamefont {Pennucci}, \citenamefont {Spiewak}, \citenamefont {Stairs},
  \citenamefont {Stovall}, \citenamefont {Swiggum},\ and\ \citenamefont
  {Zhu}}]{Cromartie_2019}%
  \BibitemOpen
  \bibfield  {author} {\bibinfo {author} {\bibfnamefont {H.~T.}\ \bibnamefont
  {Cromartie}}, \bibinfo {author} {\bibfnamefont {E.}~\bibnamefont {Fonseca}},
  \bibinfo {author} {\bibfnamefont {S.~M.}\ \bibnamefont {Ransom}}, \bibinfo
  {author} {\bibfnamefont {P.~B.}\ \bibnamefont {Demorest}}, \bibinfo {author}
  {\bibfnamefont {Z.}~\bibnamefont {Arzoumanian}}, \bibinfo {author}
  {\bibfnamefont {H.}~\bibnamefont {Blumer}}, \bibinfo {author} {\bibfnamefont
  {P.~R.}\ \bibnamefont {Brook}}, \bibinfo {author} {\bibfnamefont {M.~E.}\
  \bibnamefont {DeCesar}}, \bibinfo {author} {\bibfnamefont {T.}~\bibnamefont
  {Dolch}}, \bibinfo {author} {\bibfnamefont {J.~A.}\ \bibnamefont {Ellis}},
  \bibinfo {author} {\bibfnamefont {R.~D.}\ \bibnamefont {Ferdman}}, \bibinfo
  {author} {\bibfnamefont {E.~C.}\ \bibnamefont {Ferrara}}, \bibinfo {author}
  {\bibfnamefont {N.}~\bibnamefont {Garver-Daniels}}, \bibinfo {author}
  {\bibfnamefont {P.~A.}\ \bibnamefont {Gentile}}, \bibinfo {author}
  {\bibfnamefont {M.~L.}\ \bibnamefont {Jones}}, \bibinfo {author}
  {\bibfnamefont {M.~T.}\ \bibnamefont {Lam}}, \bibinfo {author} {\bibfnamefont
  {D.~R.}\ \bibnamefont {Lorimer}}, \bibinfo {author} {\bibfnamefont {R.~S.}\
  \bibnamefont {Lynch}}, \bibinfo {author} {\bibfnamefont {M.~A.}\ \bibnamefont
  {McLaughlin}}, \bibinfo {author} {\bibfnamefont {C.}~\bibnamefont {Ng}},
  \bibinfo {author} {\bibfnamefont {D.~J.}\ \bibnamefont {Nice}}, \bibinfo
  {author} {\bibfnamefont {T.~T.}\ \bibnamefont {Pennucci}}, \bibinfo {author}
  {\bibfnamefont {R.}~\bibnamefont {Spiewak}}, \bibinfo {author} {\bibfnamefont
  {I.~H.}\ \bibnamefont {Stairs}}, \bibinfo {author} {\bibfnamefont
  {K.}~\bibnamefont {Stovall}}, \bibinfo {author} {\bibfnamefont {J.~K.}\
  \bibnamefont {Swiggum}}, \ and\ \bibinfo {author} {\bibfnamefont {W.~W.}\
  \bibnamefont {Zhu}},\ }\href {\doibase 10.1038/s41550-019-0880-2} {\ \textbf
  {\bibinfo {volume} {4}},\ \bibinfo {pages} {72--76} (\bibinfo {year}
  {2019})}\BibitemShut {NoStop}%
\bibitem [{\citenamefont {Fonseca}\ \emph {et~al.}(2021)\citenamefont
  {Fonseca}, \citenamefont {Cromartie}, \citenamefont {Pennucci}, \citenamefont
  {Ray}, \citenamefont {Kirichenko}, \citenamefont {Ransom}, \citenamefont
  {Demorest}, \citenamefont {Stairs}, \citenamefont {Arzoumanian},
  \citenamefont {Guillemot}, \citenamefont {Parthasarathy}, \citenamefont
  {Kerr}, \citenamefont {Cognard}, \citenamefont {Baker}, \citenamefont
  {Blumer}, \citenamefont {Brook}, \citenamefont {DeCesar}, \citenamefont
  {Dolch}, \citenamefont {Dong}, \citenamefont {Ferrara}, \citenamefont
  {Fiore}, \citenamefont {Garver-Daniels}, \citenamefont {Good}, \citenamefont
  {Jennings}, \citenamefont {Jones}, \citenamefont {Kaspi}, \citenamefont
  {Lam}, \citenamefont {Lorimer}, \citenamefont {Luo}, \citenamefont {McEwen},
  \citenamefont {McKee}, \citenamefont {McLaughlin}, \citenamefont {McMann},
  \citenamefont {Meyers}, \citenamefont {Naidu}, \citenamefont {Ng},
  \citenamefont {Nice}, \citenamefont {Pol}, \citenamefont {Radovan},
  \citenamefont {Shapiro-Albert}, \citenamefont {Tan}, \citenamefont
  {Tendulkar}, \citenamefont {Swiggum}, \citenamefont {Wahl},\ and\
  \citenamefont {Zhu}}]{Fonseca_2021}%
  \BibitemOpen
  \bibfield  {author} {\bibinfo {author} {\bibfnamefont {E.}~\bibnamefont
  {Fonseca}}, \bibinfo {author} {\bibfnamefont {H.~T.}\ \bibnamefont
  {Cromartie}}, \bibinfo {author} {\bibfnamefont {T.~T.}\ \bibnamefont
  {Pennucci}}, \bibinfo {author} {\bibfnamefont {P.~S.}\ \bibnamefont {Ray}},
  \bibinfo {author} {\bibfnamefont {A.~Yu.}\ \bibnamefont {Kirichenko}},
  \bibinfo {author} {\bibfnamefont {S.~M.}\ \bibnamefont {Ransom}}, \bibinfo
  {author} {\bibfnamefont {P.~B.}\ \bibnamefont {Demorest}}, \bibinfo {author}
  {\bibfnamefont {I.~H.}\ \bibnamefont {Stairs}}, \bibinfo {author}
  {\bibfnamefont {Z.}~\bibnamefont {Arzoumanian}}, \bibinfo {author}
  {\bibfnamefont {L.}~\bibnamefont {Guillemot}}, \bibinfo {author}
  {\bibfnamefont {A.}~\bibnamefont {Parthasarathy}}, \bibinfo {author}
  {\bibfnamefont {M.}~\bibnamefont {Kerr}}, \bibinfo {author} {\bibfnamefont
  {I.}~\bibnamefont {Cognard}}, \bibinfo {author} {\bibfnamefont {P.~T.}\
  \bibnamefont {Baker}}, \bibinfo {author} {\bibfnamefont {H.}~\bibnamefont
  {Blumer}}, \bibinfo {author} {\bibfnamefont {P.~R.}\ \bibnamefont {Brook}},
  \bibinfo {author} {\bibfnamefont {M.}~\bibnamefont {DeCesar}}, \bibinfo
  {author} {\bibfnamefont {T.}~\bibnamefont {Dolch}}, \bibinfo {author}
  {\bibfnamefont {F.~A.}\ \bibnamefont {Dong}}, \bibinfo {author}
  {\bibfnamefont {E.~C.}\ \bibnamefont {Ferrara}}, \bibinfo {author}
  {\bibfnamefont {W.}~\bibnamefont {Fiore}}, \bibinfo {author} {\bibfnamefont
  {N.}~\bibnamefont {Garver-Daniels}}, \bibinfo {author} {\bibfnamefont
  {D.~C.}\ \bibnamefont {Good}}, \bibinfo {author} {\bibfnamefont
  {R.}~\bibnamefont {Jennings}}, \bibinfo {author} {\bibfnamefont {M.~L.}\
  \bibnamefont {Jones}}, \bibinfo {author} {\bibfnamefont {V.~M.}\ \bibnamefont
  {Kaspi}}, \bibinfo {author} {\bibfnamefont {M.~T.}\ \bibnamefont {Lam}},
  \bibinfo {author} {\bibfnamefont {D.~R.}\ \bibnamefont {Lorimer}}, \bibinfo
  {author} {\bibfnamefont {J.}~\bibnamefont {Luo}}, \bibinfo {author}
  {\bibfnamefont {A.}~\bibnamefont {McEwen}}, \bibinfo {author} {\bibfnamefont
  {J.~W.}\ \bibnamefont {McKee}}, \bibinfo {author} {\bibfnamefont {M.~A.}\
  \bibnamefont {McLaughlin}}, \bibinfo {author} {\bibfnamefont
  {N.}~\bibnamefont {McMann}}, \bibinfo {author} {\bibfnamefont {B.~W.}\
  \bibnamefont {Meyers}}, \bibinfo {author} {\bibfnamefont {A.}~\bibnamefont
  {Naidu}}, \bibinfo {author} {\bibfnamefont {C.}~\bibnamefont {Ng}}, \bibinfo
  {author} {\bibfnamefont {D.~J.}\ \bibnamefont {Nice}}, \bibinfo {author}
  {\bibfnamefont {N.}~\bibnamefont {Pol}}, \bibinfo {author} {\bibfnamefont
  {H.~A.}\ \bibnamefont {Radovan}}, \bibinfo {author} {\bibfnamefont
  {B.}~\bibnamefont {Shapiro-Albert}}, \bibinfo {author} {\bibfnamefont
  {C.~M.}\ \bibnamefont {Tan}}, \bibinfo {author} {\bibfnamefont {S.~P.}\
  \bibnamefont {Tendulkar}}, \bibinfo {author} {\bibfnamefont {J.~K.}\
  \bibnamefont {Swiggum}}, \bibinfo {author} {\bibfnamefont {H.~M.}\
  \bibnamefont {Wahl}}, \ and\ \bibinfo {author} {\bibfnamefont {W.~W.}\
  \bibnamefont {Zhu}},\ }\href {\doibase 10.3847/2041-8213/ac03b8} {\ \textbf
  {\bibinfo {volume} {915}},\ \bibinfo {pages} {L12} (\bibinfo {year}
  {2021})}\BibitemShut {NoStop}%
\bibitem [{\citenamefont {Miller}\ \emph {et~al.}(2021)\citenamefont {Miller},
  \citenamefont {Lamb}, \citenamefont {Dittmann}, \citenamefont {Bogdanov},
  \citenamefont {Arzoumanian}, \citenamefont {Gendreau}, \citenamefont
  {Guillot}, \citenamefont {Ho}, \citenamefont {Lattimer}, \citenamefont
  {Loewenstein}, \citenamefont {Morsink}, \citenamefont {Ray}, \citenamefont
  {Wolff}, \citenamefont {Baker}, \citenamefont {Cazeau}, \citenamefont
  {Manthripragada}, \citenamefont {Markwardt}, \citenamefont {Okajima},
  \citenamefont {Pollard}, \citenamefont {Cognard}, \citenamefont {Cromartie},
  \citenamefont {Fonseca}, \citenamefont {Guillemot}, \citenamefont {Kerr},
  \citenamefont {Parthasarathy}, \citenamefont {Pennucci}, \citenamefont
  {Ransom},\ and\ \citenamefont {Stairs}}]{Miller_2021}%
  \BibitemOpen
  \bibfield  {author} {\bibinfo {author} {\bibfnamefont {M.~C.}\ \bibnamefont
  {Miller}}, \bibinfo {author} {\bibfnamefont {F.~K.}\ \bibnamefont {Lamb}},
  \bibinfo {author} {\bibfnamefont {A.~J.}\ \bibnamefont {Dittmann}}, \bibinfo
  {author} {\bibfnamefont {S.}~\bibnamefont {Bogdanov}}, \bibinfo {author}
  {\bibfnamefont {Z.}~\bibnamefont {Arzoumanian}}, \bibinfo {author}
  {\bibfnamefont {K.~C.}\ \bibnamefont {Gendreau}}, \bibinfo {author}
  {\bibfnamefont {S.}~\bibnamefont {Guillot}}, \bibinfo {author} {\bibfnamefont
  {W.~C.~G.}\ \bibnamefont {Ho}}, \bibinfo {author} {\bibfnamefont {J.~M.}\
  \bibnamefont {Lattimer}}, \bibinfo {author} {\bibfnamefont {M.}~\bibnamefont
  {Loewenstein}}, \bibinfo {author} {\bibfnamefont {S.~M.}\ \bibnamefont
  {Morsink}}, \bibinfo {author} {\bibfnamefont {P.~S.}\ \bibnamefont {Ray}},
  \bibinfo {author} {\bibfnamefont {M.~T.}\ \bibnamefont {Wolff}}, \bibinfo
  {author} {\bibfnamefont {C.~L.}\ \bibnamefont {Baker}}, \bibinfo {author}
  {\bibfnamefont {T.}~\bibnamefont {Cazeau}}, \bibinfo {author} {\bibfnamefont
  {S.}~\bibnamefont {Manthripragada}}, \bibinfo {author} {\bibfnamefont
  {C.~B.}\ \bibnamefont {Markwardt}}, \bibinfo {author} {\bibfnamefont
  {T.}~\bibnamefont {Okajima}}, \bibinfo {author} {\bibfnamefont
  {S.}~\bibnamefont {Pollard}}, \bibinfo {author} {\bibfnamefont
  {I.}~\bibnamefont {Cognard}}, \bibinfo {author} {\bibfnamefont {H.~T.}\
  \bibnamefont {Cromartie}}, \bibinfo {author} {\bibfnamefont {E.}~\bibnamefont
  {Fonseca}}, \bibinfo {author} {\bibfnamefont {L.}~\bibnamefont {Guillemot}},
  \bibinfo {author} {\bibfnamefont {M.}~\bibnamefont {Kerr}}, \bibinfo {author}
  {\bibfnamefont {A.}~\bibnamefont {Parthasarathy}}, \bibinfo {author}
  {\bibfnamefont {T.~T.}\ \bibnamefont {Pennucci}}, \bibinfo {author}
  {\bibfnamefont {S.}~\bibnamefont {Ransom}}, \ and\ \bibinfo {author}
  {\bibfnamefont {I.}~\bibnamefont {Stairs}},\ }\href {\doibase
  10.3847/2041-8213/ac089b} {\ \textbf {\bibinfo {volume} {918}},\ \bibinfo
  {pages} {L28} (\bibinfo {year} {2021})}\BibitemShut {NoStop}%
\bibitem [{\citenamefont {Riley}\ \emph {et~al.}(2021)\citenamefont {Riley},
  \citenamefont {Watts}, \citenamefont {Ray}, \citenamefont {Bogdanov},
  \citenamefont {Guillot}, \citenamefont {Morsink}, \citenamefont {Bilous},
  \citenamefont {Arzoumanian}, \citenamefont {Choudhury}, \citenamefont
  {Deneva}, \citenamefont {Gendreau}, \citenamefont {Harding}, \citenamefont
  {Ho}, \citenamefont {Lattimer}, \citenamefont {Loewenstein}, \citenamefont
  {Ludlam}, \citenamefont {Markwardt}, \citenamefont {Okajima}, \citenamefont
  {Prescod-Weinstein}, \citenamefont {Remillard}, \citenamefont {Wolff},
  \citenamefont {Fonseca}, \citenamefont {Cromartie}, \citenamefont {Kerr},
  \citenamefont {Pennucci}, \citenamefont {Parthasarathy}, \citenamefont
  {Ransom}, \citenamefont {Stairs}, \citenamefont {Guillemot},\ and\
  \citenamefont {Cognard}}]{Riley_2021}%
  \BibitemOpen
  \bibfield  {author} {\bibinfo {author} {\bibfnamefont {Thomas~E.}\
  \bibnamefont {Riley}}, \bibinfo {author} {\bibfnamefont {Anna~L.}\
  \bibnamefont {Watts}}, \bibinfo {author} {\bibfnamefont {Paul~S.}\
  \bibnamefont {Ray}}, \bibinfo {author} {\bibfnamefont {Slavko}\ \bibnamefont
  {Bogdanov}}, \bibinfo {author} {\bibfnamefont {Sebastien}\ \bibnamefont
  {Guillot}}, \bibinfo {author} {\bibfnamefont {Sharon~M.}\ \bibnamefont
  {Morsink}}, \bibinfo {author} {\bibfnamefont {Anna~V.}\ \bibnamefont
  {Bilous}}, \bibinfo {author} {\bibfnamefont {Zaven}\ \bibnamefont
  {Arzoumanian}}, \bibinfo {author} {\bibfnamefont {Devarshi}\ \bibnamefont
  {Choudhury}}, \bibinfo {author} {\bibfnamefont {Julia~S.}\ \bibnamefont
  {Deneva}}, \bibinfo {author} {\bibfnamefont {Keith~C.}\ \bibnamefont
  {Gendreau}}, \bibinfo {author} {\bibfnamefont {Alice~K.}\ \bibnamefont
  {Harding}}, \bibinfo {author} {\bibfnamefont {Wynn C.~G.}\ \bibnamefont
  {Ho}}, \bibinfo {author} {\bibfnamefont {James~M.}\ \bibnamefont {Lattimer}},
  \bibinfo {author} {\bibfnamefont {Michael}\ \bibnamefont {Loewenstein}},
  \bibinfo {author} {\bibfnamefont {Renee~M.}\ \bibnamefont {Ludlam}}, \bibinfo
  {author} {\bibfnamefont {Craig~B.}\ \bibnamefont {Markwardt}}, \bibinfo
  {author} {\bibfnamefont {Takashi}\ \bibnamefont {Okajima}}, \bibinfo {author}
  {\bibfnamefont {Chanda}\ \bibnamefont {Prescod-Weinstein}}, \bibinfo {author}
  {\bibfnamefont {Ronald~A.}\ \bibnamefont {Remillard}}, \bibinfo {author}
  {\bibfnamefont {Michael~T.}\ \bibnamefont {Wolff}}, \bibinfo {author}
  {\bibfnamefont {Emmanuel}\ \bibnamefont {Fonseca}}, \bibinfo {author}
  {\bibfnamefont {H.~Thankful}\ \bibnamefont {Cromartie}}, \bibinfo {author}
  {\bibfnamefont {Matthew}\ \bibnamefont {Kerr}}, \bibinfo {author}
  {\bibfnamefont {Timothy~T.}\ \bibnamefont {Pennucci}}, \bibinfo {author}
  {\bibfnamefont {Aditya}\ \bibnamefont {Parthasarathy}}, \bibinfo {author}
  {\bibfnamefont {Scott}\ \bibnamefont {Ransom}}, \bibinfo {author}
  {\bibfnamefont {Ingrid}\ \bibnamefont {Stairs}}, \bibinfo {author}
  {\bibfnamefont {Lucas}\ \bibnamefont {Guillemot}}, \ and\ \bibinfo {author}
  {\bibfnamefont {Ismael}\ \bibnamefont {Cognard}},\ }\href {\doibase
  10.3847/2041-8213/ac0a81} {\ \textbf {\bibinfo {volume} {918}},\ \bibinfo
  {pages} {L27} (\bibinfo {year} {2021})}\BibitemShut {NoStop}%
\bibitem [{\citenamefont {Lattimer}\ and\ \citenamefont
  {Steiner}(2014)}]{Lattimer:2014sga}%
  \BibitemOpen
  \bibfield  {author} {\bibinfo {author} {\bibfnamefont {James~M.}\
  \bibnamefont {Lattimer}}\ and\ \bibinfo {author} {\bibfnamefont {Andrew~W.}\
  \bibnamefont {Steiner}},\ }\bibfield  {title} {\enquote {\bibinfo {title}
  {{Constraints on the symmetry energy using the mass-radius relation of
  neutron stars}},}\ }\href {\doibase 10.1140/epja/i2014-14040-y} {\bibfield
  {journal} {\bibinfo  {journal} {Eur. Phys. J. A}\ }\textbf {\bibinfo {volume}
  {50}},\ \bibinfo {pages} {40} (\bibinfo {year} {2014})},\ \Eprint
  {http://arxiv.org/abs/1403.1186} {arXiv:1403.1186 [nucl-th]} \BibitemShut
  {NoStop}%
\bibitem [{\citenamefont {Miller}\ \emph {et~al.}(2019)\citenamefont {Miller},
  \citenamefont {Lamb}, \citenamefont {Dittmann}, \citenamefont {Bogdanov},
  \citenamefont {Arzoumanian}, \citenamefont {Gendreau}, \citenamefont
  {Guillot}, \citenamefont {Harding}, \citenamefont {Ho}, \citenamefont
  {Lattimer}, \citenamefont {Ludlam}, \citenamefont {Mahmoodifar},
  \citenamefont {Morsink}, \citenamefont {Ray}, \citenamefont {Strohmayer},
  \citenamefont {Wood}, \citenamefont {Enoto}, \citenamefont {Foster},
  \citenamefont {Okajima}, \citenamefont {Prigozhin},\ and\ \citenamefont
  {Soong}}]{Miller_2019}%
  \BibitemOpen
  \bibfield  {author} {\bibinfo {author} {\bibfnamefont {M.~C.}\ \bibnamefont
  {Miller}}, \bibinfo {author} {\bibfnamefont {F.~K.}\ \bibnamefont {Lamb}},
  \bibinfo {author} {\bibfnamefont {A.~J.}\ \bibnamefont {Dittmann}}, \bibinfo
  {author} {\bibfnamefont {S.}~\bibnamefont {Bogdanov}}, \bibinfo {author}
  {\bibfnamefont {Z.}~\bibnamefont {Arzoumanian}}, \bibinfo {author}
  {\bibfnamefont {K.~C.}\ \bibnamefont {Gendreau}}, \bibinfo {author}
  {\bibfnamefont {S.}~\bibnamefont {Guillot}}, \bibinfo {author} {\bibfnamefont
  {A.~K.}\ \bibnamefont {Harding}}, \bibinfo {author} {\bibfnamefont
  {W.~C.~G.}\ \bibnamefont {Ho}}, \bibinfo {author} {\bibfnamefont {J.~M.}\
  \bibnamefont {Lattimer}}, \bibinfo {author} {\bibfnamefont {R.~M.}\
  \bibnamefont {Ludlam}}, \bibinfo {author} {\bibfnamefont {S.}~\bibnamefont
  {Mahmoodifar}}, \bibinfo {author} {\bibfnamefont {S.~M.}\ \bibnamefont
  {Morsink}}, \bibinfo {author} {\bibfnamefont {P.~S.}\ \bibnamefont {Ray}},
  \bibinfo {author} {\bibfnamefont {T.~E.}\ \bibnamefont {Strohmayer}},
  \bibinfo {author} {\bibfnamefont {K.~S.}\ \bibnamefont {Wood}}, \bibinfo
  {author} {\bibfnamefont {T.}~\bibnamefont {Enoto}}, \bibinfo {author}
  {\bibfnamefont {R.}~\bibnamefont {Foster}}, \bibinfo {author} {\bibfnamefont
  {T.}~\bibnamefont {Okajima}}, \bibinfo {author} {\bibfnamefont
  {G.}~\bibnamefont {Prigozhin}}, \ and\ \bibinfo {author} {\bibfnamefont
  {Y.}~\bibnamefont {Soong}},\ }\href {\doibase 10.3847/2041-8213/ab50c5} {\
  \textbf {\bibinfo {volume} {887}},\ \bibinfo {pages} {L24} (\bibinfo {year}
  {2019})}\BibitemShut {NoStop}%
\bibitem [{\citenamefont {Riley}\ \emph {et~al.}(2019)\citenamefont {Riley},
  \citenamefont {Watts}, \citenamefont {Bogdanov}, \citenamefont {Ray},
  \citenamefont {Ludlam}, \citenamefont {Guillot}, \citenamefont {Arzoumanian},
  \citenamefont {Baker}, \citenamefont {Bilous}, \citenamefont {Chakrabarty},
  \citenamefont {Gendreau}, \citenamefont {Harding}, \citenamefont {Ho},
  \citenamefont {Lattimer}, \citenamefont {Morsink},\ and\ \citenamefont
  {Strohmayer}}]{Riley_2019}%
  \BibitemOpen
  \bibfield  {author} {\bibinfo {author} {\bibfnamefont {T.~E.}\ \bibnamefont
  {Riley}}, \bibinfo {author} {\bibfnamefont {A.~L.}\ \bibnamefont {Watts}},
  \bibinfo {author} {\bibfnamefont {S.}~\bibnamefont {Bogdanov}}, \bibinfo
  {author} {\bibfnamefont {P.~S.}\ \bibnamefont {Ray}}, \bibinfo {author}
  {\bibfnamefont {R.~M.}\ \bibnamefont {Ludlam}}, \bibinfo {author}
  {\bibfnamefont {S.}~\bibnamefont {Guillot}}, \bibinfo {author} {\bibfnamefont
  {Z.}~\bibnamefont {Arzoumanian}}, \bibinfo {author} {\bibfnamefont {C.~L.}\
  \bibnamefont {Baker}}, \bibinfo {author} {\bibfnamefont {A.~V.}\ \bibnamefont
  {Bilous}}, \bibinfo {author} {\bibfnamefont {D.}~\bibnamefont {Chakrabarty}},
  \bibinfo {author} {\bibfnamefont {K.~C.}\ \bibnamefont {Gendreau}}, \bibinfo
  {author} {\bibfnamefont {A.~K.}\ \bibnamefont {Harding}}, \bibinfo {author}
  {\bibfnamefont {W.~C.~G.}\ \bibnamefont {Ho}}, \bibinfo {author}
  {\bibfnamefont {J.~M.}\ \bibnamefont {Lattimer}}, \bibinfo {author}
  {\bibfnamefont {S.~M.}\ \bibnamefont {Morsink}}, \ and\ \bibinfo {author}
  {\bibfnamefont {T.~E.}\ \bibnamefont {Strohmayer}},\ }\href {\doibase
  10.3847/2041-8213/ab481c} {\ \textbf {\bibinfo {volume} {887}},\ \bibinfo
  {pages} {L21} (\bibinfo {year} {2019})}\BibitemShut {NoStop}%
\bibitem [{\citenamefont {Abbott}(2018)}]{PhysRevLett.121.161101}%
  \BibitemOpen
  \bibfield  {author} {\bibinfo {author} {\bibfnamefont {B.~P. et~al.}\
  \bibnamefont {Abbott}} (\bibinfo {collaboration} {The LIGO Scientific
  Collaboration and the Virgo Collaboration}),\ }\bibfield  {title} {\enquote
  {\bibinfo {title} {Gw170817: Measurements of neutron star radii and equation
  of state},}\ }\href {\doibase 10.1103/PhysRevLett.121.161101} {\bibfield
  {journal} {\bibinfo  {journal} {Phys. Rev. Lett.}\ }\textbf {\bibinfo
  {volume} {121}},\ \bibinfo {pages} {161101} (\bibinfo {year}
  {2018})}\BibitemShut {NoStop}%
\bibitem [{\citenamefont {Margueron}\ \emph
  {et~al.}(2018{\natexlab{b}})\citenamefont {Margueron}, \citenamefont
  {Hoffmann~Casali},\ and\ \citenamefont {Gulminelli}}]{Margueron:2017eqc}%
  \BibitemOpen
  \bibfield  {author} {\bibinfo {author} {\bibfnamefont {J\'er\^ome}\
  \bibnamefont {Margueron}}, \bibinfo {author} {\bibfnamefont {Rudiney}\
  \bibnamefont {Hoffmann~Casali}}, \ and\ \bibinfo {author} {\bibfnamefont
  {Francesca}\ \bibnamefont {Gulminelli}},\ }\bibfield  {title} {\enquote
  {\bibinfo {title} {{Equation of state for dense nucleonic matter from
  metamodeling. I. Foundational aspects}},}\ }\href {\doibase
  10.1103/PhysRevC.97.025805} {\bibfield  {journal} {\bibinfo  {journal} {Phys.
  Rev. C}\ }\textbf {\bibinfo {volume} {97}},\ \bibinfo {pages} {025805}
  (\bibinfo {year} {2018}{\natexlab{b}})},\ \Eprint
  {http://arxiv.org/abs/1708.06894} {arXiv:1708.06894 [nucl-th]} \BibitemShut
  {NoStop}%
\bibitem [{\citenamefont {Xie}\ and\ \citenamefont {Li}(2019)}]{Xie:2019sqb}%
  \BibitemOpen
  \bibfield  {author} {\bibinfo {author} {\bibfnamefont {Wen-Jie}\ \bibnamefont
  {Xie}}\ and\ \bibinfo {author} {\bibfnamefont {Bao-An}\ \bibnamefont {Li}},\
  }\bibfield  {title} {\enquote {\bibinfo {title} {{Bayesian Inference of
  High-density Nuclear Symmetry Energy from Radii of Canonical Neutron
  Stars}},}\ }\href {\doibase 10.3847/1538-4357/ab3f37} {\bibfield  {journal}
  {\bibinfo  {journal} {Astrophys. J.}\ }\textbf {\bibinfo {volume} {883}},\
  \bibinfo {pages} {174} (\bibinfo {year} {2019})},\ \Eprint
  {http://arxiv.org/abs/1907.10741} {arXiv:1907.10741 [astro-ph.HE]}
  \BibitemShut {NoStop}%
\bibitem [{\citenamefont {Xie}\ and\ \citenamefont {Li}(2020)}]{Xie:2020tdo}%
  \BibitemOpen
  \bibfield  {author} {\bibinfo {author} {\bibfnamefont {Wen-Jie}\ \bibnamefont
  {Xie}}\ and\ \bibinfo {author} {\bibfnamefont {Bao-An}\ \bibnamefont {Li}},\
  }\bibfield  {title} {\enquote {\bibinfo {title} {{Bayesian Inference of the
  Symmetry Energy of Superdense Neutron-rich Matter from Future Radius
  Measurements of Massive Neutron Stars}},}\ }\href {\doibase
  10.3847/1538-4357/aba271} {\bibfield  {journal} {\bibinfo  {journal}
  {Astrophys. J.}\ }\textbf {\bibinfo {volume} {899}},\ \bibinfo {pages} {4}
  (\bibinfo {year} {2020})},\ \Eprint {http://arxiv.org/abs/2005.07216}
  {arXiv:2005.07216 [astro-ph.HE]} \BibitemShut {NoStop}%
\bibitem [{\citenamefont {Read}\ \emph {et~al.}(2009)\citenamefont {Read},
  \citenamefont {Markakis}, \citenamefont {Shibata}, \citenamefont {Uryu},
  \citenamefont {Creighton},\ and\ \citenamefont {Friedman}}]{Read:2009yp}%
  \BibitemOpen
  \bibfield  {author} {\bibinfo {author} {\bibfnamefont {Jocelyn~S.}\
  \bibnamefont {Read}}, \bibinfo {author} {\bibfnamefont {Charalampos}\
  \bibnamefont {Markakis}}, \bibinfo {author} {\bibfnamefont {Masaru}\
  \bibnamefont {Shibata}}, \bibinfo {author} {\bibfnamefont {Koji}\
  \bibnamefont {Uryu}}, \bibinfo {author} {\bibfnamefont {Jolien D.~E.}\
  \bibnamefont {Creighton}}, \ and\ \bibinfo {author} {\bibfnamefont {John~L.}\
  \bibnamefont {Friedman}},\ }\bibfield  {title} {\enquote {\bibinfo {title}
  {{Measuring the neutron star equation of state with gravitational wave
  observations}},}\ }\href {\doibase 10.1103/PhysRevD.79.124033} {\bibfield
  {journal} {\bibinfo  {journal} {Phys. Rev. D}\ }\textbf {\bibinfo {volume}
  {79}},\ \bibinfo {pages} {124033} (\bibinfo {year} {2009})},\ \Eprint
  {http://arxiv.org/abs/0901.3258} {arXiv:0901.3258 [gr-qc]} \BibitemShut
  {NoStop}%
\bibitem [{\citenamefont {Ozel}\ and\ \citenamefont
  {Psaltis}(2009)}]{Ozel:2009da}%
  \BibitemOpen
  \bibfield  {author} {\bibinfo {author} {\bibfnamefont {Feryal}\ \bibnamefont
  {Ozel}}\ and\ \bibinfo {author} {\bibfnamefont {Dimitrios}\ \bibnamefont
  {Psaltis}},\ }\bibfield  {title} {\enquote {\bibinfo {title} {{Reconstructing
  the Neutron-Star Equation of State from Astrophysical Measurements}},}\
  }\href {\doibase 10.1103/PhysRevD.80.103003} {\bibfield  {journal} {\bibinfo
  {journal} {Phys. Rev. D}\ }\textbf {\bibinfo {volume} {80}},\ \bibinfo
  {pages} {103003} (\bibinfo {year} {2009})},\ \Eprint
  {http://arxiv.org/abs/0905.1959} {arXiv:0905.1959 [astro-ph.HE]} \BibitemShut
  {NoStop}%
\bibitem [{\citenamefont {Steiner}\ \emph {et~al.}(2010)\citenamefont
  {Steiner}, \citenamefont {Lattimer},\ and\ \citenamefont
  {Brown}}]{Steiner:2010fz}%
  \BibitemOpen
  \bibfield  {author} {\bibinfo {author} {\bibfnamefont {Andrew~W.}\
  \bibnamefont {Steiner}}, \bibinfo {author} {\bibfnamefont {James~M.}\
  \bibnamefont {Lattimer}}, \ and\ \bibinfo {author} {\bibfnamefont
  {Edward~F.}\ \bibnamefont {Brown}},\ }\bibfield  {title} {\enquote {\bibinfo
  {title} {{The Equation of State from Observed Masses and Radii of Neutron
  Stars}},}\ }\href {\doibase 10.1088/0004-637X/722/1/33} {\bibfield  {journal}
  {\bibinfo  {journal} {Astrophys. J.}\ }\textbf {\bibinfo {volume} {722}},\
  \bibinfo {pages} {33--54} (\bibinfo {year} {2010})},\ \Eprint
  {http://arxiv.org/abs/1005.0811} {arXiv:1005.0811 [astro-ph.HE]} \BibitemShut
  {NoStop}%
\bibitem [{\citenamefont {Greif}\ \emph {et~al.}(2020)\citenamefont {Greif},
  \citenamefont {Hebeler}, \citenamefont {Lattimer}, \citenamefont {Pethick},\
  and\ \citenamefont {Schwenk}}]{Greif:2020pju}%
  \BibitemOpen
  \bibfield  {author} {\bibinfo {author} {\bibfnamefont {S.~K.}\ \bibnamefont
  {Greif}}, \bibinfo {author} {\bibfnamefont {K.}~\bibnamefont {Hebeler}},
  \bibinfo {author} {\bibfnamefont {J.~M.}\ \bibnamefont {Lattimer}}, \bibinfo
  {author} {\bibfnamefont {C.~J.}\ \bibnamefont {Pethick}}, \ and\ \bibinfo
  {author} {\bibfnamefont {A.}~\bibnamefont {Schwenk}},\ }\bibfield  {title}
  {\enquote {\bibinfo {title} {{Equation of state constraints from nuclear
  physics, neutron star masses, and future moment of inertia measurements}},}\
  }\href {\doibase 10.3847/1538-4357/abaf55} {\bibfield  {journal} {\bibinfo
  {journal} {Astrophys. J.}\ }\textbf {\bibinfo {volume} {901}},\ \bibinfo
  {pages} {155} (\bibinfo {year} {2020})},\ \Eprint
  {http://arxiv.org/abs/2005.14164} {arXiv:2005.14164 [astro-ph.HE]}
  \BibitemShut {NoStop}%
\bibitem [{\citenamefont {Al-Mamun}\ \emph {et~al.}(2021)\citenamefont
  {Al-Mamun}, \citenamefont {Steiner}, \citenamefont {N\"attil\"a},
  \citenamefont {Lange}, \citenamefont {O'Shaughnessy}, \citenamefont {Tews},
  \citenamefont {Gandolfi}, \citenamefont {Heinke},\ and\ \citenamefont
  {Han}}]{Al-Mamun:2020vzu}%
  \BibitemOpen
  \bibfield  {author} {\bibinfo {author} {\bibfnamefont {Mohammad}\
  \bibnamefont {Al-Mamun}}, \bibinfo {author} {\bibfnamefont {Andrew~W.}\
  \bibnamefont {Steiner}}, \bibinfo {author} {\bibfnamefont {Joonas}\
  \bibnamefont {N\"attil\"a}}, \bibinfo {author} {\bibfnamefont {Jacob}\
  \bibnamefont {Lange}}, \bibinfo {author} {\bibfnamefont {Richard}\
  \bibnamefont {O'Shaughnessy}}, \bibinfo {author} {\bibfnamefont {Ingo}\
  \bibnamefont {Tews}}, \bibinfo {author} {\bibfnamefont {Stefano}\
  \bibnamefont {Gandolfi}}, \bibinfo {author} {\bibfnamefont {Craig}\
  \bibnamefont {Heinke}}, \ and\ \bibinfo {author} {\bibfnamefont {Sophia}\
  \bibnamefont {Han}},\ }\bibfield  {title} {\enquote {\bibinfo {title}
  {{Combining Electromagnetic and Gravitational-Wave Constraints on
  Neutron-Star Masses and Radii}},}\ }\href {\doibase
  10.1103/PhysRevLett.126.061101} {\bibfield  {journal} {\bibinfo  {journal}
  {Phys. Rev. Lett.}\ }\textbf {\bibinfo {volume} {126}},\ \bibinfo {pages}
  {061101} (\bibinfo {year} {2021})},\ \Eprint
  {http://arxiv.org/abs/2008.12817} {arXiv:2008.12817 [astro-ph.HE]}
  \BibitemShut {NoStop}%
\bibitem [{\citenamefont {Tews}\ \emph {et~al.}(2018)\citenamefont {Tews},
  \citenamefont {Carlson}, \citenamefont {Gandolfi},\ and\ \citenamefont
  {Reddy}}]{Tews:2018kmu}%
  \BibitemOpen
  \bibfield  {author} {\bibinfo {author} {\bibfnamefont {Ingo}\ \bibnamefont
  {Tews}}, \bibinfo {author} {\bibfnamefont {Joseph}\ \bibnamefont {Carlson}},
  \bibinfo {author} {\bibfnamefont {Stefano}\ \bibnamefont {Gandolfi}}, \ and\
  \bibinfo {author} {\bibfnamefont {Sanjay}\ \bibnamefont {Reddy}},\ }\bibfield
   {title} {\enquote {\bibinfo {title} {{Constraining the speed of sound inside
  neutron stars with chiral effective field theory interactions and
  observations}},}\ }\href {\doibase 10.3847/1538-4357/aac267} {\bibfield
  {journal} {\bibinfo  {journal} {Astrophys. J.}\ }\textbf {\bibinfo {volume}
  {860}},\ \bibinfo {pages} {149} (\bibinfo {year} {2018})},\ \Eprint
  {http://arxiv.org/abs/1801.01923} {arXiv:1801.01923 [nucl-th]} \BibitemShut
  {NoStop}%
\bibitem [{\citenamefont {Annala}\ \emph {et~al.}(2020)\citenamefont {Annala},
  \citenamefont {Gorda}, \citenamefont {Kurkela}, \citenamefont {N\"attil\"a},\
  and\ \citenamefont {Vuorinen}}]{Annala:2019puf}%
  \BibitemOpen
  \bibfield  {author} {\bibinfo {author} {\bibfnamefont {Eemeli}\ \bibnamefont
  {Annala}}, \bibinfo {author} {\bibfnamefont {Tyler}\ \bibnamefont {Gorda}},
  \bibinfo {author} {\bibfnamefont {Aleksi}\ \bibnamefont {Kurkela}}, \bibinfo
  {author} {\bibfnamefont {Joonas}\ \bibnamefont {N\"attil\"a}}, \ and\
  \bibinfo {author} {\bibfnamefont {Aleksi}\ \bibnamefont {Vuorinen}},\
  }\bibfield  {title} {\enquote {\bibinfo {title} {{Evidence for quark-matter
  cores in massive neutron stars}},}\ }\href {\doibase
  10.1038/s41567-020-0914-9} {\bibfield  {journal} {\bibinfo  {journal} {Nature
  Phys.}\ }\textbf {\bibinfo {volume} {16}},\ \bibinfo {pages} {907--910}
  (\bibinfo {year} {2020})},\ \Eprint {http://arxiv.org/abs/1903.09121}
  {arXiv:1903.09121 [astro-ph.HE]} \BibitemShut {NoStop}%
\bibitem [{\citenamefont {Lindblom}(2010)}]{Lindblom:2010bb}%
  \BibitemOpen
  \bibfield  {author} {\bibinfo {author} {\bibfnamefont {Lee}\ \bibnamefont
  {Lindblom}},\ }\bibfield  {title} {\enquote {\bibinfo {title} {{Spectral
  Representations of Neutron-Star Equations of State}},}\ }\href {\doibase
  10.1103/PhysRevD.82.103011} {\bibfield  {journal} {\bibinfo  {journal} {Phys.
  Rev. D}\ }\textbf {\bibinfo {volume} {82}},\ \bibinfo {pages} {103011}
  (\bibinfo {year} {2010})},\ \Eprint {http://arxiv.org/abs/1009.0738}
  {arXiv:1009.0738 [astro-ph.HE]} \BibitemShut {NoStop}%
\bibitem [{\citenamefont {Essick}\ \emph {et~al.}(2021)\citenamefont {Essick},
  \citenamefont {Tews}, \citenamefont {Landry},\ and\ \citenamefont
  {Schwenk}}]{Essick:2021kjb}%
  \BibitemOpen
  \bibfield  {author} {\bibinfo {author} {\bibfnamefont {Reed}\ \bibnamefont
  {Essick}}, \bibinfo {author} {\bibfnamefont {Ingo}\ \bibnamefont {Tews}},
  \bibinfo {author} {\bibfnamefont {Philippe}\ \bibnamefont {Landry}}, \ and\
  \bibinfo {author} {\bibfnamefont {Achim}\ \bibnamefont {Schwenk}},\
  }\bibfield  {title} {\enquote {\bibinfo {title} {{Astrophysical Constraints
  on the Symmetry Energy and the Neutron Skin of Pb208 with Minimal Modeling
  Assumptions}},}\ }\href {\doibase 10.1103/PhysRevLett.127.192701} {\bibfield
  {journal} {\bibinfo  {journal} {Phys. Rev. Lett.}\ }\textbf {\bibinfo
  {volume} {127}},\ \bibinfo {pages} {192701} (\bibinfo {year} {2021})},\
  \Eprint {http://arxiv.org/abs/2102.10074} {arXiv:2102.10074 [nucl-th]}
  \BibitemShut {NoStop}%
\bibitem [{\citenamefont {Lim}\ and\ \citenamefont {Holt}(2018)}]{Lim:2018bkq}%
  \BibitemOpen
  \bibfield  {author} {\bibinfo {author} {\bibfnamefont {Yeunhwan}\
  \bibnamefont {Lim}}\ and\ \bibinfo {author} {\bibfnamefont {Jeremy~W.}\
  \bibnamefont {Holt}},\ }\bibfield  {title} {\enquote {\bibinfo {title}
  {{Neutron star tidal deformabilities constrained by nuclear theory and
  experiment}},}\ }\href {\doibase 10.1103/PhysRevLett.121.062701} {\bibfield
  {journal} {\bibinfo  {journal} {Phys. Rev. Lett.}\ }\textbf {\bibinfo
  {volume} {121}},\ \bibinfo {pages} {062701} (\bibinfo {year} {2018})},\
  \Eprint {http://arxiv.org/abs/1803.02803} {arXiv:1803.02803 [nucl-th]}
  \BibitemShut {NoStop}%
\bibitem [{\citenamefont {Raaijmakers}\ \emph {et~al.}(2021)\citenamefont
  {Raaijmakers}, \citenamefont {Greif}, \citenamefont {Hebeler}, \citenamefont
  {Hinderer}, \citenamefont {Nissanke}, \citenamefont {Schwenk}, \citenamefont
  {Riley}, \citenamefont {Watts}, \citenamefont {Lattimer},\ and\ \citenamefont
  {Ho}}]{Raaijmakers:2021uju}%
  \BibitemOpen
  \bibfield  {author} {\bibinfo {author} {\bibfnamefont {G.}~\bibnamefont
  {Raaijmakers}}, \bibinfo {author} {\bibfnamefont {S.~K.}\ \bibnamefont
  {Greif}}, \bibinfo {author} {\bibfnamefont {K.}~\bibnamefont {Hebeler}},
  \bibinfo {author} {\bibfnamefont {T.}~\bibnamefont {Hinderer}}, \bibinfo
  {author} {\bibfnamefont {S.}~\bibnamefont {Nissanke}}, \bibinfo {author}
  {\bibfnamefont {A.}~\bibnamefont {Schwenk}}, \bibinfo {author} {\bibfnamefont
  {T.~E.}\ \bibnamefont {Riley}}, \bibinfo {author} {\bibfnamefont {A.~L.}\
  \bibnamefont {Watts}}, \bibinfo {author} {\bibfnamefont {J.~M.}\ \bibnamefont
  {Lattimer}}, \ and\ \bibinfo {author} {\bibfnamefont {W.~C.~G.}\ \bibnamefont
  {Ho}},\ }\bibfield  {title} {\enquote {\bibinfo {title} {{Constraints on the
  Dense Matter Equation of State and Neutron Star Properties from
  NICER\textquoteright{}s Mass\textendash{}Radius Estimate of PSR J0740+6620
  and Multimessenger Observations}},}\ }\href {\doibase
  10.3847/2041-8213/ac089a} {\bibfield  {journal} {\bibinfo  {journal}
  {Astrophys. J. Lett.}\ }\textbf {\bibinfo {volume} {918}},\ \bibinfo {pages}
  {L29} (\bibinfo {year} {2021})},\ \Eprint {http://arxiv.org/abs/2105.06981}
  {arXiv:2105.06981 [astro-ph.HE]} \BibitemShut {NoStop}%
\bibitem [{\citenamefont {Malik}\ \emph {et~al.}(2022)\citenamefont {Malik},
  \citenamefont {Ferreira}, \citenamefont {Agrawal},\ and\ \citenamefont
  {Provid\^encia}}]{Malik:2022zol}%
  \BibitemOpen
  \bibfield  {author} {\bibinfo {author} {\bibfnamefont {Tuhin}\ \bibnamefont
  {Malik}}, \bibinfo {author} {\bibfnamefont {M\'arcio}\ \bibnamefont
  {Ferreira}}, \bibinfo {author} {\bibfnamefont {B.~K.}\ \bibnamefont
  {Agrawal}}, \ and\ \bibinfo {author} {\bibfnamefont {Constan\c{c}a}\
  \bibnamefont {Provid\^encia}},\ }\bibfield  {title} {\enquote {\bibinfo
  {title} {{Relativistic Description of Dense Matter Equation of State and
  Compatibility with Neutron Star Observables: A Bayesian Approach}},}\ }\href
  {\doibase 10.3847/1538-4357/ac5d3c} {\bibfield  {journal} {\bibinfo
  {journal} {Astrophys. J.}\ }\textbf {\bibinfo {volume} {930}},\ \bibinfo
  {pages} {17} (\bibinfo {year} {2022})},\ \Eprint
  {http://arxiv.org/abs/2201.12552} {arXiv:2201.12552 [nucl-th]} \BibitemShut
  {NoStop}%
\bibitem [{\citenamefont {Patra}\ \emph {et~al.}(2022)\citenamefont {Patra},
  \citenamefont {Imam}, \citenamefont {Agrawal}, \citenamefont {Mukherjee},\
  and\ \citenamefont {Malik}}]{Patra:2022yqc}%
  \BibitemOpen
  \bibfield  {author} {\bibinfo {author} {\bibfnamefont {N.~K.}\ \bibnamefont
  {Patra}}, \bibinfo {author} {\bibfnamefont {Sk~Md~Adil}\ \bibnamefont
  {Imam}}, \bibinfo {author} {\bibfnamefont {B.~K.}\ \bibnamefont {Agrawal}},
  \bibinfo {author} {\bibfnamefont {Arunava}\ \bibnamefont {Mukherjee}}, \ and\
  \bibinfo {author} {\bibfnamefont {Tuhin}\ \bibnamefont {Malik}},\ }\bibfield
  {title} {\enquote {\bibinfo {title} {{Nearly model-independent constraints on
  dense matter equation of state in a Bayesian approach}},}\ }\href {\doibase
  10.1103/PhysRevD.106.043024} {\bibfield  {journal} {\bibinfo  {journal}
  {Phys. Rev. D}\ }\textbf {\bibinfo {volume} {106}},\ \bibinfo {pages}
  {043024} (\bibinfo {year} {2022})},\ \Eprint
  {http://arxiv.org/abs/2203.08521} {arXiv:2203.08521 [nucl-th]} \BibitemShut
  {NoStop}%
\bibitem [{\citenamefont {Chen}\ \emph {et~al.}(2010)\citenamefont {Chen},
  \citenamefont {Ko}, \citenamefont {Li},\ and\ \citenamefont
  {Xu}}]{Chen:2010qx}%
  \BibitemOpen
  \bibfield  {author} {\bibinfo {author} {\bibfnamefont {Lie-Wen}\ \bibnamefont
  {Chen}}, \bibinfo {author} {\bibfnamefont {Che~Ming}\ \bibnamefont {Ko}},
  \bibinfo {author} {\bibfnamefont {Bao-An}\ \bibnamefont {Li}}, \ and\
  \bibinfo {author} {\bibfnamefont {Jun}\ \bibnamefont {Xu}},\ }\bibfield
  {title} {\enquote {\bibinfo {title} {{Density slope of the nuclear symmetry
  energy from the neutron skin thickness of heavy nuclei}},}\ }\href {\doibase
  10.1103/PhysRevC.82.024321} {\bibfield  {journal} {\bibinfo  {journal} {Phys.
  Rev. C}\ }\textbf {\bibinfo {volume} {82}},\ \bibinfo {pages} {024321}
  (\bibinfo {year} {2010})},\ \Eprint {http://arxiv.org/abs/1004.4672}
  {arXiv:1004.4672 [nucl-th]} \BibitemShut {NoStop}%
\bibitem [{\citenamefont {Xu}\ and\ \citenamefont
  {Papakonstantinou}(2022)}]{PhysRevC.105.044305}%
  \BibitemOpen
  \bibfield  {author} {\bibinfo {author} {\bibfnamefont {Jun}\ \bibnamefont
  {Xu}}\ and\ \bibinfo {author} {\bibfnamefont {Panagiota}\ \bibnamefont
  {Papakonstantinou}},\ }\bibfield  {title} {\enquote {\bibinfo {title}
  {Bayesian inference of finite-nuclei observables based on the kids model},}\
  }\href {\doibase 10.1103/PhysRevC.105.044305} {\bibfield  {journal} {\bibinfo
   {journal} {Phys. Rev. C}\ }\textbf {\bibinfo {volume} {105}},\ \bibinfo
  {pages} {044305} (\bibinfo {year} {2022})}\BibitemShut {NoStop}%
\bibitem [{\citenamefont {Chen}\ and\ \citenamefont
  {Piekarewicz}(2014)}]{PhysRevC.90.044305}%
  \BibitemOpen
  \bibfield  {author} {\bibinfo {author} {\bibfnamefont {Wei-Chia}\
  \bibnamefont {Chen}}\ and\ \bibinfo {author} {\bibfnamefont {J.}~\bibnamefont
  {Piekarewicz}},\ }\bibfield  {title} {\enquote {\bibinfo {title} {Building
  relativistic mean field models for finite nuclei and neutron stars},}\ }\href
  {\doibase 10.1103/PhysRevC.90.044305} {\bibfield  {journal} {\bibinfo
  {journal} {Phys. Rev. C}\ }\textbf {\bibinfo {volume} {90}},\ \bibinfo
  {pages} {044305} (\bibinfo {year} {2014})}\BibitemShut {NoStop}%
\bibitem [{\citenamefont {Papakonstantinou}\ \emph {et~al.}(2018)\citenamefont
  {Papakonstantinou}, \citenamefont {Park}, \citenamefont {Lim},\ and\
  \citenamefont {Hyun}}]{PhysRevC.97.014312}%
  \BibitemOpen
  \bibfield  {author} {\bibinfo {author} {\bibfnamefont {Panagiota}\
  \bibnamefont {Papakonstantinou}}, \bibinfo {author} {\bibfnamefont {Tae-Sun}\
  \bibnamefont {Park}}, \bibinfo {author} {\bibfnamefont {Yeunhwan}\
  \bibnamefont {Lim}}, \ and\ \bibinfo {author} {\bibfnamefont {Chang~Ho}\
  \bibnamefont {Hyun}},\ }\bibfield  {title} {\enquote {\bibinfo {title}
  {Density dependence of the nuclear energy-density functional},}\ }\href
  {\doibase 10.1103/PhysRevC.97.014312} {\bibfield  {journal} {\bibinfo
  {journal} {Phys. Rev. C}\ }\textbf {\bibinfo {volume} {97}},\ \bibinfo
  {pages} {014312} (\bibinfo {year} {2018})}\BibitemShut {NoStop}%
\bibitem [{\citenamefont {Li}\ \emph {et~al.}(2018)\citenamefont {Li},
  \citenamefont {Cai}, \citenamefont {Chen},\ and\ \citenamefont
  {Xu}}]{LI201829}%
  \BibitemOpen
  \bibfield  {author} {\bibinfo {author} {\bibfnamefont {Bao-An}\ \bibnamefont
  {Li}}, \bibinfo {author} {\bibfnamefont {Bao-Jun}\ \bibnamefont {Cai}},
  \bibinfo {author} {\bibfnamefont {Lie-Wen}\ \bibnamefont {Chen}}, \ and\
  \bibinfo {author} {\bibfnamefont {Jun}\ \bibnamefont {Xu}},\ }\bibfield
  {title} {\enquote {\bibinfo {title} {Nucleon effective masses in neutron-rich
  matter},}\ }\href {\doibase https://doi.org/10.1016/j.ppnp.2018.01.001}
  {\bibfield  {journal} {\bibinfo  {journal} {Progress in Particle and Nuclear
  Physics}\ }\textbf {\bibinfo {volume} {99}},\ \bibinfo {pages} {29--119}
  (\bibinfo {year} {2018})}\BibitemShut {NoStop}%
\bibitem [{\citenamefont {Dutra}\ \emph {et~al.}(2014)\citenamefont {Dutra},
  \citenamefont {Louren\ifmmode~\mbox{\c{c}}\else \c{c}\fi{}o}, \citenamefont
  {Avancini}, \citenamefont {Carlson}, \citenamefont {Delfino}, \citenamefont
  {Menezes}, \citenamefont {Provid\^encia}, \citenamefont {Typel},\ and\
  \citenamefont {Stone}}]{PhysRevC.90.055203}%
  \BibitemOpen
  \bibfield  {author} {\bibinfo {author} {\bibfnamefont {M.}~\bibnamefont
  {Dutra}}, \bibinfo {author} {\bibfnamefont {O.}~\bibnamefont
  {Louren\ifmmode~\mbox{\c{c}}\else \c{c}\fi{}o}}, \bibinfo {author}
  {\bibfnamefont {S.~S.}\ \bibnamefont {Avancini}}, \bibinfo {author}
  {\bibfnamefont {B.~V.}\ \bibnamefont {Carlson}}, \bibinfo {author}
  {\bibfnamefont {A.}~\bibnamefont {Delfino}}, \bibinfo {author} {\bibfnamefont
  {D.~P.}\ \bibnamefont {Menezes}}, \bibinfo {author} {\bibfnamefont
  {C.}~\bibnamefont {Provid\^encia}}, \bibinfo {author} {\bibfnamefont
  {S.}~\bibnamefont {Typel}}, \ and\ \bibinfo {author} {\bibfnamefont {J.~R.}\
  \bibnamefont {Stone}},\ }\bibfield  {title} {\enquote {\bibinfo {title}
  {Relativistic mean-field hadronic models under nuclear matter constraints},}\
  }\href {\doibase 10.1103/PhysRevC.90.055203} {\bibfield  {journal} {\bibinfo
  {journal} {Phys. Rev. C}\ }\textbf {\bibinfo {volume} {90}},\ \bibinfo
  {pages} {055203} (\bibinfo {year} {2014})}\BibitemShut {NoStop}%
\bibitem [{\citenamefont {Xu}\ \emph {et~al.}(2009{\natexlab{a}})\citenamefont
  {Xu}, \citenamefont {Chen}, \citenamefont {Li},\ and\ \citenamefont
  {Ma}}]{Xu:2009vi}%
  \BibitemOpen
  \bibfield  {author} {\bibinfo {author} {\bibfnamefont {Jun}\ \bibnamefont
  {Xu}}, \bibinfo {author} {\bibfnamefont {Lie-Wen}\ \bibnamefont {Chen}},
  \bibinfo {author} {\bibfnamefont {Bao-An}\ \bibnamefont {Li}}, \ and\
  \bibinfo {author} {\bibfnamefont {Hong-Ru}\ \bibnamefont {Ma}},\ }\bibfield
  {title} {\enquote {\bibinfo {title} {{Nuclear constraints on properties of
  neutron star crusts}},}\ }\href {\doibase 10.1088/0004-637X/697/2/1549}
  {\bibfield  {journal} {\bibinfo  {journal} {Astrophys. J.}\ }\textbf
  {\bibinfo {volume} {697}},\ \bibinfo {pages} {1549--1568} (\bibinfo {year}
  {2009}{\natexlab{a}})},\ \Eprint {http://arxiv.org/abs/0901.2309}
  {arXiv:0901.2309 [astro-ph.SR]} \BibitemShut {NoStop}%
\bibitem [{\citenamefont {Xu}\ \emph {et~al.}(2009{\natexlab{b}})\citenamefont
  {Xu}, \citenamefont {Chen}, \citenamefont {Li},\ and\ \citenamefont
  {Ma}}]{PhysRevC.79.035802}%
  \BibitemOpen
  \bibfield  {author} {\bibinfo {author} {\bibfnamefont {Jun}\ \bibnamefont
  {Xu}}, \bibinfo {author} {\bibfnamefont {Lie-Wen}\ \bibnamefont {Chen}},
  \bibinfo {author} {\bibfnamefont {Bao-An}\ \bibnamefont {Li}}, \ and\
  \bibinfo {author} {\bibfnamefont {Hong-Ru}\ \bibnamefont {Ma}},\ }\bibfield
  {title} {\enquote {\bibinfo {title} {Locating the inner edge of the neutron
  star crust using terrestrial nuclear laboratory data},}\ }\href {\doibase
  10.1103/PhysRevC.79.035802} {\bibfield  {journal} {\bibinfo  {journal} {Phys.
  Rev. C}\ }\textbf {\bibinfo {volume} {79}},\ \bibinfo {pages} {035802}
  (\bibinfo {year} {2009}{\natexlab{b}})}\BibitemShut {NoStop}%
\bibitem [{\citenamefont {Link}\ \emph {et~al.}(1999)\citenamefont {Link},
  \citenamefont {Epstein},\ and\ \citenamefont
  {Lattimer}}]{PhysRevLett.83.3362}%
  \BibitemOpen
  \bibfield  {author} {\bibinfo {author} {\bibfnamefont {Bennett}\ \bibnamefont
  {Link}}, \bibinfo {author} {\bibfnamefont {Richard~I.}\ \bibnamefont
  {Epstein}}, \ and\ \bibinfo {author} {\bibfnamefont {James~M.}\ \bibnamefont
  {Lattimer}},\ }\bibfield  {title} {\enquote {\bibinfo {title} {Pulsar
  constraints on neutron star structure and equation of state},}\ }\href
  {\doibase 10.1103/PhysRevLett.83.3362} {\bibfield  {journal} {\bibinfo
  {journal} {Phys. Rev. Lett.}\ }\textbf {\bibinfo {volume} {83}},\ \bibinfo
  {pages} {3362--3365} (\bibinfo {year} {1999})}\BibitemShut {NoStop}%
\bibitem [{\citenamefont {Lattimer}\ and\ \citenamefont
  {Prakash}(2000)}]{LATTIMER2000121}%
  \BibitemOpen
  \bibfield  {author} {\bibinfo {author} {\bibfnamefont {James~M.}\
  \bibnamefont {Lattimer}}\ and\ \bibinfo {author} {\bibfnamefont {Madappa}\
  \bibnamefont {Prakash}},\ }\bibfield  {title} {\enquote {\bibinfo {title}
  {Nuclear matter and its role in supernovae, neutron stars and compact object
  binary mergers},}\ }\href {\doibase
  https://doi.org/10.1016/S0370-1573(00)00019-3} {\bibfield  {journal}
  {\bibinfo  {journal} {Physics Reports}\ }\textbf {\bibinfo {volume}
  {333-334}},\ \bibinfo {pages} {121--146} (\bibinfo {year}
  {2000})}\BibitemShut {NoStop}%
\bibitem [{\citenamefont {{Lattimer}}\ and\ \citenamefont
  {{Prakash}}(2001)}]{2001ApJ...550..426L}%
  \BibitemOpen
  \bibfield  {author} {\bibinfo {author} {\bibfnamefont {J.~M.}\ \bibnamefont
  {{Lattimer}}}\ and\ \bibinfo {author} {\bibfnamefont {M.}~\bibnamefont
  {{Prakash}}},\ }\bibfield  {title} {\enquote {\bibinfo {title} {{Neutron Star
  Structure and the Equation of State}},}\ }\href {\doibase 10.1086/319702}
  {\bibfield  {journal} {\bibinfo  {journal} {\apj}\ }\textbf {\bibinfo
  {volume} {550}},\ \bibinfo {pages} {426--442} (\bibinfo {year} {2001})},\
  \Eprint {http://arxiv.org/abs/astro-ph/0002232} {arXiv:astro-ph/0002232
  [astro-ph]} \BibitemShut {NoStop}%
\bibitem [{\citenamefont {{Baym}}\ \emph {et~al.}(1971)\citenamefont {{Baym}},
  \citenamefont {{Pethick}},\ and\ \citenamefont
  {{Sutherland}}}]{1971ApJ...170..299B}%
  \BibitemOpen
  \bibfield  {author} {\bibinfo {author} {\bibfnamefont {Gordon}\ \bibnamefont
  {{Baym}}}, \bibinfo {author} {\bibfnamefont {Christopher}\ \bibnamefont
  {{Pethick}}}, \ and\ \bibinfo {author} {\bibfnamefont {Peter}\ \bibnamefont
  {{Sutherland}}},\ }\bibfield  {title} {\enquote {\bibinfo {title} {{The
  Ground State of Matter at High Densities: Equation of State and Stellar
  Models}},}\ }\href {\doibase 10.1086/151216} {\bibfield  {journal} {\bibinfo
  {journal} {\apj}\ }\textbf {\bibinfo {volume} {170}},\ \bibinfo {pages} {299}
  (\bibinfo {year} {1971})}\BibitemShut {NoStop}%
\bibitem [{\citenamefont {Iida}\ and\ \citenamefont {Sato}(1997)}]{Iida_1997}%
  \BibitemOpen
  \bibfield  {author} {\bibinfo {author} {\bibfnamefont {Kei}\ \bibnamefont
  {Iida}}\ and\ \bibinfo {author} {\bibfnamefont {Katsuhiko}\ \bibnamefont
  {Sato}},\ }\bibfield  {title} {\enquote {\bibinfo {title} {Spin-down of
  neutron stars and compositional transitions in the cold crustal matter},}\
  }\href {\doibase 10.1086/303685} {\bibfield  {journal} {\bibinfo  {journal}
  {The Astrophysical Journal}\ }\textbf {\bibinfo {volume} {477}},\ \bibinfo
  {pages} {294--312} (\bibinfo {year} {1997})}\BibitemShut {NoStop}%
\bibitem [{\citenamefont {Tews}(2017)}]{Tews:2016ofv}%
  \BibitemOpen
  \bibfield  {author} {\bibinfo {author} {\bibfnamefont {Ingo}\ \bibnamefont
  {Tews}},\ }\bibfield  {title} {\enquote {\bibinfo {title} {{Spectrum of shear
  modes in the neutron-star crust: Estimating the nuclear-physics
  uncertainties}},}\ }\href {\doibase 10.1103/PhysRevC.95.015803} {\bibfield
  {journal} {\bibinfo  {journal} {Phys. Rev. C}\ }\textbf {\bibinfo {volume}
  {95}},\ \bibinfo {pages} {015803} (\bibinfo {year} {2017})},\ \Eprint
  {http://arxiv.org/abs/1607.06998} {arXiv:1607.06998 [nucl-th]} \BibitemShut
  {NoStop}%
\bibitem [{\citenamefont {Lim}\ and\ \citenamefont {Holt}(2017)}]{Lim:2017luh}%
  \BibitemOpen
  \bibfield  {author} {\bibinfo {author} {\bibfnamefont {Yeunhwan}\
  \bibnamefont {Lim}}\ and\ \bibinfo {author} {\bibfnamefont {Jeremy~W.}\
  \bibnamefont {Holt}},\ }\bibfield  {title} {\enquote {\bibinfo {title}
  {{Structure of neutron star crusts from new Skyrme effective interactions
  constrained by chiral effective field theory}},}\ }\href {\doibase
  10.1103/PhysRevC.95.065805} {\bibfield  {journal} {\bibinfo  {journal} {Phys.
  Rev. C}\ }\textbf {\bibinfo {volume} {95}},\ \bibinfo {pages} {065805}
  (\bibinfo {year} {2017})},\ \Eprint {http://arxiv.org/abs/1702.02898}
  {arXiv:1702.02898 [nucl-th]} \BibitemShut {NoStop}%
\bibitem [{\citenamefont {Carreau}\ \emph {et~al.}(2019)\citenamefont
  {Carreau}, \citenamefont {Gulminelli},\ and\ \citenamefont
  {Margueron}}]{Carreau:2019zdy}%
  \BibitemOpen
  \bibfield  {author} {\bibinfo {author} {\bibfnamefont {Thomas}\ \bibnamefont
  {Carreau}}, \bibinfo {author} {\bibfnamefont {Francesca}\ \bibnamefont
  {Gulminelli}}, \ and\ \bibinfo {author} {\bibfnamefont {J\'er\^ome}\
  \bibnamefont {Margueron}},\ }\bibfield  {title} {\enquote {\bibinfo {title}
  {{Bayesian analysis of the crust-core transition with a compressible
  liquid-drop model}},}\ }\href {\doibase 10.1140/epja/i2019-12884-1}
  {\bibfield  {journal} {\bibinfo  {journal} {Eur. Phys. J. A}\ }\textbf
  {\bibinfo {volume} {55}},\ \bibinfo {pages} {188} (\bibinfo {year} {2019})},\
  \Eprint {http://arxiv.org/abs/1902.07032} {arXiv:1902.07032 [nucl-th]}
  \BibitemShut {NoStop}%
\bibitem [{\citenamefont {Newton}\ \emph
  {et~al.}(2022{\natexlab{a}})\citenamefont {Newton}, \citenamefont {Preston},
  \citenamefont {Balliet},\ and\ \citenamefont {Ross}}]{Newton:2021rni}%
  \BibitemOpen
  \bibfield  {author} {\bibinfo {author} {\bibfnamefont {William~G.}\
  \bibnamefont {Newton}}, \bibinfo {author} {\bibfnamefont {Rebecca}\
  \bibnamefont {Preston}}, \bibinfo {author} {\bibfnamefont {Lauren}\
  \bibnamefont {Balliet}}, \ and\ \bibinfo {author} {\bibfnamefont {Michael}\
  \bibnamefont {Ross}},\ }\bibfield  {title} {\enquote {\bibinfo {title} {{From
  neutron skins and neutron matter to the neutron star crust}},}\ }\href
  {\doibase 10.1016/j.physletb.2022.137481} {\bibfield  {journal} {\bibinfo
  {journal} {Phys. Lett. B}\ }\textbf {\bibinfo {volume} {834}},\ \bibinfo
  {pages} {137481} (\bibinfo {year} {2022}{\natexlab{a}})},\ \Eprint
  {http://arxiv.org/abs/2111.07969} {arXiv:2111.07969 [nucl-th]} \BibitemShut
  {NoStop}%
\bibitem [{\citenamefont {Hinderer}(2008)}]{Hinderer:2007mb}%
  \BibitemOpen
  \bibfield  {author} {\bibinfo {author} {\bibfnamefont {Tanja}\ \bibnamefont
  {Hinderer}},\ }\bibfield  {title} {\enquote {\bibinfo {title} {{Tidal Love
  numbers of neutron stars}},}\ }\href {\doibase 10.1086/533487} {\bibfield
  {journal} {\bibinfo  {journal} {Astrophys. J.}\ }\textbf {\bibinfo {volume}
  {677}},\ \bibinfo {pages} {1216--1220} (\bibinfo {year} {2008})},\ \Eprint
  {http://arxiv.org/abs/0711.2420} {arXiv:0711.2420 [astro-ph]} \BibitemShut
  {NoStop}%
\bibitem [{\citenamefont {Hinderer}(2009)}]{Hinderer_2009}%
  \BibitemOpen
  \bibfield  {author} {\bibinfo {author} {\bibfnamefont {Tanja}\ \bibnamefont
  {Hinderer}},\ }\bibfield  {title} {\enquote {\bibinfo {title} {{ERRATUM}:
  {\textquotedblleft}{TIDAL} {LOVE} {NUMBERS} {OF} {NEUTRON}
  {STARS}{\textquotedblright} (2008, {ApJ}, 677, 1216)},}\ }\href {\doibase
  10.1088/0004-637x/697/1/964} {\bibfield  {journal} {\bibinfo  {journal} {The
  Astrophysical Journal}\ }\textbf {\bibinfo {volume} {697}},\ \bibinfo {pages}
  {964--964} (\bibinfo {year} {2009})}\BibitemShut {NoStop}%
\bibitem [{\citenamefont {Postnikov}\ \emph {et~al.}(2010)\citenamefont
  {Postnikov}, \citenamefont {Prakash},\ and\ \citenamefont
  {Lattimer}}]{PhysRevD.82.024016}%
  \BibitemOpen
  \bibfield  {author} {\bibinfo {author} {\bibfnamefont {Sergey}\ \bibnamefont
  {Postnikov}}, \bibinfo {author} {\bibfnamefont {Madappa}\ \bibnamefont
  {Prakash}}, \ and\ \bibinfo {author} {\bibfnamefont {James~M.}\ \bibnamefont
  {Lattimer}},\ }\bibfield  {title} {\enquote {\bibinfo {title} {Tidal love
  numbers of neutron and self-bound quark stars},}\ }\href {\doibase
  10.1103/PhysRevD.82.024016} {\bibfield  {journal} {\bibinfo  {journal} {Phys.
  Rev. D}\ }\textbf {\bibinfo {volume} {82}},\ \bibinfo {pages} {024016}
  (\bibinfo {year} {2010})}\BibitemShut {NoStop}%
\bibitem [{\citenamefont {van Dalen}\ \emph {et~al.}(2005)\citenamefont {van
  Dalen}, \citenamefont {Fuchs},\ and\ \citenamefont
  {Faessler}}]{PhysRevLett.95.022302}%
  \BibitemOpen
  \bibfield  {author} {\bibinfo {author} {\bibfnamefont {E.~N.~E.}\
  \bibnamefont {van Dalen}}, \bibinfo {author} {\bibfnamefont {C.}~\bibnamefont
  {Fuchs}}, \ and\ \bibinfo {author} {\bibfnamefont {Amand}\ \bibnamefont
  {Faessler}},\ }\bibfield  {title} {\enquote {\bibinfo {title} {Effective
  nucleon masses in symmetric and asymmetric nuclear matter},}\ }\href
  {\doibase 10.1103/PhysRevLett.95.022302} {\bibfield  {journal} {\bibinfo
  {journal} {Phys. Rev. Lett.}\ }\textbf {\bibinfo {volume} {95}},\ \bibinfo
  {pages} {022302} (\bibinfo {year} {2005})}\BibitemShut {NoStop}%
\bibitem [{\citenamefont {Tews}\ \emph {et~al.}(2017)\citenamefont {Tews},
  \citenamefont {Lattimer}, \citenamefont {Ohnishi},\ and\ \citenamefont
  {Kolomeitsev}}]{Tews_2017}%
  \BibitemOpen
  \bibfield  {author} {\bibinfo {author} {\bibfnamefont {Ingo}\ \bibnamefont
  {Tews}}, \bibinfo {author} {\bibfnamefont {James~M.}\ \bibnamefont
  {Lattimer}}, \bibinfo {author} {\bibfnamefont {Akira}\ \bibnamefont
  {Ohnishi}}, \ and\ \bibinfo {author} {\bibfnamefont {Evgeni~E.}\ \bibnamefont
  {Kolomeitsev}},\ }\bibfield  {title} {\enquote {\bibinfo {title} {Symmetry
  parameter constraints from a lower bound on neutron-matter energy},}\ }\href
  {\doibase 10.3847/1538-4357/aa8db9} {\bibfield  {journal} {\bibinfo
  {journal} {The Astrophysical Journal}\ }\textbf {\bibinfo {volume} {848}},\
  \bibinfo {pages} {105} (\bibinfo {year} {2017})}\BibitemShut {NoStop}%
\bibitem [{\citenamefont {Zhang}\ \emph {et~al.}(2017)\citenamefont {Zhang},
  \citenamefont {Cai}, \citenamefont {Li}, \citenamefont {Newton},\ and\
  \citenamefont {Xu}}]{Zhang:2017ncy}%
  \BibitemOpen
  \bibfield  {author} {\bibinfo {author} {\bibfnamefont {Nai-Bo}\ \bibnamefont
  {Zhang}}, \bibinfo {author} {\bibfnamefont {Bao-Jun}\ \bibnamefont {Cai}},
  \bibinfo {author} {\bibfnamefont {Bao-An}\ \bibnamefont {Li}}, \bibinfo
  {author} {\bibfnamefont {William~G.}\ \bibnamefont {Newton}}, \ and\ \bibinfo
  {author} {\bibfnamefont {Jun}\ \bibnamefont {Xu}},\ }\bibfield  {title}
  {\enquote {\bibinfo {title} {{How tightly is the nuclear symmetry energy
  constrained by a unitary Fermi gas?}}}\ }\href {\doibase
  10.1007/s41365-017-0336-2} {\bibfield  {journal} {\bibinfo  {journal} {Nucl.
  Sci. Tech.}\ }\textbf {\bibinfo {volume} {28}},\ \bibinfo {pages} {181}
  (\bibinfo {year} {2017})},\ \Eprint {http://arxiv.org/abs/1704.02687}
  {arXiv:1704.02687 [nucl-th]} \BibitemShut {NoStop}%
\bibitem [{\citenamefont {Gil}\ \emph {et~al.}(2019)\citenamefont {Gil},
  \citenamefont {Papakonstantinou}, \citenamefont {Hyun},\ and\ \citenamefont
  {Oh}}]{Gil:2018yah}%
  \BibitemOpen
  \bibfield  {author} {\bibinfo {author} {\bibfnamefont {Hana}\ \bibnamefont
  {Gil}}, \bibinfo {author} {\bibfnamefont {Panagiota}\ \bibnamefont
  {Papakonstantinou}}, \bibinfo {author} {\bibfnamefont {Chang~Ho}\
  \bibnamefont {Hyun}}, \ and\ \bibinfo {author} {\bibfnamefont {Yongseok}\
  \bibnamefont {Oh}},\ }\bibfield  {title} {\enquote {\bibinfo {title} {{From
  homogeneous matter to finite nuclei: Role of the effective mass}},}\ }\href
  {\doibase 10.1103/PhysRevC.99.064319} {\bibfield  {journal} {\bibinfo
  {journal} {Phys. Rev. C}\ }\textbf {\bibinfo {volume} {99}},\ \bibinfo
  {pages} {064319} (\bibinfo {year} {2019})},\ \Eprint
  {http://arxiv.org/abs/1805.11321} {arXiv:1805.11321 [nucl-th]} \BibitemShut
  {NoStop}%
\bibitem [{\citenamefont {Gil}\ \emph {et~al.}(2021)\citenamefont {Gil},
  \citenamefont {Kim}, \citenamefont {Papakonstantinou},\ and\ \citenamefont
  {Hyun}}]{Gil:2020wqs}%
  \BibitemOpen
  \bibfield  {author} {\bibinfo {author} {\bibfnamefont {Hana}\ \bibnamefont
  {Gil}}, \bibinfo {author} {\bibfnamefont {Young-Min}\ \bibnamefont {Kim}},
  \bibinfo {author} {\bibfnamefont {Panagiota}\ \bibnamefont
  {Papakonstantinou}}, \ and\ \bibinfo {author} {\bibfnamefont {Chang~Ho}\
  \bibnamefont {Hyun}},\ }\bibfield  {title} {\enquote {\bibinfo {title}
  {{Constraining the density dependence of the symmetry energy with nuclear
  data and astronomical observations in the Korea-IBS-Daegu-SKKU framework}},}\
  }\href {\doibase 10.1103/PhysRevC.103.034330} {\bibfield  {journal} {\bibinfo
   {journal} {Phys. Rev. C}\ }\textbf {\bibinfo {volume} {103}},\ \bibinfo
  {pages} {034330} (\bibinfo {year} {2021})},\ \Eprint
  {http://arxiv.org/abs/2010.13354} {arXiv:2010.13354 [nucl-th]} \BibitemShut
  {NoStop}%
\bibitem [{\citenamefont {Gil}\ \emph {et~al.}(2022)\citenamefont {Gil},
  \citenamefont {Papakonstantinou},\ and\ \citenamefont {Hyun}}]{Gil:2021ols}%
  \BibitemOpen
  \bibfield  {author} {\bibinfo {author} {\bibfnamefont {Hana}\ \bibnamefont
  {Gil}}, \bibinfo {author} {\bibfnamefont {Panagiota}\ \bibnamefont
  {Papakonstantinou}}, \ and\ \bibinfo {author} {\bibfnamefont {Chang~Ho}\
  \bibnamefont {Hyun}},\ }\bibfield  {title} {\enquote {\bibinfo {title}
  {{Constraints on the curvature of nuclear symmetry energy from recent
  astronomical data within the KIDS framework}},}\ }\href {\doibase
  10.1142/S0218301322500136} {\bibfield  {journal} {\bibinfo  {journal} {Int.
  J. Mod. Phys. E}\ }\textbf {\bibinfo {volume} {31}},\ \bibinfo {pages}
  {2250013} (\bibinfo {year} {2022})},\ \Eprint
  {http://arxiv.org/abs/2110.09802} {arXiv:2110.09802 [nucl-th]} \BibitemShut
  {NoStop}%
\bibitem [{\citenamefont {Yue}\ \emph {et~al.}(2022)\citenamefont {Yue},
  \citenamefont {Chen}, \citenamefont {Zhang},\ and\ \citenamefont
  {Zhou}}]{Yue:2021yfx}%
  \BibitemOpen
  \bibfield  {author} {\bibinfo {author} {\bibfnamefont {Tong-Gang}\
  \bibnamefont {Yue}}, \bibinfo {author} {\bibfnamefont {Lie-Wen}\ \bibnamefont
  {Chen}}, \bibinfo {author} {\bibfnamefont {Zhen}\ \bibnamefont {Zhang}}, \
  and\ \bibinfo {author} {\bibfnamefont {Ying}\ \bibnamefont {Zhou}},\
  }\bibfield  {title} {\enquote {\bibinfo {title} {{Constraints on the symmetry
  energy from PREX-II in the multimessenger era}},}\ }\href {\doibase
  10.1103/PhysRevResearch.4.L022054} {\bibfield  {journal} {\bibinfo  {journal}
  {Phys. Rev. Res.}\ }\textbf {\bibinfo {volume} {4}},\ \bibinfo {pages}
  {L022054} (\bibinfo {year} {2022})},\ \Eprint
  {http://arxiv.org/abs/2102.05267} {arXiv:2102.05267 [nucl-th]} \BibitemShut
  {NoStop}%
\bibitem [{\citenamefont {Newton}\ \emph
  {et~al.}(2022{\natexlab{b}})\citenamefont {Newton}, \citenamefont {Balliet},
  \citenamefont {Budimir}, \citenamefont {Crocombe}, \citenamefont {Douglas},
  \citenamefont {Head}, \citenamefont {Langford}, \citenamefont {Rivera},\ and\
  \citenamefont {Sanford}}]{Newton:2021yru}%
  \BibitemOpen
  \bibfield  {author} {\bibinfo {author} {\bibfnamefont {William~G.}\
  \bibnamefont {Newton}}, \bibinfo {author} {\bibfnamefont {Lauren}\
  \bibnamefont {Balliet}}, \bibinfo {author} {\bibfnamefont {Srdan}\
  \bibnamefont {Budimir}}, \bibinfo {author} {\bibfnamefont {Gabriel}\
  \bibnamefont {Crocombe}}, \bibinfo {author} {\bibfnamefont {Brianna}\
  \bibnamefont {Douglas}}, \bibinfo {author} {\bibfnamefont {Thomas~Blake}\
  \bibnamefont {Head}}, \bibinfo {author} {\bibfnamefont {Zach}\ \bibnamefont
  {Langford}}, \bibinfo {author} {\bibfnamefont {Luis}\ \bibnamefont {Rivera}},
  \ and\ \bibinfo {author} {\bibfnamefont {Josh}\ \bibnamefont {Sanford}},\
  }\bibfield  {title} {\enquote {\bibinfo {title} {{Ensembles of unified crust
  and core equations of state in a nuclear-multimessenger astrophysics
  environment}},}\ }\href {\doibase 10.1140/epja/s10050-022-00710-0} {\bibfield
   {journal} {\bibinfo  {journal} {Eur. Phys. J. A}\ }\textbf {\bibinfo
  {volume} {58}},\ \bibinfo {pages} {69} (\bibinfo {year}
  {2022}{\natexlab{b}})},\ \Eprint {http://arxiv.org/abs/2112.12108}
  {arXiv:2112.12108 [astro-ph.HE]} \BibitemShut {NoStop}%
\bibitem [{\citenamefont {Zhu}\ \emph {et~al.}(2023)\citenamefont {Zhu},
  \citenamefont {Li},\ and\ \citenamefont {Liu}}]{Zhu:2022ibs}%
  \BibitemOpen
  \bibfield  {author} {\bibinfo {author} {\bibfnamefont {Zhenyu}\ \bibnamefont
  {Zhu}}, \bibinfo {author} {\bibfnamefont {Ang}\ \bibnamefont {Li}}, \ and\
  \bibinfo {author} {\bibfnamefont {Tong}\ \bibnamefont {Liu}},\ }\bibfield
  {title} {\enquote {\bibinfo {title} {{A Bayesian Inference of a Relativistic
  Mean-field Model of Neutron Star Matter from Observations of NICER and
  GW170817/AT2017gfo}},}\ }\href {\doibase 10.3847/1538-4357/acac1f} {\bibfield
   {journal} {\bibinfo  {journal} {Astrophys. J.}\ }\textbf {\bibinfo {volume}
  {943}},\ \bibinfo {pages} {163} (\bibinfo {year} {2023})},\ \Eprint
  {http://arxiv.org/abs/2211.02007} {arXiv:2211.02007 [astro-ph.HE]}
  \BibitemShut {NoStop}%
\end{thebibliography}%
\end{document}